\documentclass[prd, aps, superscriptaddress, showpacs]{revtex4}

\usepackage{latexsym, graphicx, color}

\begin{document}

\title{Quantum teleportation between moving detectors} 
\author{Shih-Yuin Lin}
\email{sylin@cc.ncue.edu.tw}
\affiliation{Department of Physics, National Changhua University of Education, Changhua 50007, Taiwan}
\affiliation{Department of Physics and Astronomy, University of Waterloo, Waterloo, Ontario, Canada N2L 3G1}
\author{Chung-Hsien Chou}
\email{chouch@mail.ncku.edu.tw}
\affiliation{Department of Physics, National Cheng Kung University, Tainan 701, Taiwan}
\author{B. L. Hu}
\email{blhu@umd.edu}
\affiliation{Joint Quantum Institute and Maryland Center for Fundamental Physics, \\
University of Maryland, College Park, Maryland 20742-4111, USA}
\date{February 11, 2015}

\begin{abstract}
It is commonly believed that the fidelity of quantum teleportation using localized quantum objects with one party or both accelerated in vacuum would be degraded due to the heat up by the Unruh effect. In this paper we point out that the Unruh effect is not the whole story in accounting for all the relativistic effects in quantum teleportation. First, there could be degradation of fidelity by a common field environment even when both quantum objects are in inertial motion. Second, relativistic effects entering the description of the dynamics such as frame dependence, time dilation, and Doppler shift, already existent in inertial motion, can compete with or even overwhelm the effect due to uniform acceleration in a quantum field. We show it is not true that larger acceleration of an object would necessarily lead to a faster degradation of fidelity. These claims are based on four cases of quantum teleportation we studied using two Unruh--DeWitt detectors coupled via a common quantum field initially in the Minkowski vacuum. We find the quantum entanglement evaluated around the light cone, rather than the conventional ones evaluated on the Minkowski time slices, is the necessary condition for the averaged fidelity of quantum teleportation beating the classical one. These results are useful as a guide to making judicious choices of states and parameter ranges and estimation of the efficiency of quantum teleportation in relativistic quantum systems under environmental influences.
\end{abstract}


\maketitle

\section{Introduction}

Quantum teleportation (QT) is by now quite well recognized as a feature process in the application  of quantum information
\cite{NC00, QTelepExpt1, QTelepExpt2}.
A novel and exclusively quantum process QT is also of basic theoretical interest because it necessitates a proper treatment of quantum measurement and entanglement dynamics in realistic physical conditions, such as  environmental influences. The advent of a new era of quantum sciences and engineering demands more precise understanding and further clarification of such fundamental issues. This includes quantum information and classical information, quantum nonlocality and relativistic locality, and spacelike correlations and causality. The study of these issues in a relativistic setting now belongs to a new field called relativistic quantum information \cite{rqi}.

The first scheme of QT was proposed by Bennett {\it et al.} 
(BBCJPW) \cite{BBCJPW93}, in which an {\it unknown} state of a qubit $C$ is teleported from one spatially localized agent Alice to 
another agent Bob using an entangled pair of qubits $A$ and $B$ prepared in one of the Bell states and shared by Alice and Bob, 
respectively. Such an idea was then adapted to the systems with continuous variables such as harmonic oscillators (HOs) by Vaidman 
\cite{Va94}, who introduced an ideal Einstein-Podolsky-Rosen (EPR) state \cite{EPR35} for the shared entangled pair to teleport an 
unknown coherent state. Braunstein and Kimble (BK) \cite{BK98} generalized Vaidman's scheme from the ideal EPR states with exact 
correlations to squeezed coherent states. In doing so the uncertainty of the measurable quantities has to be considered, which reduces
the degree of entanglement of the $AB$ pair as well as the fidelity of quantum teleportation (FiQT).

Alsing and Milburn made the first attempt of calculating the FiQT between two moving cavities in relativistic motions 
\cite{AM03}--one is at rest (Alice), the other is uniformly accelerated (Bob, called Rob in \cite{AM03} with the initial ``R" for 
``Rindler observer"; we follow this convention in Section \ref{physreal} for a similar setup)
in the Minkowski frame--to see how the fidelity is degraded by the Unruh effect (also see Refs. \cite{SU05, FM05}).
Later Landulfo and Matsas considered a complete BBCJPW QT in a two-level detector qubit model, where Rob's detector is uniformly accelerated and interacting with the quantum field only in a finite duration. They found that the FiQT in the future asymptotic region
using the out state of the entangled pair is indeed reduced by the Unruh effect experienced by Rob \cite{LM09}. Along the line of Ref.
\cite{AM03} Friis {\it et al.} \cite{FF13} studied the role of the dynamical Casimir effect in the QT between cavities in relativistic 
motions.

Alternatively, Shiokawa \cite{Tom09} considered QT in the Unruh--DeWitt (UD) detector theory \cite{Unr76, DeW79} with the agents in motions similar to those in Ref. \cite{AM03}, but based on the BK scheme {\it in the interaction region}: An unknown coherent state of a UD detector with internal HO is teleported from Alice to Bob using an entangled pair of similar UD detectors initially in a two-mode squeezed state and shared by Alice and Bob. Unfortunately, the FiQT considered in Ref. \cite{Tom09} is not the physical one.
More careful consideration is needed to get the correct results \cite{LSCH12}.

Indeed, when considering QT in a fully relativistic system, particularly in the interaction region of the localized
objects and quantum fields, one has to take all the factors listed below into account consistently.

\subsection{Relativistic effects} 

Localized objects in a relativistic system may behave differently when observed in different reference frames:

\paragraph{Frame dependence}

Since quantum entanglement between two spatially localized degrees of freedom is a kind of spacelike correlation in a
quantum state, which depends on reference frames, quantum entanglement of two localized objects separated
in space is frame dependent.

\paragraph{Time dilation}

When two localized objects in uniform motion have a nonzero relative speed, both will perceive the same time dilation of each other
in their rest frame constructed by the radar times and distances.
If one object undergoes some phase of acceleration but the other does not, e.g., the worldlines in the twin problem, then
the time dilations perceived by these two objects will be asymmetric. All these time dilation effects are included
in the proper time parametrization of the worldline of an object localized in space.

\paragraph{Relativistic Doppler shift}

Suppose Alice continuously sends a clock signal periodic in her proper time to Bob, then Bob will see Alice's clock running slower or faster than the one at rest when the received signal is redshifted or blueshifted, depending on their relative motion.

These three basic properties of relativistic quantum systems essential for the consideration of QT have not been properly recognized or explored in detail or depth.

\subsection{Environmental influences}

The qubits or detectors in question are unavoidably coupled with quantum fields, which act as an ubiquitous environment:

\paragraph{Quantum decoherence}

Each qubit or HO can be decohered by virtue of its coupling to a quantum field.  However,  mutual influences  mediated by the field between two localized qubits or HOs when placed in close range can lessen the decoherence on each.

\paragraph{Entanglement dynamics}

The entanglement between two qubits or HOs changes in time as their reduced state evolves.

\paragraph{Unruh effect}

A pointlike object such as a UD detector coupled with a quantum field and uniformly accelerated in the Minkowski vacuum of the field would experience a thermal bath of the field quanta at the Unruh temperature proportional to its proper acceleration 
\footnote{The Unruh effect is both an environmental as well as a kinematic effect,  the former referring to the interaction of a detector with a quantum field and the latter referring to its uniform acceleration (UA).  Both the quantum field and the acceleration aspects can be referred to as relativistic, as the Unruh effect is often referred to. However, in this paper, for the sake of conceptual clarity we will reserve the word ``relativistic" to refer to special relativistic effects between inertial frames, as depicted in the previous subsection plus noninertial effects as in the twin trajectories. The Unruh effect will be referred to specifically for the case of UA in the Minkowski vacuum, noninertial as it certainly is, in which thermality in the detector persists in the duration of UA.}.

\subsection{New issues in dealing with quantum teleportation}

The above factors have been considered earlier in some detail in our study on entanglement dynamics \cite{ASH, LH09, LCH08}, but there are new issues of foundational value that need be included in the consideration of QT.
Below we mention three issues related to relativistic open quantum systems:

\paragraph{Measurement in different frames}

Quantum states make sense only in a given frame in which a Hamiltonian is well defined \cite{AA81, AA84}. Two quantum states of the same system with quantum fields in different frames are directly comparable only on those totally overlapping time slices associated with some moment in each frame. By a measurement local in space, e.g., on a pointlike UD detector coupled with a quantum field, quantum states of the combined system in different frames can be interpreted as if they collapsed on different time slices passing through the same measurement event 
\footnote{We say a field state ``collapses on a hypersurface" if the quantum state of the field degrees of freedom defined on that
hypersurface is collapsed by a projective measurement.}.
Nevertheless, the postmeasurement states will evolve to the same state up to a coordinate transformation when they are compared at some time slice in the future, if the combined system respects relativistic covariance \cite{Lin11a}.

\paragraph{Consistency of entangled pair}

As indicated in the BK scheme, the FiQT could depend on (i) quantum entanglement of the entangled pair and (ii) the consistency of the quantum state of the entangled pair with their initial state. Both would be reduced by the coupling with an environment, and applying an improved protocol of QT may suppress the deflection of (ii).

\paragraph{Comparing FiQT and entanglement}

It is easy to modify the BBCJPW scheme to see that the FiQT of qubits in pure states can be either 1 or 0, depending only on whether the qubit pair is entangled or not. In contrast to qubits in pure states, the best possible FiQT in the BK scheme depends on how strong the HO pair is entangled \cite{MV09}. To compare the degree of entanglement of the entangled pair and the FiQT applying them in relativistic systems, Shiokawa considered a ``pseudofidelity" of QT evaluated on the same time slice for the degree of entanglement by imagining that right at the moment Alice has just performed the joint measurement, Bob gets the information of the outcome from Alice instantaneously and immediately performs the proper local operations on his part of the entangled pair \cite{Tom09, LSCH12}. In reality, classical information needs some time to travel from Alice to Bob, and during the traveling time, Bob's part of the entangled pair keeps evolving, so the physical FiQT will not be equal to the ``pseudofidelity" and is thus incommensurate in general with the degree of entanglement of the entangled pair evaluated on the Minkowski time slice. This feature has been overlooked in the literature.

\subsection{Organization of this paper}

To address all the above issues consistently and thoroughly, we start with the action of a fully relativistic system.
We introduce the model in Sec. II, then derive the formula of the FiQT for our model in Sec. III,
where we discuss the relation between the fidelity and the degree of quantum entanglement of the detector pair.
In Secs. IV to VII, respectively we apply our formulation to four representative cases with Alice at rest and
1) Bob also at rest \cite{LH09},
2) Bob (Rob) uniformly accelerated in a finite period of time \cite{AM03, LCH08},
3) Bob being the traveling twin in the twin problem \cite{RHK}, and
4) Bob undergoing alternating uniform acceleration \cite{DLMH13}.
The trajectories and kinematics of each case can be found in the sample references given above.
Finally we summarize and discuss our findings in Sec. VIII.
In the Appendix we show the consistency of the reduced states of the detectors under the spatially local projective measurements.

\section{Model}
\label{model}

Consider a model with three identical Unruh--DeWitt detectors
$A$, $B$, and $C$ moving in a quantum field $\Phi(x)$ in (3+1)-dimensional Minkowski space.
The internal degrees of freedom $Q^{}_A$, $Q^{}_B$, and $Q^{}_C$ of the pointlike detectors $A$, $B$, and $C$,
respectively, behave like simple harmonic oscillators with mass $m=1$ and natural frequency $\Omega$.
The action of the combined system is given by \cite{LCH08}
\begin{eqnarray}
  S &=& -\int d^4 x \sqrt{-g} {1\over 2}\partial_\mu\Phi(x) \partial^\mu\Phi(x) +
    \sum_{{\bf d}=A,B,C}\int d\tau_{}^{\bf d}\,\, {1\over 2}\left[\left(\partial_{\bf d}Q_{\bf d}\right)^2
    -\Omega_{0}^2 Q_{\bf d}^2\right] \nonumber\\ & &
    + \sum_{{\bf d}=A,B} \lambda_0 \int d^4 x \int d\tau_{\bf d} \,\,
    Q_{\bf d}(\tau_{}^{\bf d})\Phi (x)\delta^4\left(x^{\mu}-z_{\bf d}^{\mu}(\tau_{}^{\bf d})\right),
  \label{Stot1}
\end{eqnarray}
where $\mu =0,1,2,3$; $g_{\mu\rho} = {\rm diag}(-1,1,1,1)$; $\partial^{}_{\bf d}\equiv \partial/\partial \tau_{}^{\bf d}$;
$\tau_{}^A$, $\tau_{}^B$ and $\tau_{}^C$ are proper times for $Q_A$, $Q_B$, and $Q_C$, respectively;
and the lightspeed $c\equiv 1$. The scalar field $\Phi_{\bf x}(t) \equiv \Phi(t,{\bf x}) = \Phi(x)$
is assumed to be massless, and $\lambda_0$ is the coupling constant.
Detectors $A$ and $B$ are held by Alice and Bob, respectively, who may be moving in different ways,
while detector $C$ carries the quantum state to be teleported and goes with the sender.

Suppose the initial state of the combined system defined on the $t=0$ hypersurface in the Minkowski coordinates is a product state
$\hat{\rho}^{}_{\Phi_{\bf x}}\otimes\hat{\rho}^{}_{AB}\otimes\hat{\rho}_C^{(\alpha,r_0)}$, where $\hat{\rho}^{}_{\Phi_{\bf x}} = \left| 0_M\right>\left< 0_M\right|$ is the Minkowski vacuum of the field, $\hat{\rho}^{}_{AB}$ is a two-mode squeezed state of detectors $A$ and $B$, and $\hat{\rho}_C^{(\alpha,r_0)}$ is a squeezed coherent state of detector $C$ with $\alpha=\alpha^{}_R + i\alpha^{}_I$ and $r_0$ the squeezed parameter. In the $(K, \Delta)$ representation \cite{UZ89, Lin11a} (the double Fourier transform of the usual Wigner function, namely the ``Wigner characteristic function" \cite{GZ99}), we express the last two as
\begin{eqnarray}
   & & \rho_C^{(\alpha,r_0)}(K^C, \Delta^C) = 
	\int d\Sigma^C e^{{i\over \hbar}K^C\Sigma^C} 
	 \left. \langle Q^{}_C | \hat{\rho}_C^{(\alpha, r_0)} | 
	 Q'_C\rangle \right|_{Q^{}_C,\, Q'_C=\Sigma^C {\mp}(\Delta^C/2)} 
	\nonumber\\ &=& \exp \left[ {-1\over 2\hbar}
	 {\left( {1\over 2\Omega}e^{2r_0} (K^C)^2+{\Omega\over 2}e^{-2r_0}(\Delta^C)^2 \right)}
	 +{i\over \hbar} \left(\sqrt{2\hbar\over\Omega} \alpha^{}_R K^C - \sqrt{2\hbar\Omega}\alpha^{}_I \Delta^C \right) \right],
\label{rhoCI}
\end{eqnarray}
and \begin{eqnarray}
  & & \rho^{}_{AB}(K^A, K^B, \Delta^A, \Delta^B) \nonumber\\ &=&
  \exp - {1\over 8}\left[ {1\over \bar{\beta}^2}(K^A+K^B)^2 + {\bar{\beta}^2\over\hbar^2} (\Delta^A +\Delta^B)^2
	+ {\bar{\alpha}^2\over\hbar^2} (K^A - K^B)^2+ {1\over\bar{\alpha}^2} (\Delta^A -\Delta^B)^2 \right]
\label{rhoABI}
\end{eqnarray}
with parameters $\bar{\alpha}$ and $\bar{\beta}$.
One may choose $\bar{\alpha}=e^{-r_1}\sqrt{\hbar/\Omega}$ and $\bar{\beta}=e^{-r_1}\sqrt{\hbar\Omega}$,
where $r_1$ is the squeezed parameter. As $r_1\to \infty$, $\rho^{}_{AB}$ goes to an ideal EPR state with the correlations
$\langle \hat{Q}_A-\hat{Q}_B \rangle= \langle \hat{P}_A+\hat{P}_B \rangle=0$ without uncertainty,
while $Q_A+Q_B$ and $P_A-P_B$ are totally uncertain. Here, $P_{\bf d}$ is the conjugate momentum to $Q_{\bf d}$.

In general the factors in $\rho_C^{(\alpha,r_0)}(K^C, \Delta^C)$ will vary in time. To concentrate on the best FiQT that
the entangled $AB$ pair can offer, however, we follow Ref. \cite{Tom09} and assume the dynamics of $\rho_C^{(\alpha,r_0)}$ is frozen
or, equivalently, assume $\rho_C^{(\alpha,r_0)}$ is created just before teleportation.

At $t=0$ in the Minkowski frame, the detectors $A$ and $B$ start to couple with the field,
while the detector $C$ is isolated from others.
By virtue of the linearity of the combined system $(\ref{Stot1})$, the quantum state of the combined system started
with a Gaussian state will always be Gaussian, and therefore the reduced state of the three detectors is Gaussian for all times.
In the $(K,\Delta)$ representation the reduced Wigner function at the coordinate time $x^0=T$ in the reference frame
of some observer has the form
\begin{eqnarray}
  & &\rho^{}_{ABC}({\bf K},{\bf \Delta};T) = \exp\left[ {i\over \hbar} \sum_{{\bf d}}
    \left( \langle \hat{Q}_{\bf d}(T)\rangle K^{\bf d} -
     \langle \hat{P}_{\bf d}(T)\rangle \Delta^{\bf d}\right)   \right.\nonumber\\  & & \left.
  -{1\over 2\hbar^2} \sum_{{\bf d},{\bf d'}} \left( K^{\bf d} {\cal Q}_{{\bf d}{\bf d'}}(T) K^{\bf d'}
  +\Delta^{\bf d} {\cal P}_{{\bf d}{\bf d'}}(T) \Delta^{\bf d'} - 2 K^{\bf d} {\cal R}_{{\bf d}{\bf d'}}(T) \Delta^{\bf d'}\right)
\right],
\label{rhoABC}
\end{eqnarray}
where ${\bf d}, {\bf d'} = A,B,C$, and the factors
\begin{eqnarray}
{\cal Q}_{{\bf d}{\bf d'}}(T) &=& \left.{\hbar\delta\over i\delta K^{\bf d}} {\hbar\delta\over i\delta K^{\bf d'}}
  \rho^{}_{ABC}
  \right|_{{\bf K}={\bf \Delta}=0} =
  \langle \delta \hat{Q}_{\bf d}(\tau_{\bf d}(T)), \delta \hat{Q}_{\bf d'}(\tau_{\bf d'}(T)) \rangle, \label{Qdd1T}\\
{\cal P}_{{\bf d}{\bf d'}}(T) &=& \left.{i\hbar\delta\over \delta \Delta^{\bf d}} {i\hbar\delta\over \delta \Delta^{\bf d'}}
  \rho^{}_{ABC}
  \right|_{{\bf K}={\bf \Delta}=0} =
  \langle \delta \hat{P}_{\bf d}(\tau_{\bf d}(T)), \delta \hat{P}_{\bf d'}(\tau_{\bf d'}(T)) \rangle, \label{Pdd1T}\\
{\cal R}_{{\bf d}{\bf d'}}(T) &=& \left. {\hbar\delta\over i\delta K^{\bf d}} {i\hbar\delta\over \delta \Delta^{\bf d'}}
  \rho^{}_{ABC}
  \right|_{{\bf K}={\bf \Delta}=0} =
  \langle \delta \hat{Q}_{\bf d}(\tau_{\bf d}(T)), \delta \hat{P}_{\bf d'}(\tau_{\bf d'}(T)) \rangle, \label{Rdd1T}
\end{eqnarray}
are actually those symmetric two-point correlators of the detectors in their covariance matrices
($\langle \hat{\cal O}, \hat{\cal O}'\rangle\equiv \langle \hat{\cal O}\hat{\cal O}'+\hat{\cal O}'\hat{\cal O}\rangle/2$
and $\delta \hat{\cal O} \equiv \hat{\cal O}-\langle\hat{\cal O}\rangle$),
which can be obtained in the Heisenberg picture by taking the expectation values of
the evolving operators with respect to the initial state defined on the fiducial time slice.

\section{Fidelity of Quantum Teleportation and Entanglement}
\label{FiQTEnLC}

For our later use, below we reexpress and generalize the definitions and calculations for QT
of a Gaussian state from Alice to Bob in Refs. \cite{CKW02, Fi02} in terms of the ($K$, $\Delta$) representation.
Suppose the reduced state of the three detectors continuously evolves to $\rho^{}_{ABC}({\bf K},{\bf \Delta};t_1)$ in the Minkowski
frame when Alice's and Bob's proper times are $\tau^A_1\equiv \tau^A(t_1)$ and $\tau^B_1\equiv\tau_{}^B(t_1)$, respectively.
At this moment Alice preforms a joint Gaussian measurement locally in space on $A$ and $C$ so that the postmeasurement
state right after $t_1$ in the Minkowski frame becomes $\tilde{\rho}^{}_{ABC}({\bf K},{\bf \Delta};t_1)= \tilde{\rho}^{(\beta)}_{AC}
(K^A, K^C,\Delta^A, \Delta^C)\tilde{\rho}^{}_{B}(K^B, \Delta^B)$,
where we assume the quantum state of detectors $A$ and $C$ becomes another two-mode squeezed state
\begin{eqnarray}
  \tilde{\rho}^{(\beta)}_{AC}(K^A, K^C,\Delta^A, \Delta^C) &=& \exp\left[
  {i\over \hbar} \left( \sqrt{2\hbar\over\Omega}\beta_R K^C -\sqrt{2\hbar\Omega}\beta_I\Delta^C\right) \right.
   \nonumber\\ & & \left. -{1\over 2\hbar^2}
  \left( K^m \tilde{\cal Q}_{mn} K^n +\Delta^m \tilde{\cal P}_{mn} \Delta^n - 2 K^m \tilde{\cal R}_{mn} \Delta^n\right)\right],
\label{PMSAC}
\end{eqnarray}
with $m,n = A,C$ so that Alice gets the outcome $\beta = \beta_R + i\beta_I$.
(Here and below, the Einstein notation of summing over repeated dummy indices is understood, and $\sum_{m,n}$ is ignored.)
Then Eq. $(\ref{PMSAC})$ yields the reduced state of detector $B$
\begin{eqnarray}
  \tilde{\rho}^{}_B({\cal K}^B) &=& N_B \int {d^2 {\cal K}^C \over 2\pi\hbar}{d^2 {\cal K}^A \over 2\pi\hbar}
  \tilde{\rho}_{AC}^{(\beta) *}({\cal K}^A, {\cal K}^C)\times \nonumber\\ 
	& & \rho^{}_{ABC}({\cal K}^A, {\cal K}^B, {\cal K}^C; t_1),
\label{rhoB}
\end{eqnarray}
right after $\tau^B_1$,
where $N_B$ is the normalization constant, ${\cal K}^{\bf d}\equiv (K^{\bf d}, \Delta^{\bf d})$, and $d^2 {\cal K}^{\bf d} \equiv
dK^{\bf d} d\Delta^{\bf d}$. If we require $1={\rm Tr}^{}_B\, \tilde{\rho}^{}_B$ ($= \tilde{\rho}^{}_B|_{K^B=\Delta^B=0}$), then $N_B$
will depend on $\beta$. Alternatively, following Ref. \cite{Tom09}, we can require $N_B$ to be independent of $\beta$, and then ${\rm Tr}^{}_B\, \tilde{\rho}^{}_B$ will be proportional to the probability $P(\beta)$ of finding detectors $A$ and $C$ in the state $(\ref{PMSAC})$.
Let ${\rm Tr}^{}_B\, \tilde{\rho}^{}_B = P(\beta)$; then, the normalization condition reads [$\vec{0}\equiv (0,0)$]
\begin{eqnarray}
  1&=&\int d^2\beta P(\beta)= \int d\beta^{}_R d\beta^{}_I \, \tilde{\rho}^{}_B( {\cal K}^B=\vec{0})\nonumber\\
  &=& N_B\int d\beta_R d\beta_I {d^2{\cal K}^A\over 2\pi \hbar}{d^2{\cal K}^C\over 2\pi \hbar} 
    \tilde{\rho}_{AC}^{(\beta) *}({\cal K}^A, {\cal K}^C)\rho^{}_{ABC}( {\cal K}^A,\vec{0}, {\cal K}^C; t_1) \nonumber\\
  &=& N_B \int {d^2{\cal K}^A\over 2\pi \hbar}{d^2{\cal K}^C\over 2\pi \hbar} \rho^{}_{ABC}({\cal K}^A,\vec{0}, {\cal K}^C; t_1)
    2\pi\delta\left(\sqrt{2\over\hbar\Omega}K^C\right)2\pi\delta\left(\sqrt{2\Omega\over\hbar}\Delta^C\right)\times
    \nonumber\\ & & \hspace{1cm}\exp \left[ -{1\over 2\hbar^2}
    \left( K^m \tilde{\cal Q}_{mn} K^n +\Delta^m \tilde{\cal P}_{mn} \Delta^n - 2 K^m \tilde{\cal R}_{mn} \Delta^n\right)\right]
    \nonumber\\
  &=& {N_B\over 2\hbar} \int d^2{\cal K}^A \exp {-1\over 2\hbar^2}
    \left[ \left({\cal Q}_{AA}^{[1]} + \tilde{\cal Q}_{AA}\right) (K^A)^2 + 
    \left({\cal P}_{AA}^{[1]} + \tilde{\cal P}_{AA}\right) (\Delta^A)^2 -
    2 K^A \left({\cal R}_{AA}^{[1]} + \tilde{\cal R}_{AA}\right) \Delta^A\right],\nonumber
\end{eqnarray}
after inserting Eqs. $(\ref{rhoABC})$ and $(\ref{PMSAC})$ into the integrand.
Here, ${\cal S}^{[n]}$ denotes the value of the factor ${\cal S} = {\cal Q}, {\cal P}$, or ${\cal R}$ being taken at
$t_n-\epsilon$ with $\epsilon\to 0+$. Thus, we have
\begin{equation}
  N_B = {1\over\pi\hbar}\sqrt{ \left({\cal Q}_{AA}^{[1]} + \tilde{\cal Q}_{AA}\right)
  \left({\cal P}_{AA}^{[1]} + \tilde{\cal P}_{AA}\right)- \left({\cal R}_{AA}^{[1]} + \tilde{\cal R}_{AA}\right)^2}.
\label{NormB}
\end{equation}

Right after the joint measurement on $A$ and $C$, Alice sends the outcome $\beta$ of the measurement to Bob by a classical signal
at the speed of light. Suppose the signal reaches Bob at 
his proper time $\tau_{}^B = \tau_1^{adv} \equiv \tau_{}^{adv}(t_1)$ 
[here, ``$adv$" stands for ``advanced" \cite{OLMH11}, and $\tau^{adv}$ is the advanced time defined by
$|z^\mu_B(\tau_{}^{adv}(t))- z^\mu_A(t)|^2 = 0$ with $z^0_B(\tau_{}^{adv}(t))> z^0_A(t)$],
when the reduced state of detector $B$ has evolved from the postmeasurement state (\ref{rhoB}) to
$\tilde{\rho}'_B$. According to the information received, Bob could choose a suitable operation on detector $B$ to turn its
quantum state to a copy of the original unknown state carried by detector $C$. In the BK scheme \cite{BK98, Tom09},
the operation Bob should perform is a displacement by $\beta$ in the phase space of detector $B$, 
namely, $\hat{\rho}^{}_{out} = \hat{D}(\beta)\hat{\tilde{\rho}''}^{}_B$,
where $\tilde{\rho}''_B$ is the reduced state of detector $B$ keeps evolving from $\tau_1^{adv}$ 
to the operation event, and $\hat{D}(\beta)$ is the displacement operator, or in the $(K,\Delta)$ representation,
\begin{equation}
  \rho_{out}({\cal K}^B) =
  \tilde{\rho}''_B({\cal K}^B) \exp {i\over\hbar}\left( \sqrt{2\hbar\over\Omega}\beta_R K^B
   -\sqrt{2\hbar\Omega}\beta_I \Delta^B\right).
\label{rhoOut}
\end{equation}
The fidelity of quantum teleportation (FiQT) from $|\alpha,r_0\rangle^{}_C$ to $|\alpha,r_0\rangle^{}_B$ is then defined as
\begin{equation}
  F(\beta) \equiv {{}^{}_B\hspace{-.05cm}\left<
  \right.\alpha, r_0\,|\hat{\rho}_{out}|\,\alpha, r_0\rangle^{}_B\over {\rm Tr}^{}_B\rho^{}_{out} }.
\end{equation}
If we have an ensemble of the distinguishable $ABC$ triplets of the detectors, the quantity we are interested in will be
the {\it averaged} FiQT
\footnote{In Refs. \cite{CKW02, Fi02, MV09} this is simply called ``the fidelity of teleportation," with the averaging understood.},
defined by
\begin{equation}
  F_{av} \equiv \int d^2 \beta P(\beta) F(\beta) = \int d\beta_R d\beta_I
		 {{\rm Tr}^{}_B\tilde{\rho}^{}_{B} \over {\rm Tr}^{}_B\tilde{\rho}''_{B}}  \,\,
   {}^{}_B\hspace{-.05cm}\langle\alpha, r_0 |\hat{\rho}_{out}|\alpha, r_0 \rangle^{}_B ,
\label{FavDef}
\end{equation}
since ${\rm Tr}^{}_B\rho^{}_{out} = \rho^{}_{out}({\cal K}^B=\vec{0})
= \tilde{\rho}''_{B}({\cal K}^B=\vec{0})= {\rm Tr}^{}_B\tilde{\rho}''_{B}$.

\subsection{Direct comparison of FiQT and entanglement}

In Ref. \cite{MV09} Mari and Vitali showed that the {\it optimal} averaged FiQT of a coherent state is bounded above by
\begin{equation}
  F_{opt} \le {1 \over 1+ (2c_-/\hbar)},
\label{MVbound}
\end{equation}
where $c_-$ is the lowest symplectic eigenvalue of the partially transposed covariance matrix in the reduced state of the entangled
$AB$ pair defined on the time slice right before the joint measurement at $t_1$ \cite{VW02, LH09}.
$c_-$ can be related to quantum entanglement of the $AB$ pair by noting that the logarithmic negativity is
given by $E_{\cal N} = {\rm max}\{ 0, -\log_2 (2c_-/\hbar) \}$.
Nevertheless, the dynamics of detector $B$ between Alice's measurement and Bob's operation have been ignored in obtaining the above inequality. In a relativistic open quantum system, (\ref{MVbound}) does not make exact sense, since the averaged FiQT on the left side of (\ref{MVbound}) is a {\it timelike} correlation connecting the joint measurement event by Alice and the operation event by Bob, while the quantity on the right side of (\ref{MVbound}) is a {\it spacelike} correlation extracted from the covariance matrix of detectors $A$ and $B$ defined on the hypersurface of simultaneity right before the wave functional collapses.

\begin{figure}
\includegraphics[width=6cm]{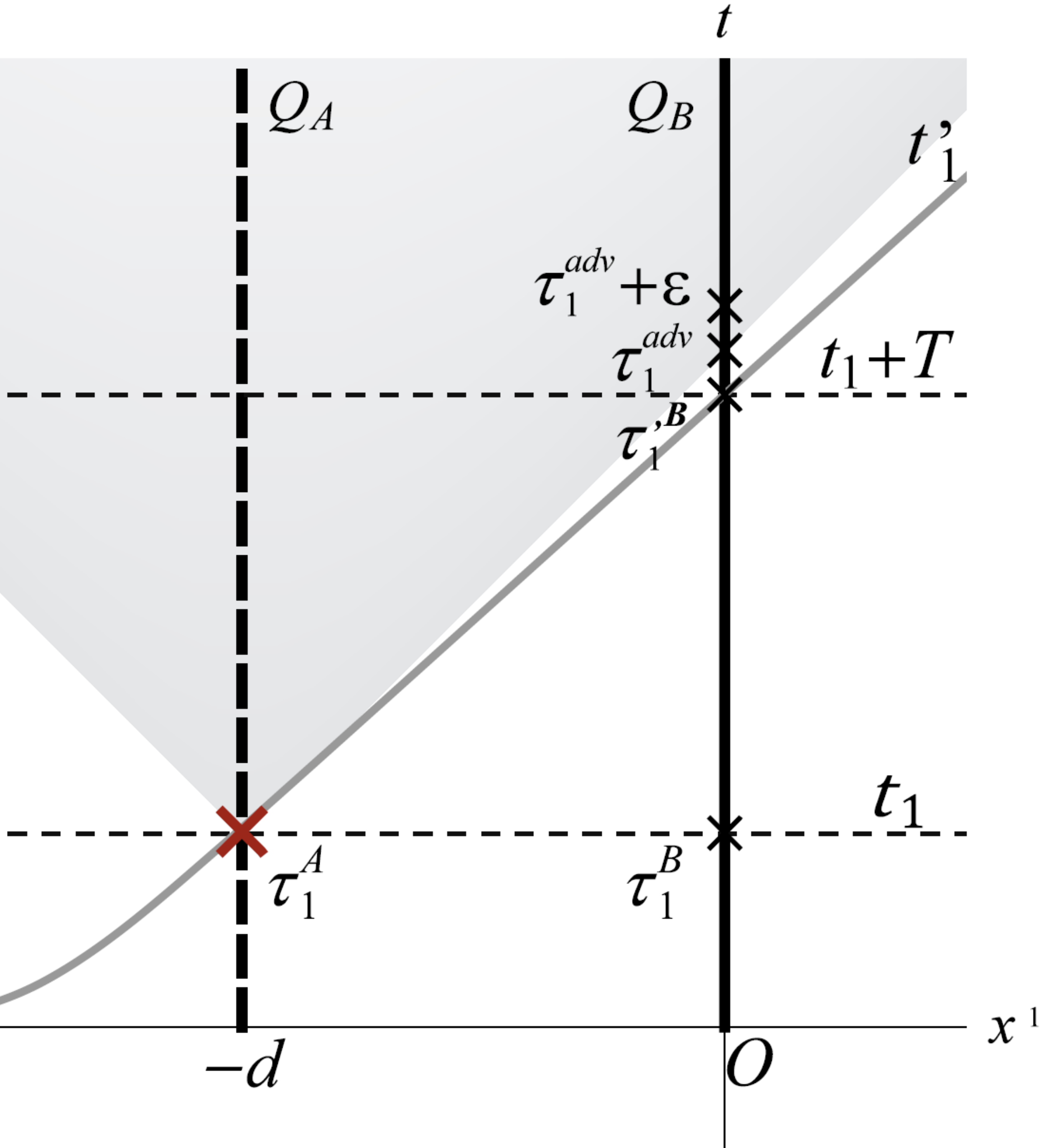}
\caption{Setup for QT from Alice (thick dashed worldline) to Bob (thick solid worldline); both are at rest in the Minskowski vacuum.
The gray solid curve represents the $t_1'$ slice in some coordinate system,
and the gray dashed horizontal lines represent the $t$ slices in the Minkowski coordinates.
The shaded region represents the future light cone of the joint measurement event on $A$ and $C$ by Alice (red cross).}
\label{IHO2}
\end{figure}

To compare the averaged FiQT directly with a function of $c_-$
defined on the $t_1$-slice in the Minkowski frame, one might imagine that Bob receives the
outcome $\beta$ and makes the proper operation on detector $B$ {\it instantaneously} at $\tau^B_1$ when
the worldline of $B$ intersects the $t_1$-slice (see Fig. \ref{IHO2}) \cite{Tom09}, which is unphysical.

A better way to make a direct comparison is to transform the combined system to a new reference frame with the fiducial time slice
overlapping with the $t=0$ hypersurface in our original setup but the time slice passing Alice's measurement event being very close to
the future light cone of the event (e.g., the gray solid curve in Fig. \ref{IHO2} joining Alice's worldline at $\tau^A_1$ and Bob's
worldline at $\tau'^B_1 = \tau_1^{adv}-\epsilon$,  $\epsilon\to 0+$).
Then the wave functional defined in this new reference frame is collapsed around the future light cone of the joint measurement event,
right after which Bob receives the signal from Alice and immediately performs the operation on detector $B$ (at $\tau_1^{adv}+\epsilon$
in Fig. \ref{IHO2}), which is still around the same future light cone and so $\tilde{\rho}''_B\approx \tilde{\rho}'_B \approx
\tilde{\rho}^{}_B$. In this way both sides of (\ref{MVbound}) are evaluated around the future light cone of Alice's measurement event,
or around the past light cone of Bob's operation event, and both sides of (\ref{MVbound}) will be independent of the reference frame in a relativistic detector-field system when $\epsilon\to 0$. In the Appendix 
we show that the reduced state of detector $B$
collapsed around the light cone of the joint measurement event on $A$ and $C$ is consistent with the reduced state initiated with
the one collapsed simultaneously with the measurement event in a conventional reference frame and then evolves to the future
light cone of the event. Actually, the reduced state of detector $B$ at the moment that Bob is crossing the future light cone of Alice's spatially local measurement event is independent of the choice of coordinates here.

Denoting by $t'$ the coordinate time of a new coordinate system such that $\tau^A(t'_1) = \tau^A(t_1) = \tau^A_1$ and $\tau^B(t'_1) = \tau_1^{adv}-\epsilon \equiv \tau'^B_1$ at $t'=t'_1$, and assuming $\rho''_B(\tau_1^{adv}+\epsilon) \approx \rho^{}_B (\tau_1^{adv}-\epsilon)$ as $\epsilon \to 0+$. Then, we can repeat the same approach described earlier in this section to reduce Eq. (\ref{FavDef}) to
\begin{equation}
  F_{av} =\int d\beta_R d\beta_I\,\,{}^{}_B\hspace{-.05cm}\langle\alpha, r_0 |\hat{\rho}_{out}|\alpha, r_0 \rangle^{}_B =
	  \int d\beta_R d\beta_I {d^2{\cal K}^B\over 2\pi \hbar} \rho_B^{(\alpha, r_0) *}({\cal K}^B) \rho_{out}({\cal K}^B),
\end{equation}
where $\hat{\rho}_B^{(\alpha, r_0)}$ in the ($K$, $\Delta$) representation 
is the same as Eq. (\ref{rhoCI}) except the index $C$ there is replaced by $B$.
From Eqs. $(\ref{rhoCI})$ and $(\ref{rhoOut})$, with the help of Eqs. $(\ref{rhoB})$, $(\ref{rhoABC})$ and $(\ref{PMSAC})$,
and with $t_1$ replaced by $t'_1$, we have
\begin{eqnarray}
  F_{av} &=& N_B \int d\beta_R d\beta_I {\prod_{\bf d} d^2{\cal K}^{\bf d} \over (2\pi \hbar)^3}
    \rho^{}_{ABC}({\bf K}, {\bf \Delta}; t'_1)\times \nonumber\\ & &
    \exp\left\{{i\over\hbar}\left[ \sqrt{2\hbar\over \Omega}(\alpha^{}_R-\beta^{}_R)(K^C-K^B)-
    \sqrt{2\hbar\Omega}(\alpha^{}_I-\beta^{}_I)(\Delta^C-\Delta^B)\right] -\right. \nonumber\\
    & &\left. {1\over 2\hbar^2}\left[ {\hbar\over 2\Omega} e^{2r_0} (K^B)^2
    +{\hbar\over 2}\Omega e^{-2r_0} (\Delta^B)^2 + K^m \tilde{\cal Q}_{mn} K^n
		+\Delta^m \tilde{\cal P}_{mn} \Delta^n - 2 K^m \tilde{\cal R}_{mn}\Delta^n\right] \right\}  \nonumber\\ &=&
  N_B \int {\prod_{\bf d} d^2{\cal K}^{\bf d}\over (2\pi \hbar)^3} \rho^{}_{ABC}({\bf K}, {\bf \Delta}; t'_1)
    (2\pi)^2 \delta\left(\sqrt{2\over\hbar \Omega}(K^C-K^B)\right)
    \delta\left(\sqrt{2\Omega \over\hbar}(\Delta^C-\Delta^B)\right) \times \nonumber\\ & & 
    \exp \left\{-{1\over 2\hbar^2}\left[ {\hbar\over 2\Omega}{e^{2r_0}} (K^B)^2+
		{\hbar\over 2}\Omega {-e^{2r_0}}(\Delta^B)^2 + 
     K^m \tilde{\cal Q}_{mn} K^n +\Delta^m \tilde{\cal P}_{mn} \Delta^n - 2 K^m \tilde{\cal R}_{mn}
    \Delta^n\right] \right\}.
\label{FavIntegral}
\end{eqnarray}
Thus,
\begin{equation}
  F_{av}  = {\hbar^2\pi N_B \over \sqrt{\det \tilde{\bf V}}} ,
\label{Favformula}
\end{equation}
{where $N_B$ is the same as Eq. (\ref{NormB}) except $t_1$ is replaced by $t'_1$ and
\begin{equation}
 \hspace{-2cm}
 \tilde{\bf V} = \left(\begin{array}{cccc}
 {\cal Q}_{AA}^{[1']}+\tilde{\cal Q}_{AA} & -{\cal R}_{AA}^{[1']}-\tilde{\cal R}_{AA} &
 {\cal Q}_{AB}^{[1']}+\tilde{\cal Q}_{AC} &  -{\cal R}_{AB}^{[1']}-\tilde{\cal R}_{AC} \\
 -{\cal R}_{AA}^{[1']}-\tilde{\cal R}_{AA} & {\cal P}_{AA}^{[1']}+\tilde{\cal P}_{AA} &
 -{\cal R}_{BA}^{[1']}-\tilde{\cal R}_{CA} &  {\cal P}_{AB}^{[1']}+\tilde{\cal P}_{AC} \\
 {\cal Q}_{AB}^{[1']}+\tilde{\cal Q}_{AC} & -{\cal R}_{BA}^{[1']}-\tilde{\cal R}_{CA} &
 {\cal Q}_{BB}^{[1']}+\tilde{\cal Q}_{CC}+ {\hbar e^{2r_0}\Omega^{-1}} & -{\cal R}_{BB}^{[1']}-\tilde{\cal R}_{CC} \\
 -{\cal R}_{AB}^{[1']}-\tilde{\cal R}_{AC} & {\cal P}_{AB}^{[1']}+\tilde{\cal P}_{AC} &
 -{\cal R}_{BB}^{[1']}-\tilde{\cal R}_{CC} & {\cal P}_{BB}^{[1']}+\tilde{\cal P}_{CC}+ {\hbar e^{-2r_0}\Omega}
 \end{array}\right).
\label{tildeV}
\end{equation}
Here the symmetric two-point correlators of the detectors, e.g., $Q_{\bf dd'}^{[1']}\equiv Q_{\bf dd'}(t'_1) = \langle
\delta \hat{Q}_{\bf d} (\tau^{\bf d}(t'_1)) \delta\hat{Q}_{\bf d'}(\tau^{\bf d'}(t'_1)\rangle$ are the expectation values
of the operators of detector $A$ at $\tau^A_1$ and the operators of detector $B$ at $\tau'^B_1 = \tau_1^{adv}-\epsilon$,
with respect to the initial state of the combined system defined on the fiducial time slice $t'=t=0$.
One can easily write down a similar formula for 
the QT from Bob to Alice by switching their roles and letting detector $C$ go with Bob.}

Note that $F_{av}$ in Eq. $(\ref{Favformula})$ is independent of $\alpha$ only if $\tilde{\rho}_{AC}^{(\beta)}$ is in the form of
Eq. $(\ref{PMSAC})$, where the $\beta$ terms are independent of $K^A$ or $\Delta^A$.
The state $(\ref{PMSAC})$ is chosen so that the analytic calculation is the simplest while the result is still interesting.
One may choose another state consistent with the ideal EPR state as the squeeze parameter $r_2\to\infty$ instead;
for example, $K^C$ and $\Delta^C$ are replaced by $(K^C-K^A)$ and $(\Delta^C+\Delta^A)$,
respectively. Then, $N_B$ and the $F_{av}$ will be more complicated and will depend on $\alpha$.
In practice, the choice of the state may depend on the experimental setting. 

Below we consider the cases with the factors in the two-mode squeezed state $(\ref{PMSAC})$ of detectors $A$ and $C$ right
after the joint measurement given by  $\tilde{\cal Q}_{AA}=\tilde{\cal Q}_{CC}= {\hbar\over 2\Omega}\cosh 2r_2$, $\tilde{\cal Q}_{AC}=
{\hbar\over 2\Omega}\sinh 2r_2$, $\tilde{\cal P}_{AA}=\tilde{\cal P}_{CC}={\hbar\over 2}\Omega\cosh 2 r_2$,
$\tilde{\cal P}_{AC}= -{\hbar\over 2}\Omega\sinh 2r_2$ with squeezed parameter $r_2$, and $\tilde{\cal R}_{mn}=0$.

If the joint measurement on detectors $A$ and $C$ is done perfectly such that $r_2\to \infty$, then
from Eqs. $(\ref{Favformula})$, $(\ref{tildeV})$, and $(\ref{NormB})$, we have
\begin{equation}
  F_{av}(\tau^A_1,\tau'^B_1) \to \hbar\left[ \left( \hbar {e^{2r_0}}\Omega^{-1}+ \langle\delta \hat{Q}_-^2\rangle\right)
  \left(\hbar {e^{-2r_0}} \Omega+  \langle\delta \hat{P}_+^2\rangle\right)
  -\left(\langle\delta \hat{Q}_-,\delta \hat{P}_+\rangle\right)^2\right]^{-1/2},
\end{equation}
where $\hat{Q}_- \equiv \hat{Q}_A(\tau^A_1)- \hat{Q}_B(\tau'^B_1)$ and $\hat{P}_+ \equiv \hat{P}_A(\tau^A_1)+ \hat{P}_B(\tau'^B_1)$.
If, in addition, the initial state $\rho^{}_{AB}$ of detectors $A$ and $B$ in Eq. $(\ref{rhoABI})$ were frozen in time and decoupled from the field, then one would have 
\begin{equation}
  F_{av}(\tau^A_1,\tau'^B_1) =F_{av}(0,0) = {1\over \sqrt{(e^{2 r_0}+e^{-2r_1})(e^{-2r_0}+e^{-2r_1})}},
\end{equation}
which implies that $F_{av}\to 1$ as $r_1 \to \infty$ when $\rho^{}_{AB}$ is nearly an ideal EPR state, while
$F_{av}\to 1/2$ for $r_0=0$ as $r_1 \to 0$ when $\rho^{}_{AB}$ is almost the coherent state of free detectors.
In the latter case $F_{av}= F_{cl}\equiv 1/2$ is known as the best fidelity of ``classical" teleportation of a coherent state
carried by detector $C$ using the coherent state of the $AB$ pair \cite{BK98} without considering the environmental influences.
This does not imply that $F_{av}$ of QT must be greater than $1/2$, though.
Once the correlations such as $\left<Q_-\right>=0$ needed in the protocol of QT become
more uncertain than the minimum quantum uncertainty, $F_{av}-F_{cl}$ will become negative.

The degrees of quantum entanglement of the $AB$ pair in their reduced state defined on the $t'_1$ slice, such as the logarithmic negativity
$E_{\cal N}$, can be evaluated by inserting the expressions for the two-point correlators of detectors $A$ and $B$ on that slice into
the conventional formula \cite{Si00, VW02, LH09}.
Those correlators measure the correlations between the operators of detector $A$ at some event (in Alice's worldline at $\tau^A_1$)
and the operators of detector $B$ at another event almost lightlike but still spacelike separated with the former
(in Bob's worldline at $\tau'^B_1$). We call the quantum entanglement evaluated in this way as the ``entanglement around the light cone"
(EnLC). While the degrees of entanglement of two detectors obtained in the conventional ways depend on the choice of reference frames
\cite{LCH08}, those for the EnLC do not. The inequality (\ref{MVbound}) implies that the EnLC between $A$ and $B$ ($c_- < \hbar/2$
or $E_{\cal N}>0$) is a necessary condition for the averaged FiQT of coherent states to be better than the
classical ones ($F_{opt} > F_{cl}$).

\subsection{Ultraweak coupling limit}
\label{infiweak}

In the ultraweak coupling limit, $\gamma\equiv \lambda_0^2/8\pi$ is so small that $\gamma\Lambda_1 \ll a, \Omega$,
where $\Lambda_1$ corresponds to the time resolution or the frequency cutoff of our model \cite{LH06}.
From Eqs. (28)--(29), (32)--(33), and (B2)--(B8)
in Ref. \cite{LCH08} with $\alpha^2 = (\hbar/\Omega)e^{-2r_1}$ and $\beta^2 = \hbar\Omega e^{-2r_1}$ there
(denoted by $\bar{\alpha}$ and $\bar{\beta}$ in this paper),
with $1 \gg (\gamma\Lambda_1/\Omega) \gg (\gamma/\Omega) \gg (\gamma\Lambda_1/\Omega)^2$, the elements of the covariance matrix for the
$AB$ pair at $t_1'$ with the initial state $(\ref{rhoABI})$ can be approximated by
\begin{eqnarray}
 {\cal Q}^{[1']}_{AA} &\approx& {\hbar C_1\over 2\Omega}e^{-2\gamma \tau^A_1}+\langle (\delta \hat{Q}_A(\tau^A_1))^2\rangle_{\rm v},\hspace{.5cm}
 {\cal P}^{[1']}_{AA} \approx {\hbar \over 2}\Omega C_1 e^{-2\gamma \tau^A_1}+\langle (\delta \hat{P}_A(\tau^A_1))^2\rangle_{\rm v},
  \label{QAAwc}\\
 {\cal Q}^{[1']}_{BB} &\approx& {\hbar C_1\over 2\Omega} e^{-2\gamma \tau'^B_1}+\langle (\delta \hat{Q}_B(\tau'^B_1))^2\rangle_{\rm v},
  \hspace{.5cm}
 {\cal P}^{[1']}_{BB} \approx {\hbar\over 2}\Omega C_1 e^{-2\gamma \tau'^B_1}+\langle (\delta \hat{P}_B(\tau'^B_1))^2\rangle_{\rm v},
  \label{QBBwc} \\
 {\cal Q}^{[1']}_{AB} &\approx& {\hbar S_1\over 2\Omega} e^{-\gamma(\tau^A_1+\tau'^B_1)}\cos\Omega(\tau^A_1+\tau'^B_1), \hspace{.5cm}
 {\cal P}^{[1']}_{AB} \approx -\Omega^2 {\cal Q}^{[1']}_{AB}, \\
 {\cal R}^{[1']}_{AB} &\approx& {\cal R}^{[1']}_{BA} \approx -{\hbar\over 2}S_1 e^{-\gamma(\tau^A_1+\tau'^B_1)}
    \sin\Omega(\tau^A_1+\tau'^B_1), \hspace{.5cm}
 {\cal R}^{[1']}_{AA} \approx {\cal R}^{[1']}_{BB} \approx 0, \label{RAAwc}
\end{eqnarray}
up to $\hbar \cdot O(\gamma/\Omega)$. Here, $C_n \equiv \cosh 2r_n$, $S_n\equiv \sinh 2r_n$, $\langle (\delta \hat{P}_j(\tau^j))^2
\rangle_{\rm v} \approx \Omega^2\langle (\delta \hat{Q}_j(\tau^j))^2\rangle_{\rm v} + \upsilon$ with $j=A,B$, and
$\upsilon \equiv 2\hbar\gamma\Lambda_1/\pi$. For simplicity, let us consider the cases with $r_0=0$ here. Then, Eq. (\ref{tildeV}) becomes
\begin{equation}
  \hspace{-2cm}\tilde{\bf V} =
  \left( \begin{array}{cccc}
    {\hbar\over 2\Omega} {\cal A}(\tau^A_1) & 0 & {\hbar\over 2\Omega}  {\cal X}(\tau^A_1,\tau'^B_1)
		& {\hbar\over 2} {\cal Y}(\tau^A_1,\tau'^B_1) \\
  0 & {\hbar\over 2} \Omega {\cal A}(\tau^A_1)+\upsilon &
    {\hbar\over 2} {\cal Y}(\tau^A_1,\tau'^B_1) & -{\hbar\over 2}\Omega {\cal X}(\tau^A_1,\tau'^B_1)\\ 	
  {\hbar\over 2\Omega} {\cal X}(\tau^A_1,\tau'^B_1) & {\hbar\over 2} {\cal Y}(\tau^A_1,\tau'^B_1) &
	    {\hbar\over 2\Omega} {\cal B}(\tau'^B_1) & 0 \\
  {\hbar\over 2} {\cal Y}(\tau^A_1,\tau'^B_1) & -{\hbar\over 2}\Omega {\cal X}(\tau^A_1,\tau'^B_1) & 0 &
    {\hbar\over 2} \Omega {\cal B}(\tau'^B_1)+\upsilon
\end{array}\right) + \hbar^4 O(\gamma/\Omega),
\label{tildeVwc}
\end{equation}
where
\begin{eqnarray}
  {\cal A}(\tau^A_1) &\equiv& C_2 +e^{-2\gamma \tau^A_1} C_1+2\Omega\hbar^{-1}\langle (\delta\hat{Q}_A(\tau^A_1))^2\rangle_{\rm v},
	    \label{Aoft}\\
  {\cal B}(\tau'^B_1) &\equiv& 2+C_2 + e^{-2\gamma\tau'^B_1}C_1 +2\Omega\hbar^{-1}\langle (\delta\hat{Q}_B(\tau'^B_1))^2\rangle_{\rm v},
	  	\label{BofT}\\
  {\cal X}(\tau^A_1,\tau'^B_1) &\equiv& S_2 + e^{-\gamma(\tau^A_1+\tau'^B_1)} \cos\Omega(\tau^A_1+\tau'^B_1)\, S_1,\\
  {\cal Y}(\tau^A_1,\tau'^B_1) &\equiv& e^{-\gamma(\tau^A_1+\tau'^B_1)}\sin\Omega(\tau^A_1+\tau'^B_1)\,S_1 .
\end{eqnarray}
So the averaged fidelity in the ultraweak coupling limit can be written in a simple form:
\begin{equation}
  F_{av}(\tau^A_1, \tau'^B_1) = {2 {\cal A}\over {\cal AB}-({\cal X}^2+{\cal Y}^2)} + O(\gamma\Lambda_1/\Omega). \label{weakFav}
\end{equation}
Usually, $\langle (\delta \hat{Q}_j)^2(\tau)\rangle_{\rm v} \sim  (\pm e^{-2\gamma \tau} +$ constant$)$ evolve smoothly in this limit,
while
\begin{equation}
  {\cal X}^2+{\cal Y}^2 = S_2^2+ S_1^2\, e^{-2\gamma(\tau^A_1+\tau'^B_1)}+2S_1 S_2 \,
	    e^{-\gamma(\tau^A_1+\tau'^B_1)}\cos\Omega(\tau^A_1+\tau'^B_1)
\label{X2Y2}
\end{equation}
is oscillating in $\tau^A_1+\tau'^B_1$ due to the natural squeeze-antisqueeze oscillation of
the two-mode squeezed state of detectors $A$ and $B$ at frequency $\Omega$ \cite{LSCH12}.
The maximum (minimum) values of $F_{av}$, denoted by $F^+_{av}$ ($F^-_{av}$), occur at
$\cos\Omega(\tau^A_1+\tau'^B_1)\approx 1$ ($-1$), when ${\cal Y}=0$ and
\begin{equation}
  F_{av}^{\pm}(\tau^A_1, \tau'^B_1) \approx {2{\cal A}\over {\cal AB} - \left[S_2\pm S_1 \, e^{-\gamma(\tau^A_1+\tau'^B_1)}\right]^2}.
\label{maxminFav}
\end{equation}
We call $F_{av}^+$ the best averaged FiQT from Alice to Bob.

\subsection{Improved protocol}

Similar to the function of the local oscillators in the optical experiments of QT,
if we perform a counter-rotation to $\tilde{\rho}^{}_B$ in the phase space of $(Q_B, P_B)$
to undo the $\cos \Omega(\tau^A_1+\tau'^B_1)$ or $\sin\Omega(\tau^A_1+\tau'^B_1)$ factors before displacement, namely,
$\hat{\rho}^{}_{out} = \hat{D}(\beta)\hat{R}(\Omega(\tau^A_1+\tau'^B_1))\hat{\tilde{\rho}}^{}_B$,
we will obtain the best averaged FiQT $F_{av}^+$ in the ultraweak coupling limit.

Mathematically, this can be done by transforming $(K_B, \Delta_B)$ to $(C_\Omega K_B + \Omega^{-1}S_\Omega \Delta_B,
C_\Omega \Delta_B -\Omega S_\Omega K_B )$ in Eq. (\ref{rhoB}) for $\tilde{\rho}_B$, where $C_\Omega\equiv\cos\Omega(\tau^A_1+\tau'^B_1)$
and $S_\Omega \equiv \sin \Omega(\tau^A_1+\tau'^B_1)$ \cite{FF13}. Since the detectors $B$ and $C$ are not directly correlated in
$\rho^{}_{ABC}$, the operation of this counter-rotation on detector $B$ commutes with the joint projective measurement on $A$ and $C$.

Physically, this may be realized by having Alice continuously send classical signals periodic in her proper time
to Bob during the whole history, analogous to the local oscillators in optics, so that Bob can determine what $\tau^A_1$ was when the joint measurement on $A$ and $C$ was done, accordingly Bob can counter-rotate detector $B$ for a proper angle $\Omega(\tau^A_1+\tau'^B_1)$ mod $2\pi$ with $\tau'^B_1$ input from his own clock.

Our numerical results show that this improved protocol is almost the optimal according to (\ref{MVbound}),
though in some cases we have to introduce a further squeezing to the coherent state to be teleported
in order to optimize the fidelity [see Fig. \ref{ENFav134} (lower-left)].

After introducing the notations and formalism for QT in relativistic consideration, we will now
examine carefully the special-relativistic effects and the Unruh effect in each of the following four cases.

\section{Case 1---Alice and Bob both at rest: Two inertial detectors}

Let us apply our formulation to the first case, with both Alice and Bob 
at rest in the Minkowski space and separated at a distance $d$, as the setup in Fig. \ref{IHO2}.

\subsection{Late-time behavior}
\label{LT2IHO}

The late-time steady state of detectors $A$ and $B$ is simple, in the sense that there is no natural oscillation in time.
The late-time two-point correlators on the same Minkowski time slice for two UD detectors at rest have been given in Eqs.
(48)--(51) of Ref. \cite{LH09}. In these expressions the mutual influences of detectors $A$ and $B$ to all orders
(more on the mutual influences; see Sec. \ref{ET2IHO}) are included.
From the discussion above Eq. (58) in Ref. \cite{LH09}, one sees that if detectors $A$ and $B$ are close enough ($d < d_{\rm ent}$
with the entanglement distance $d_{\rm ent}$ defined in Ref. \cite{LH09}), at late times these two detectors will have
\begin{equation}
 \langle (\delta\hat{Q}_A - \delta\hat{Q}_B)^2\rangle  \langle (\delta\hat{P}_A +\delta\hat{P}_B)^2 \rangle < \hbar^2,
\end{equation}
with the operators $\hat{Q}_A(t)$, $\hat{P}_A(t)$, $\hat{Q}_B(t)$, and $\hat{P}_B(t)$ at the same Minkowski time $t$
\footnote{The left-hand side of Eq.(58) in Ref. \cite{LH09} and the corresponding expression in the
statements above it should be corrected to $16 {\rm Re} {\cal F}_{0-} {\rm Re} {\cal F}_{2+} - \hbar^2/4$.}.
This implies that the $AB$ pair is in a steady two-mode squeezed state with a phase of $\pi/4$ in the
$Q_AQ_B$ subspace of the phase space, and so we may be allowed to apply the protocol in Sec. \ref{FiQTEnLC} to obtain an
averaged FiQT of a coherent state from Alice to Bob or from Bob to Alice,
\begin{equation}
 F_{av} \approx {1\over 1 + 2 \hbar^{-1} \sqrt{\langle (\delta\hat{Q}_-)^2\rangle
  \langle (\delta\hat{P}_+)^2 \rangle /4 }} > F_{cl} \equiv {1\over 2}
\end{equation}
in the weak coupling limit according to (\ref{MVbound}) and beat the classical fidelity $F_{cl}$.

\begin{figure}
\includegraphics[width=6.5cm]{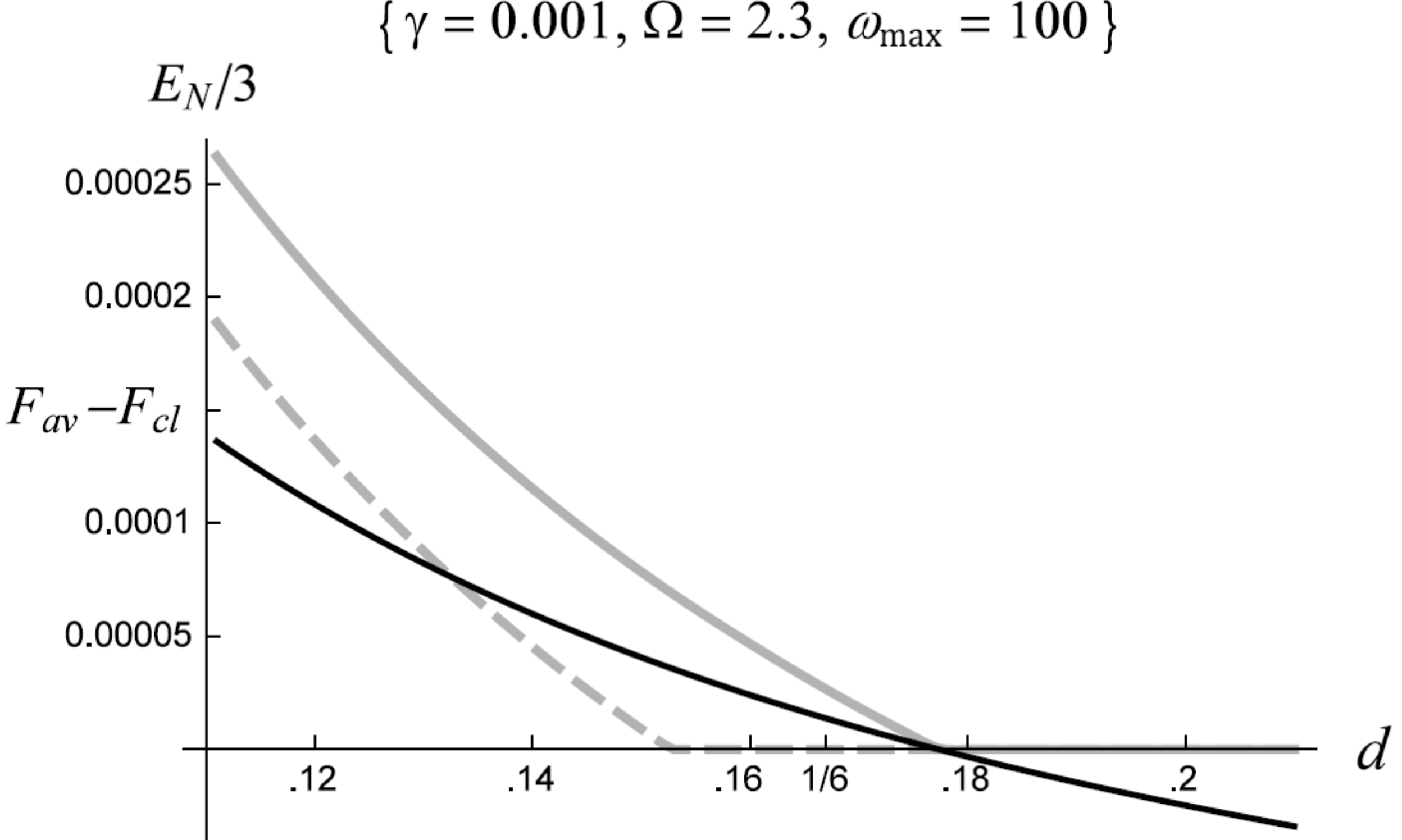}
\includegraphics[width=6.5cm]{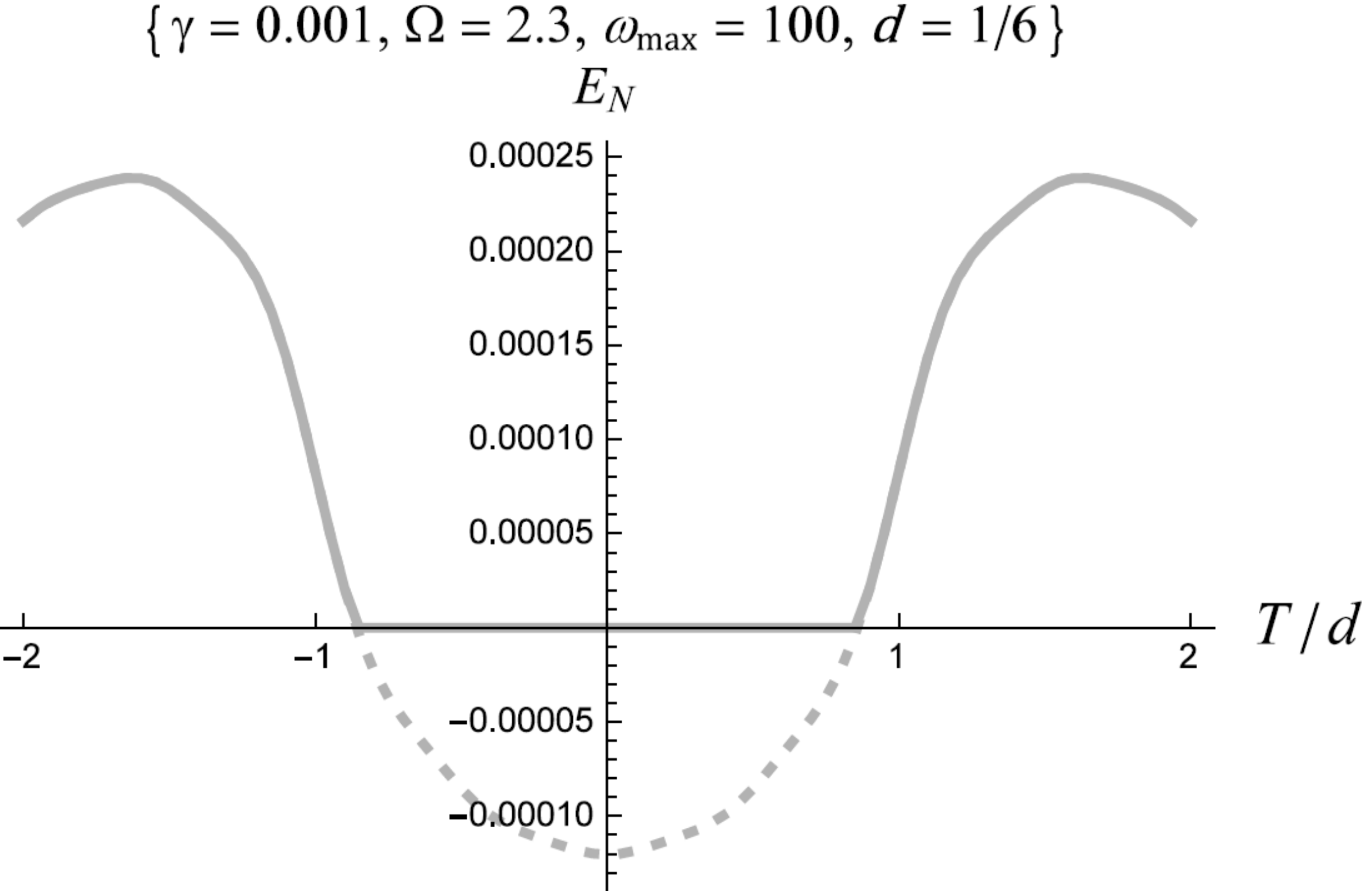}
\caption{(Left) The late-time logarithmic negativity $E_{\cal N}$ (scaled by 3) of two inertial HOs separated at a distance $d$,
with values taken on the future or past light cones for one of the two HOs at $t=t_1\gg 1/\gamma$ (EnLC, gray solid) and on
the $t_1$ slice in the Minkowski coordinates (EnSM, gray dashed), and the averaged fidelity $F_{av}$ of QT in both teleporting
directions subtracted by $F_{cl}$ (black). Here, $\omega_{\rm max} =100$ is the UV cutoff in Eq. (\ref{Fcpm}).
(Right) $E_{\cal N}$ of EnLC evaluated using the cross-correlators $\langle {\cal R}_A(t_1), {\cal R}_B (t_1 \pm T)\rangle$,
${\cal R} = Q, P$ at fixed separation $d=1/6\approx 0.167$, with other parameters the same.
(The dotted curve represents those negative symplectic eigenvalues that do not count in the definition of $E_{\cal N}$.)
While the two detectors have been disentangled according to $E_{\cal N}$ evaluated on the Minkowski time slices ($T=0$,
EnSM) at this distance, they are still entangled around the future and past light cones ($T/d = +1$ and $-1$, respectively;
cf. Fig. \ref{IHO2}).}
\label{ENlate2I}
\end{figure}

To look at this possibility more closely, one needs the correlators around the light cone instead of the equal-time correlators in the
Minkowski coordinates given in Ref. \cite{LH09}. First, generalize the expressions (52) in Ref. \cite{LH09} to
\begin{equation}
   {\cal F}_{c\pm}(d, T)\equiv {\hbar i\over 4\pi}\int_0^{\omega_{\rm max}} d\omega {\omega^c \cos \omega T\over
       \omega^2+2i\gamma\omega -\Omega_r^2\pm {2\gamma\over d}e^{i\omega d}}.
\label{Fcpm}
\end{equation}
For a given  UV cutoff $\omega_{\rm max}$, the late-time correlators with detectors $A$ and $B$ at different times,
$\langle \delta \hat{Q}_A^2(t)\rangle|_{\gamma t\gg 1} =\langle \delta \hat{Q}_B^2(t+T)\rangle|_{\gamma t\gg 1}=
    2 {\rm Re} [ {\cal F}_{0+}(d,0)+{\cal F}_{0-}(d,0) ]$,
$\langle \delta \hat{Q}_A (t) \delta\hat{Q}_B(t+T)\rangle|_{\gamma t\gg 1} = 2 {\rm Re} [ {\cal F}_{0+}(d,T) -{\cal F}_{0-}(d,T) ]$,
$\langle \delta\hat{P}_A^2(t)\rangle|_{\gamma t\gg 1} =\langle \delta\hat{P}_B^2(t+T)\rangle|_{\gamma t\gg 1}=
    2{\rm Re} [ {\cal F}_{2+}(d,0)+{\cal F}_{2-}(d,0) ]$, and
$\langle \delta\hat{P}_A (t)\delta\hat{P}_B(t+T)\rangle|_{\gamma t\gg 1} = 2 {\rm Re} [{\cal F}_{2+}(d,T) -{\cal F}_{2-}(d,T)]$,
can be calculated numerically. Using them, one obtains
the logarithmic negativity for the EnLC and the averaged FiQT between the two detectors
by setting $t=t_1$ and $T=\pm(d-\epsilon)$ in the above expressions such that $(\tau^A_1, \tau'^B_1)= (t_1, t_1 + d -\epsilon)$ or
$(\tau'^A_1, \tau^B_1) = (t_1,t_1-d+\epsilon)$ and then taking the limit $\epsilon\to 0+$.
An example is shown in Fig. \ref{ENlate2I}. It turns out that the late-time EnLC of the $AB$ pair is stronger than the entanglement evaluated on the same Minkowski time slice ($\tau^A_1 = \tau^B_1 = t_1$).
This implies that the entanglement distance $d_{\rm ent}$ for the EnLC is greater than the one we expected according to our old results of entanglement evaluated on the hypersurfaces of simultaneity in the Minkowski coordinates (call this EnSM) in Ref. \cite{LH09}.
As one can see in Fig. \ref{ENlate2I}, for the detectors separated at a distance $d$ in the range between the entanglement distances
for the EnLC and EnSM [$0.153 < d < 0.176$ in Fig. \ref{ENlate2I} (left)], the averaged FiQT can beat the classical fidelity
at late times ($F_{av}-F_{cl}>0$) while the detectors are disentangled in view of the EnSM ($E_{\cal N} = 0$ for $T=0$).
In this range the inequality (\ref{MVbound}) appears to be violated in view of the EnSM but it still holds in terms of the EnLC.
Together with the fact that the degree of the EnLC is independent of the choice of reference frames and invariant under coordinate
transformation, we conclude that the EnLC, rather than the EnSM, is essential in relativistic open systems
with the ``system" consisting of spatially localized objects.

\subsection{Early-time behavior}
\label{ET2IHO}

At early times, once Bob enters the future light cone of the spacetime event where detector $A$ started to couple to the field, detector $B$ will be affected by the retarded field of $A$. We call this mutual influence of the first order. Detector $B$ will respond to this influence with its backreaction to the field which in turn affects detector $A$, which is called mutual influence of the second order. The subsequent backreaction from $A$ propagates and affects $B$ again, which constitutes a mutual influence of the third order, and so on.
When the detector-field coupling is not weak enough or the spatial separation between the two detectors is not large enough, the higher-order mutual influences can get complicated and become very important soon.
Fortunately, in the Alice-Rob problem and the quantum twin problem to be introduced later, we are working in the weak coupling limit and the retarded distance \cite{OLMH11} between the two detectors will be very large in most of the history, so the mutual influences are not significant there. To compare with those results, assuming that the separation $d$ is large enough, the zero-order result without considering any mutual influences in this case would have already been a good approximation at early times.

We have obtained the evolution in $t_1$ of the logarithmic negativity $E_{\cal N}$ of the EnLC and the best averaged FiQT
$F_{av}^+$ in the weak coupling limit, as shown in Fig. \ref{ENFav134} (blue curves) for later comparison.
The evolution curves are roughly exponential decays with small oscillations on top of it at a frequency about twice
the natural frequency $\Omega$ of detectors in the weak coupling limit.

\begin{figure}
\includegraphics[width=5.5cm]{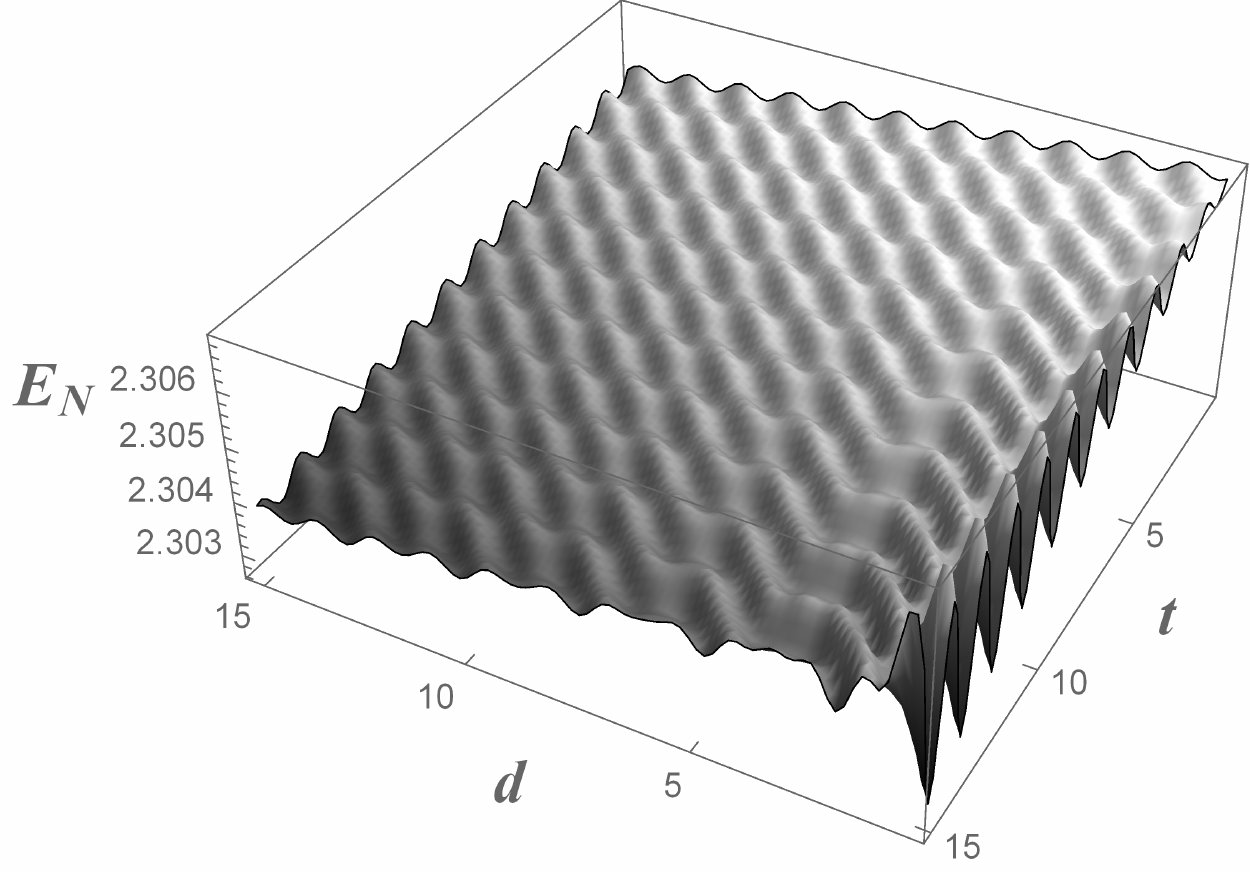}
\includegraphics[width=5.5cm]{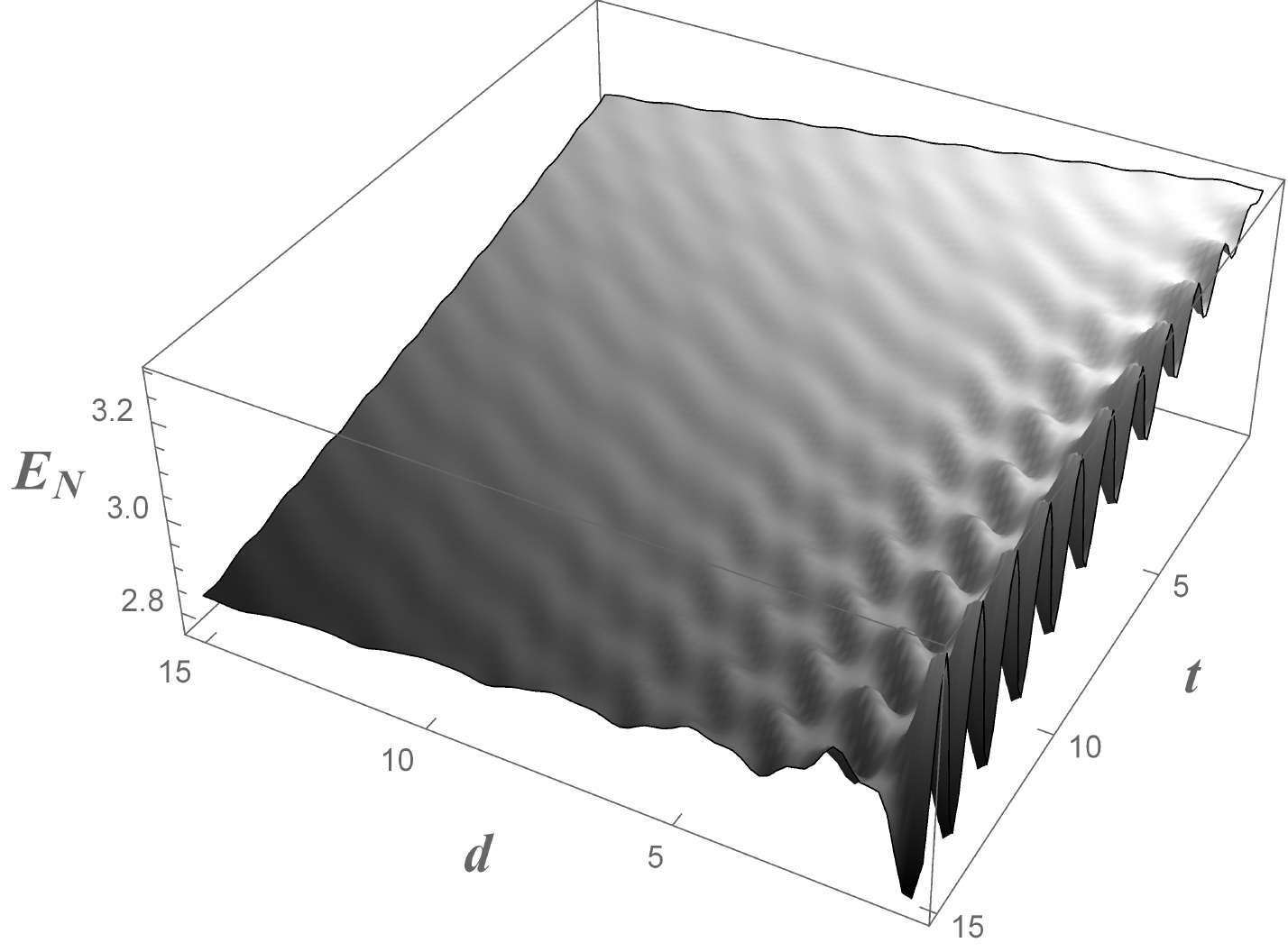}
\includegraphics[width=5.5cm]{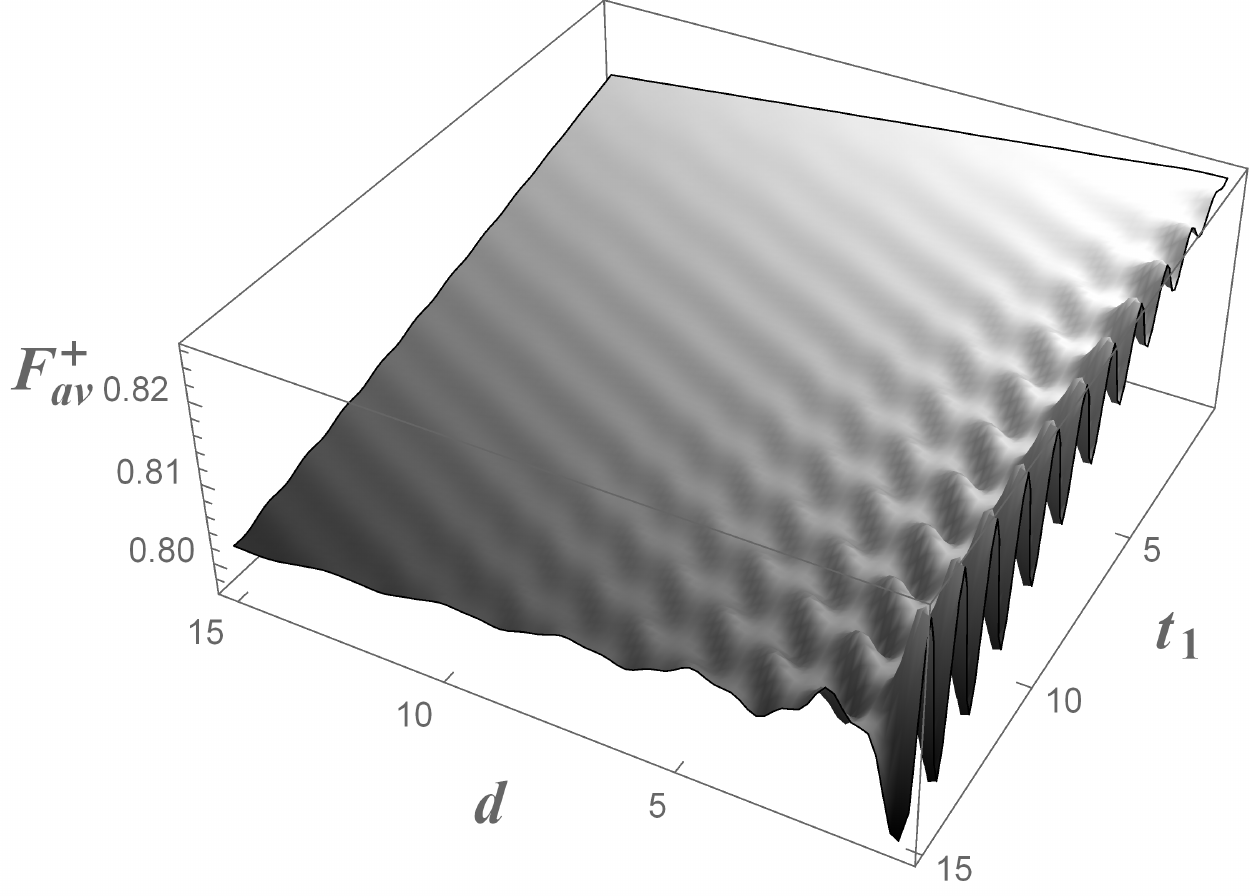}
\caption{Spatial and temporal dependence of the logarithmic negativity of EnLC and the best averaged FiQT
between Alice and Bob. The left plot is for comparison with Fig. 1 in Ref. \cite{LH09}, with the same parameters there.
For the middle and the right plots, we set $\gamma=0.001$, $\Omega=2.3$, $\Lambda_0=\Lambda_1=20$, $r_2=1.1$, and
$(\bar{\alpha}, \bar{\beta})= (e^{-r_1}/\sqrt{\Omega}, e^{-r_1}\sqrt{\Omega})$ with $r_1=1.2$.}
\label{SPDENFav}
\end{figure}

Note that the separation $d$ is also the retarded distance for the classical light signals from Alice to Bob or from Bob to Alice.
In Ref. \cite{LH09}, we see the spatial dependence of the entanglement dynamics: The evolution of quantum entanglement
between the two inertial detectors, and thus the disentanglement times, depend on $d$.
It is therefore not surprising that the evolution of the EnLC and the best averaged FiQT would show a similar
dependence on $d$ in Fig. \ref{SPDENFav}. The main differences from the EnSM results are the following. First,
for the same initial state of the $AB$ pair, if the separation $d$ is large enough, one expects that the larger $d$ is, the smaller the ``initial" (when $\tau_1^A = t =0 + \epsilon$ and $\tau'^B_1=d$) EnLC due to the longer time of decoherence of detector $B$ before entering the future light cone of Alice emitted at $t=\epsilon$, and thus the shorter is the disentanglement time of the EnLC. Second, the disentanglement {\it rate} of the EnLC is roughly the same for $t<d$ and $t>d$, while
in Ref. \cite{LH09}, we see that the degradation rate of the EnSM at early times has nontrivial $d$-dependence when $t>d$.

\section{Case 2---The Alice-Rob Problem: one inertial, one uniformly accelerated detector} 
\label{physreal}

\begin{figure}
\includegraphics[width=6cm]{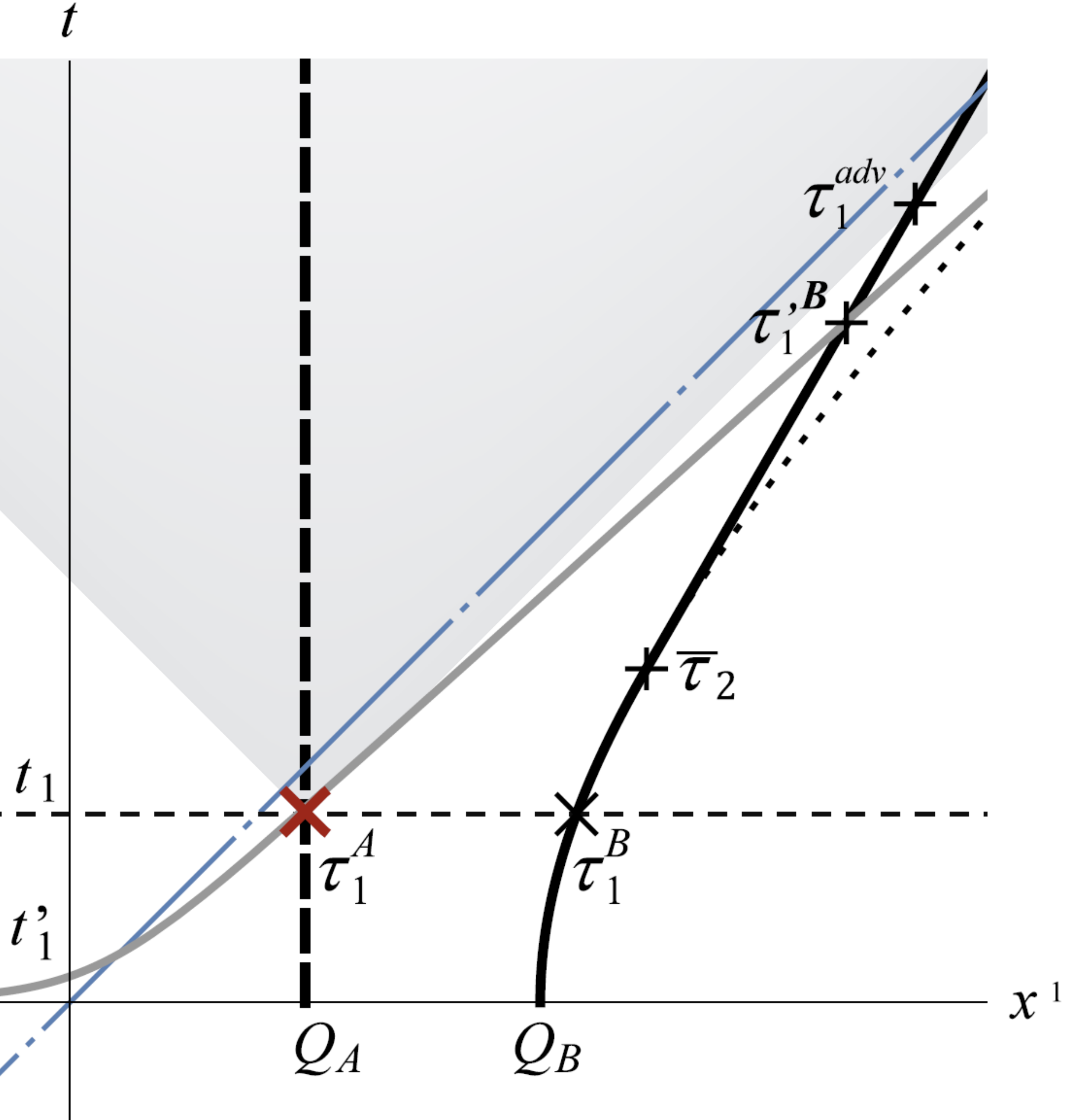}
\caption{Setup for QT from Alice (thick dashed worldline) at rest to Rob (thick solid worldline)
accelerated uniformly from $0$ to $\bar{\tau}^{}_2$ in his proper time then turning to inertial motion.
The hypersurface $t=x^1$ (blue dot-dashed line) will be the event horizon of Rob if $\bar{\tau}^{}_2 \to\infty$.}
\label{AR}
\end{figure}

Our second example has a setup slightly modified from the one in the ``Alice-Rob problem" \cite{AM03, LCH08}. It has been claimed that the Unruh effect experienced by Rob (Bob) in uniform acceleration would degrade the FiQT in this setup \cite{AM03}. This is the case in the detector models with the durations of Rob's constant linear acceleration and the duration of the detector-field interaction being the same and finite, while the teleportation is performed in the future asymptotic region when the detectors have been decoupled from the environment \cite{LM09}.  In this section we will examine how sound this claim is in our model in which the detectors are never decoupled from the fields and the QT process is performed in the interaction region. If Rob is uniformly accelerated, however, there will be an event horizon for him, beyond which no classical information can reach Rob (see Fig. \ref{AR}). To guarantee the signals emitted by Alice at all times can reach Rob to complete a QT from Alice to Rob, we still limit our considerations to the finite duration of acceleration, thus no event horizon for Rob, which for all practical purposes is a physically reasonable assumption, too.

Let us consider the setup with Alice at rest along the worldline $(t, a^{-1}-d,0,0)$ with the parameters $0 <(a^{-1}-d)<a^{-1}$
and Rob being constantly accelerated in a finite duration $0 \le \tau\le \bar{\tau}^{}_2$ then switched to inertial motion
(see Fig. \ref{AR}). In the acceleration phase Rob is going along the worldline $z_B^\mu = (a^{-1}\sinh a\tau, a^{-1}\cosh a\tau,0,0)$
the same as the one for a uniformly accelerated detector with proper acceleration $a$,
and after the moment $\tau=\bar{\tau}^{}_2$, or $\bar{t}^{}_2 = a^{-1}\sinh a\bar{\tau}^{}_2$ in the Minkowski time,
Rob moves with constant velocity along the worldline $( (\tau-\bar{\tau}^{}_2)\cosh a \bar{\tau}^{}_2+a^{-1}\sinh a\bar{\tau}^{}_2,
(\tau-\bar{\tau}^{}_2)\sinh a \bar{\tau}^{}_2 + a^{-1}\cosh a\bar{\tau}^{}_2, 0,0)$ in the Minkowski coordinates.
Here, the Minkowski time $t$ and the parameter $\tau$ are the proper times of Alice and Rob, namely, $\tau_{}^A=t$ and $\tau_{}^B=\tau$.

Suppose detector $C$ is moving with Alice and its quantum state to be teleported is created right before $t=t_1$, when Alice performs a joint measurement on detectors $A$ and $C$. Then, Alice sends out the outcome carried by a classical light signal right after $t_1$,
and Rob will receive the signal at his proper time:
\begin{equation}
  \tau^{adv}_1 \equiv \tau_{}^{adv}(t^{}_1)=\left\{ \begin{array}{lll}
   -a^{-1}\ln a\left( a^{-1}-d 
	- t_1\right)  & \,\,\,{\rm if}\,\,\,  t_1 < (1-e^{-a \bar{\tau}^{}_2})/a-d, \\
   \left(t_1 - a^{-1} + d \right)e^{a\bar{\tau}^{}_2} + a^{-1} + \bar{\tau}^{}_2 & \,\,\,{\rm otherwise.}    
	\end{array}  \right.
\label{tau1adv}
\end{equation}
Accordingly, Rob performs the local operation at $\tau^B=\tau^{adv}_1 + \epsilon$ with $\epsilon\to 0+$.

In the opposite direction, one can also consider the case with detector $C$ moving with Rob, who performs a joint measurement on
$B$ and $C$ at his proper time $\tau^B=\tau_1$ and sends the outcome to Alice by classical channel immediately.
Then, Alice will receive the message at her proper time,
\begin{equation}
  t^{adv}_1 \equiv t_{}^{adv}(\tau^{}_1)=\left\{ \begin{array}{lll}
    d+ a^{-1}\left( e^{a \tau_1}-1\right)  & \,\,\,{\rm if}\,\,\,  \tau_1 < \bar{\tau}^{}_2, \\
    d+ a^{-1}\left( e^{a\bar{\tau}^{}_2}-1 \right) + (\tau_1 - \bar{\tau}^{}_2 )e^{a\bar{\tau}^{}_2} & \,\,\,{\rm otherwise.}    
  \end{array}  \right.
\label{t1adv}
\end{equation}
and perform the local operation at $\tau^A=t^{adv}_1 + \epsilon$. Similar to $\tau^{adv}$, here $t^{adv}$ is the advanced time defined by 
$|z^\mu_A(t_{}^{adv}(\tau))- z^\mu_B(\tau)|^2=0$ with $z^0_A(t_{}^{adv}(\tau))> z^0_B(\tau)$.

\subsection{Dynamics of correlators} 

Since Rob stops accelerating at the moment $\bar{\tau}^{}_2$, 
the acceleration of detector $B$ is not really uniform. The dynamics of the correlators (\ref{Qdd1T})--(\ref{Rdd1T}) for
nonuniformly accelerated detectors in similar worldlines have been studied in Refs. \cite{OLMH11, DLMH13}.
In the weak coupling limit with a not-too-short duration of nearly constant acceleration
the behavior of such a detector is similar to a harmonic oscillator in contact with a heat bath at a time-varying ``temperature" corresponding to the proper acceleration of the detector.
Analogous to the results in Ref. \cite{OLMH11}, the dynamics of entanglement here will be dominated by the zeroth-order results of
the ``a-parts" of the self and cross-correlators \cite{LH06, LH07} and the ``v-parts" of the self-correlators of detectors $A$ and $B$.
The ``v-parts" of the cross-correlators are negligible.
The higher-order corrections by the mutual influences are also negligible in the weak coupling limit with
large initial entanglement and large spatial separation between the detectors.

For larger initial accelerations of detector $B$, the changes of the v-parts of its self-correlators during and after the
transition of the proper acceleration of detector $B$ from $a$ to $0$ are more significant. 
Consider the cases with the changing rate of the proper acceleration of detector $B$ from a finite $a$ to $0$ is fast enough so that
we can approximate the proper acceleration of detector $B$ as a step function of time, but not too fast to produce significant
nonadiabatic oscillation on top of the smooth variation.
According to the results in Refs. \cite{OLMH11} and \cite{LinDICE10},
for $\bar{\tau}^{}_2$ sufficiently large, the v-part of the self-correlators of detector $B$ behave like
\begin{eqnarray}
  & &\langle(\delta \hat{Q}_B(\tau))^2\rangle_{\rm v} \approx
    \langle (\delta \hat{Q}_B(\tau))^2\rangle_{\rm v}^{\{a\}} +
    \theta(\tau-\bar{\tau}^{}_2)\left[ -{\gamma\hbar a^2 e^{-2\gamma (\tau-\bar{\tau}^{}_2)}\over 6\pi m_0(\gamma^2+\Omega^2)^2} +
    \right.\nonumber\\ & & \hspace{1cm}\left.
    \left( \langle (\delta \hat{Q}_B(\infty))^2\rangle_{\rm v}^{\{0\}}
    -\langle (\delta \hat{Q}_B(\infty))^2\rangle_{\rm v}^{\{a\}}\right)
    \left(1-e^{-2\gamma (\tau-\bar{\tau}^{}_2)}\right)\right],
    \label{QB2NUAD} \\
  & &\langle (\delta \hat{P}_B(\tau))^2\rangle_{\rm v} \approx
    \langle (\delta \hat{P}_B(\tau))^2\rangle_{\rm v}^{\{a\}} +
    \theta(\tau-\bar{\tau}^{}_2)\times \nonumber\\ & & \hspace{1cm}\left[
    \left( \langle (\delta \hat{P}_B(\infty))^2\rangle_{\rm v}^{\{0\}}
    -\langle (\delta \hat{P}_B(\infty))^2\rangle_{\rm v}^{\{a\}}\right)
    \left(1-e^{-2\gamma (\tau-\bar{\tau}^{}_2)}\right)\right],
    \label{PB2NUAD}
\end{eqnarray}
where the superscripts $\{a\}$ and $\{0\}$ denote the self-correlators of a UD detector with the same parameters and
initial state except that it is uniformly accelerated with $a_\mu a^\mu=a^2$ and $0$, respectively, and
$\langle(\delta \hat{Q}_B(\infty))^2\rangle_{\rm v}^{\{a\},\{0\}}$ and
$\langle (\delta \hat{P}_B(\infty))^2\rangle_{\rm v}^{\{a\},\{0\}}$
are those self-correlators in steady state at late times (see Ref. \cite{LH07}).
These approximated behaviors have been verified by numerical calculations (see Figs. 3(right) and 4(right) in Ref. \cite{LinDICE10}).
Note that the $\gamma\hbar a^2$ term in 
Eq. $(\ref{QB2NUAD})$ is actually $O(\gamma/\Omega)$, so
$\langle (\delta \hat{P}_B(\tau^B_1))^2\rangle_{\rm v} \approx \Omega^2\langle (\delta \hat{Q}_B(\tau^B_1))^2\rangle_{\rm v} + \upsilon$,
and Eqs. (\ref{QAAwc})--(\ref{RAAwc}) are still good approximations up to $O(\gamma/\Omega)$, and we can keep using Eq. (\ref{weakFav}) 
here for $r_0=0$. Below, we apply these approximations to calculate the averaged FiQT in the ultraweak coupling limit.

\subsection{Averaged FiQT in ultraweak coupling limit}

\begin{figure}
\includegraphics[width=7.5cm]{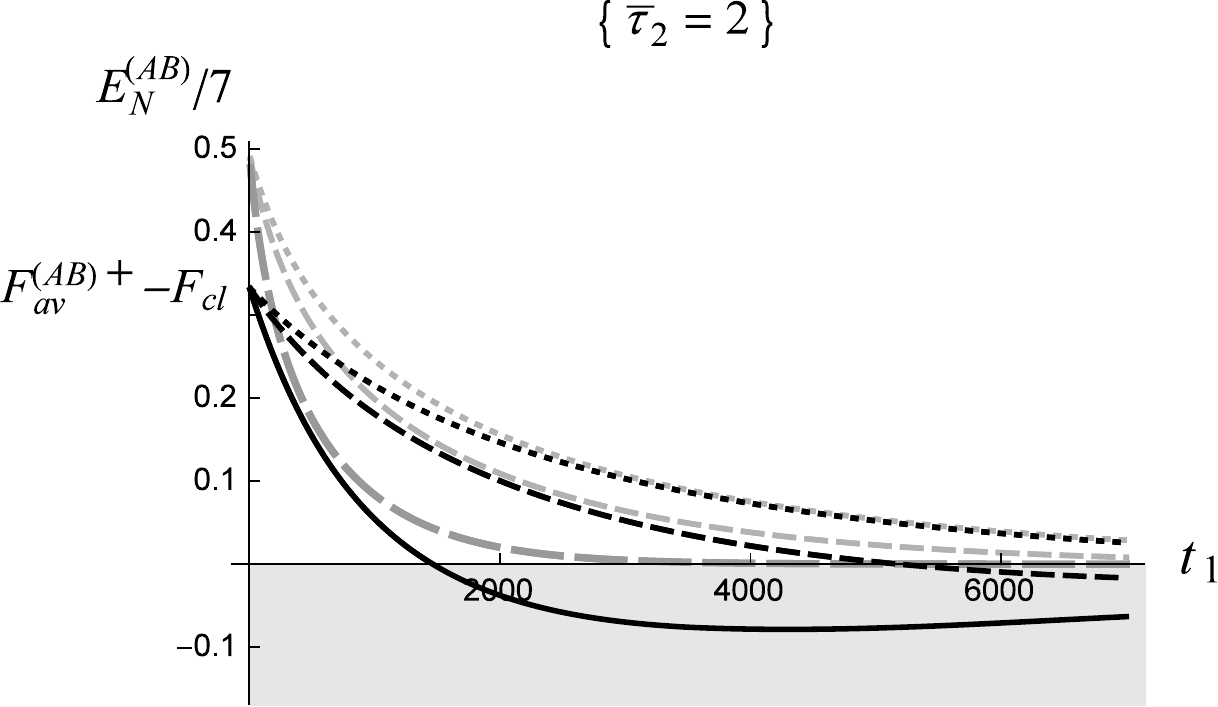}
\includegraphics[width=7.5cm]{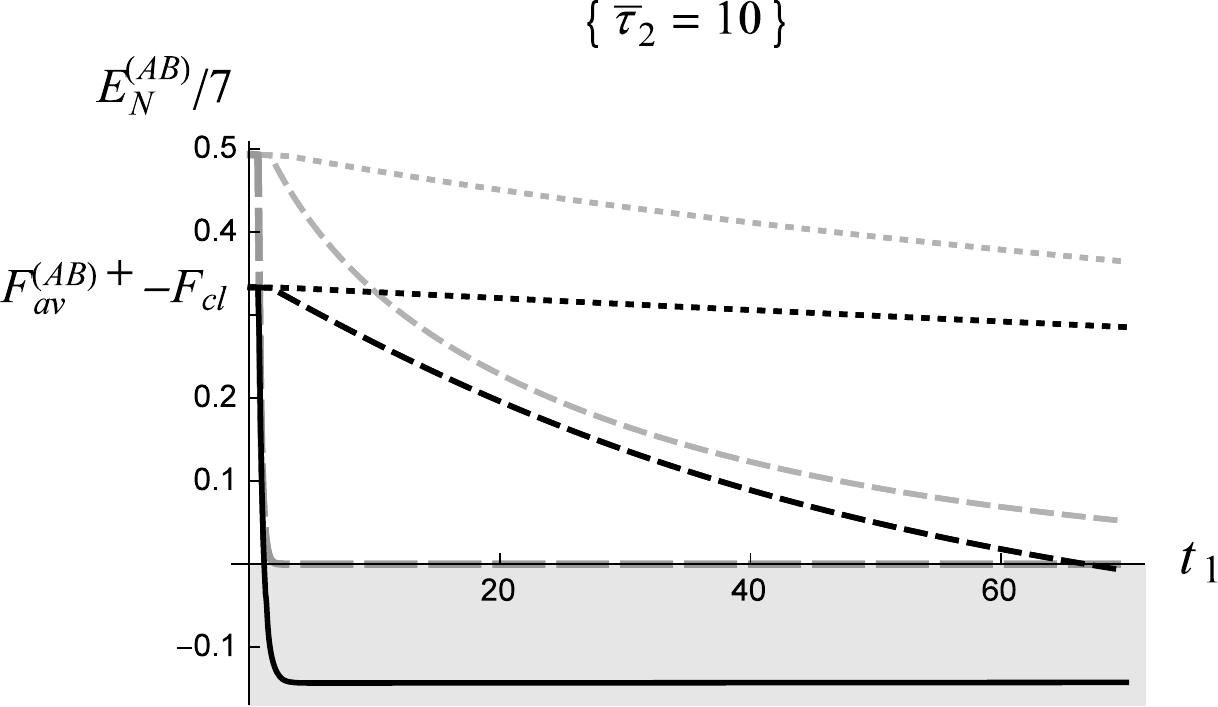}
\includegraphics[width=7.5cm]{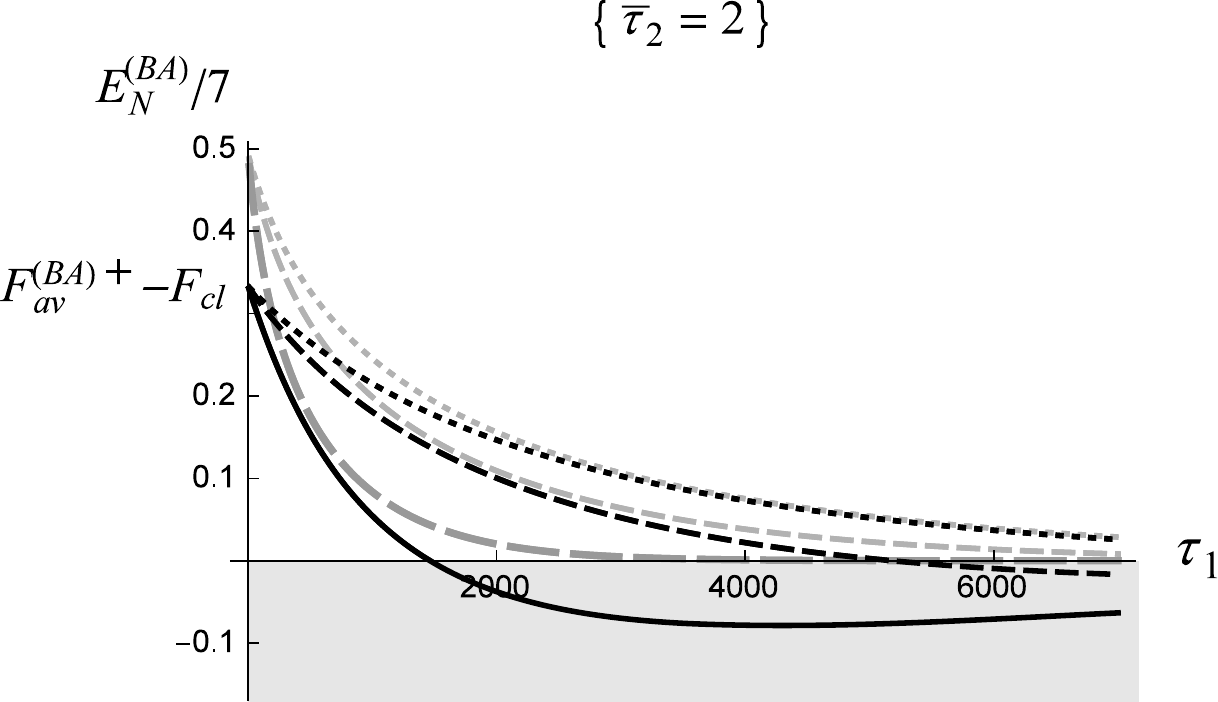}  
\includegraphics[width=7.5cm]{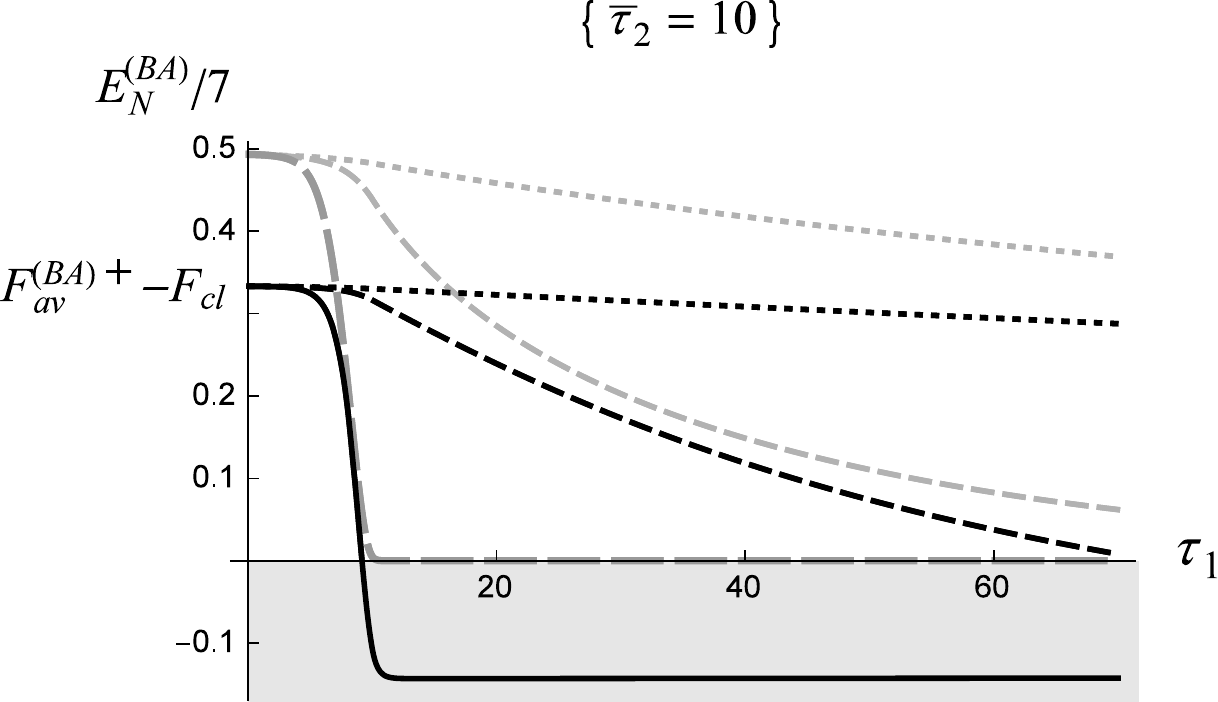} 
\caption{Comparison of the best averaged FiQT $F_{av}^+-F_{cl}$ (black curves) and the logarithmic negativities $E_{\cal N}$ (gray) of the
EnLC from Alice to Rob [$(AB)$, upper plots] and from Rob to Alice [$(BA)$, lower plots], as functions of the moment of the joint 
measurement $t_1$ by Alice ($\tau_1$ by Rob) with $\bar{\tau}^{}_2 =2$ (left) and $10$ (right), in the weak coupling limit.
Here, $a=1/4$ (dotted curves), $1/2$ (dashed), and $1$ (long-dashed gray and solid black). Other parameters are
$d=1/4$, $\gamma = 0.0001$, $\Omega = 2.3$, $\hbar=1$, $r_1=1.2$, $r_2=1.1$, $(\bar{\alpha},\bar{\beta}) = (e^{-r_1}/\sqrt{\Omega},
e^{-r_1} \sqrt{\Omega})$, and $\Lambda_0=\Lambda_1=20$.
In the upper-right plot, Rob is in the acceleration phase when receiving Alice's signal emitted at
$t_1 \le (1-e^{-a \bar{\tau}^{}_2})/a-d \approx 3.42$, $1.74$, $0.75$ for $a=1/4$, $1/2$, $1$, respectively from Eq. (\ref{tau1adv}).}
\label{FavPlus}
\end{figure}

\begin{figure}
\includegraphics[width=7.5cm]{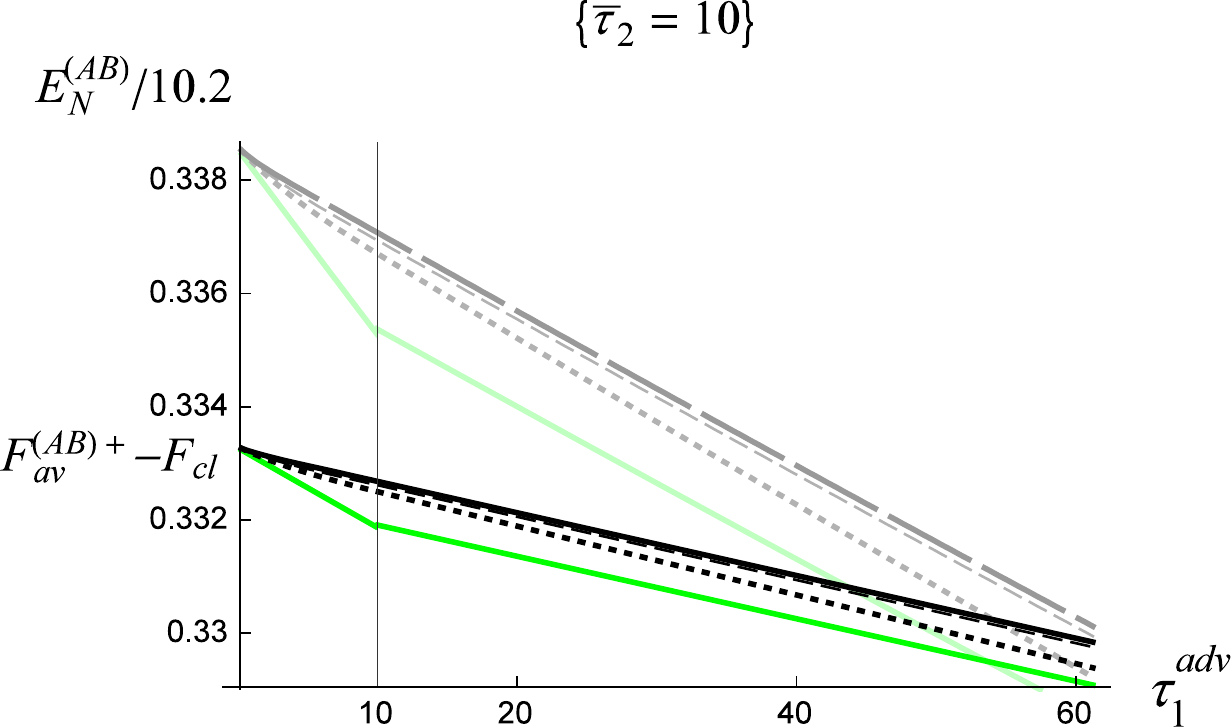}
\includegraphics[width=6.3cm]{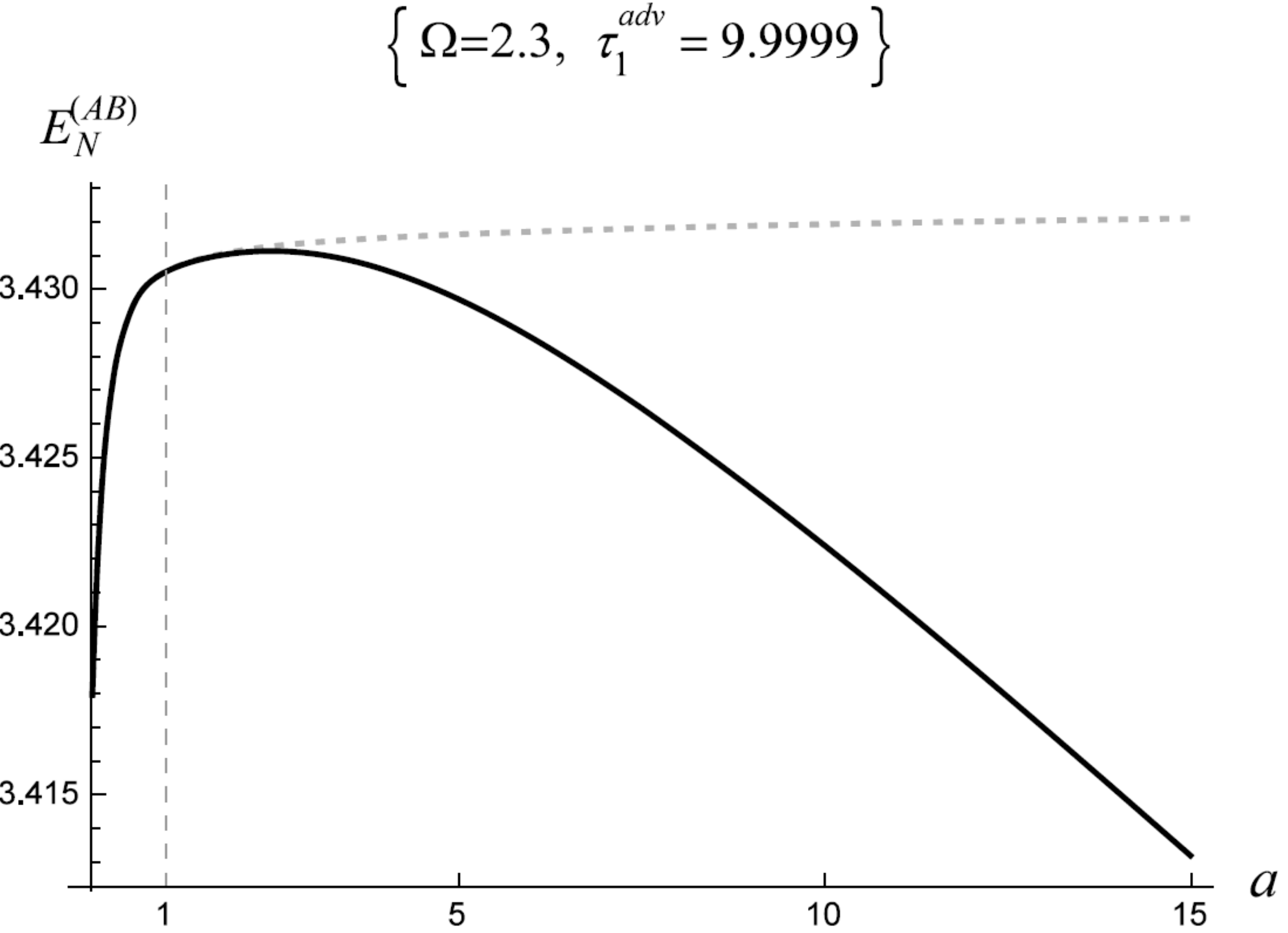}
\includegraphics[width=7.5cm]{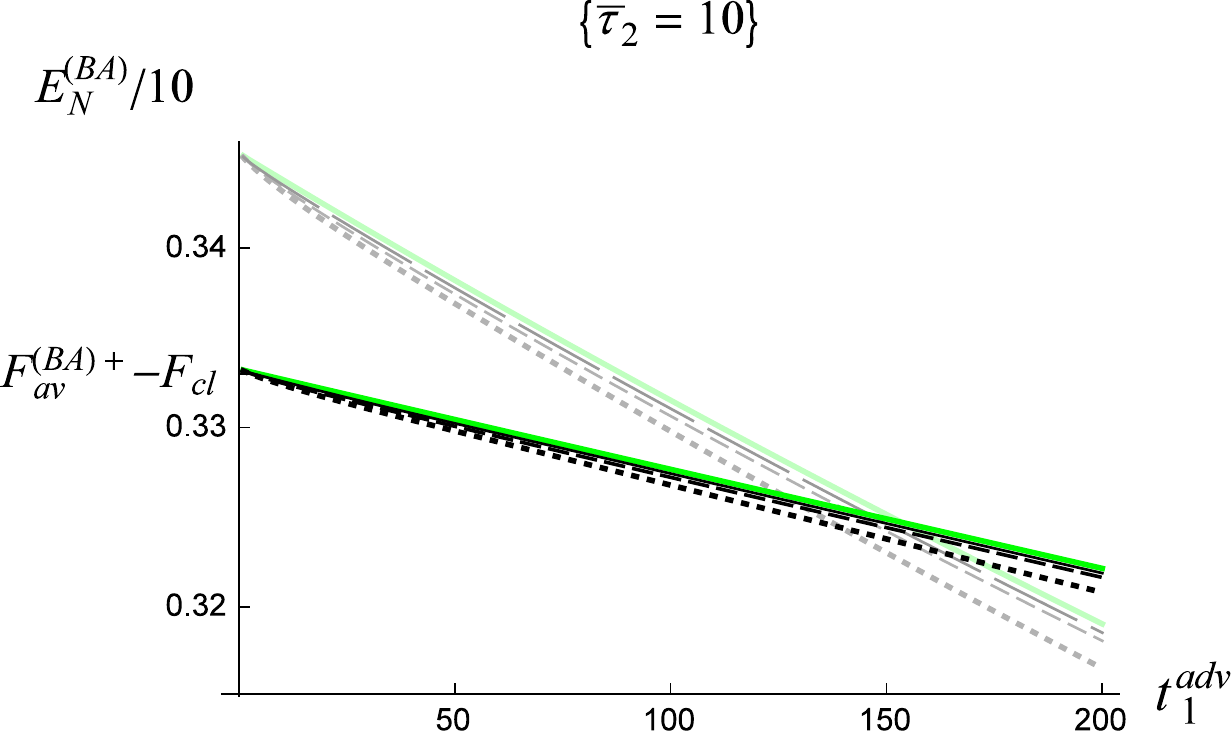}
\includegraphics[width=6.3cm]{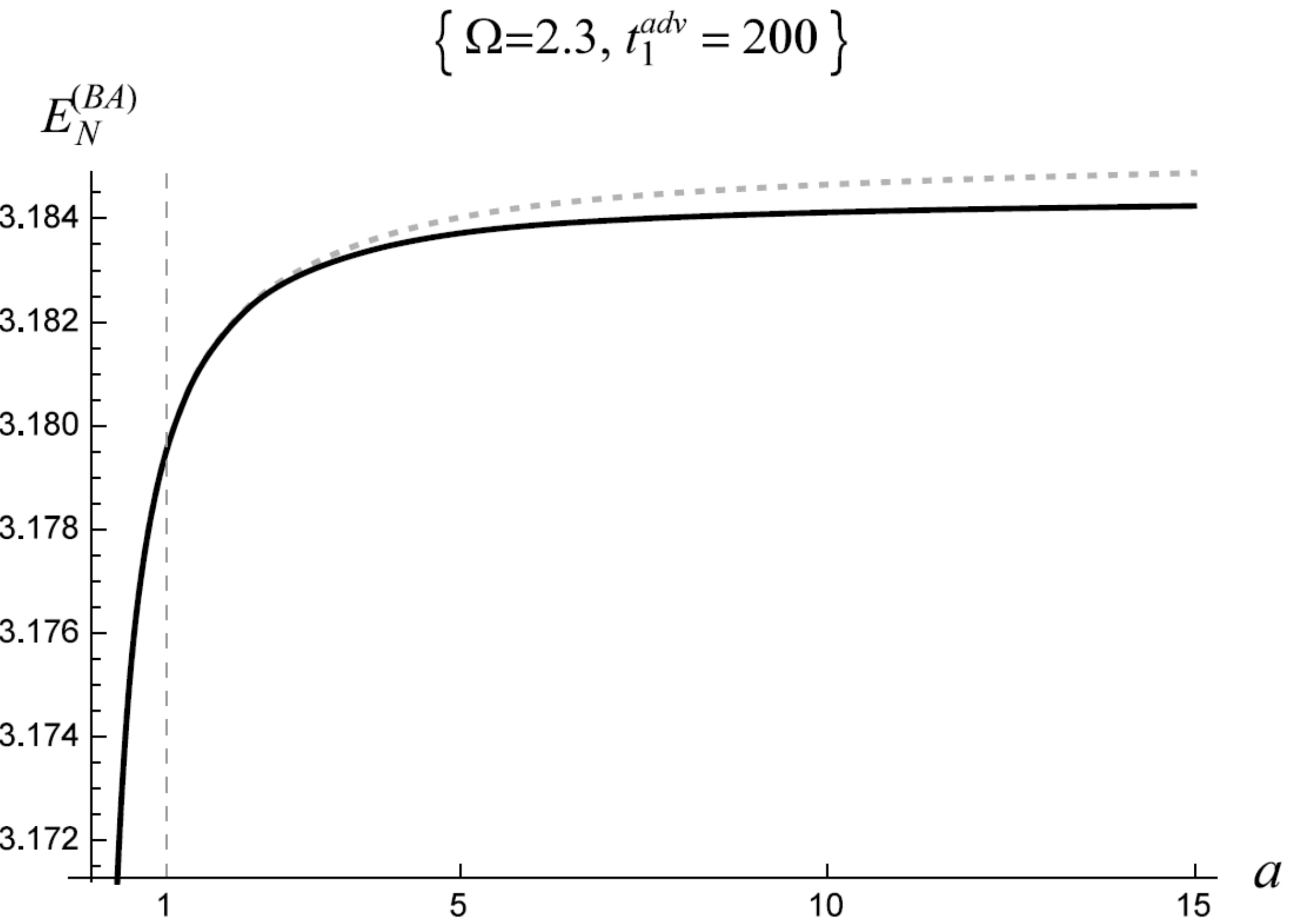}
\caption{(Left) The black and gray curves are the same results as those in the right plots of Fig. \ref{FavPlus} but now
against the moments $\tau_1^{adv}$ and $t_1^{adv}$ at which Rob and Alice receive the classical signal, respectively.
The green and light-green curves represent $F^+_{av}$ and $E_{\cal N}$, respectively, for $a=15$ and $d=[(2a)^4 + 4^4]^{-1/4}
\approx 0.033$. In the upper-left plot when $\tau_1^{adv}$ gets large enough the curves for the same quantity may cross each other
(not shown). From Eq. (\ref{t1adv}), Alice will receive the signal at $t^{adv}_1$ with $d < t^{adv}_1 < t^{adv}_{}(\bar{\tau}^{}_2)
\approx 44.98$, $295.08$, $22025.7$ for $a=1/4$, $1/2$, and $1$ if Rob emits the classical signal in his acceleration phase.
(Right) $E^{(AB)}_{\cal N}$ and $E^{(BA)}_{\cal N}$ for the EnLC 
at fixed moments $\tau^{adv}_1=9.9999$ and $t^{adv}_1=200$ in Rob's and Alice's points of view, respectively, as functions of $a$ (black).
The gray dotted curves are the same quantities with the Unruh effect removed from the self-correlators of detector $B$.
Here, $d=[(2a)^4 + 4^4]^{-1/4}$, and $\bar{\tau}_2=10$, so that, in the lower-right plot, if $a \agt 0.45$, Rob will be in the acceleration
phase when he performs the joint measurement as the sender. Other parameters are the same as those in the previous figure.}
\label{EnLCvsa}
\end{figure}

Inserting $(t_1, \tau^{adv}_1)$ in Eq. (\ref{tau1adv}) and $(t^{adv}_1, \tau_1)$ in Eq. (\ref{t1adv}) into $(\tau^A_1, \tau'^B_1)$ in
Eq. $(\ref{tildeVwc})$ and its counterpart for the opposite teleporting direction, respectively,
with the v-parts of the self-correlators $(\ref{QB2NUAD})$, $(\ref{PB2NUAD})$ and other correlators in the 
approximated form given by Eqs. (\ref{QAAwc})--(\ref{RAAwc}), we obtain the EnLC and the best averaged FiQT 
from Alice to Rob ($E_{\cal N}^{(AB)}$ and $F_{av}^{(AB)+}$, upper row) and from Rob to Alice
($E_{\cal N}^{(BA)}$ and $F_{av}^{(BA)+}$, lower row) in the sender's clock in Fig. \ref{FavPlus}
and in the receiver's point of view (observed along the past light cones) in the left plots of Fig. \ref{EnLCvsa}, respectively.

The quantities in each plot of Fig. \ref{FavPlus} do degrade faster as Rob's proper acceleration $a$ gets larger and the corresponding Unruh temperature gets higher.
However, one has to be cautious at such small accelerations ($a=1/4$ to $1$ here); none of these results can be taken as evidence of the Unruh effect. This is not only because Rob does not accelerate in a good part of the
histories shown in Fig. \ref{FavPlus} but because, more importantly, after the curves in the right plots of Fig. \ref{FavPlus} are translated to the receiver's point of view, shown in the left plots in Fig. \ref{EnLCvsa}, a larger proper acceleration of Rob turns out to give slower degradations of the best averaged FiQT and the EnLC in both teleporting directions even in Rob's acceleration phase.
In fact, one can remove the Unruh effect in the calculation by replacing the self-correlators of detector $B$ with the Unruh temperature by those for a detector at rest in the Minkowski vacuum, and one will still obtain similar curves and the same tendency of the degradation rates against the proper acceleration as those in Fig. \ref{FavPlus} and the corresponding curves in the left plots of Fig. \ref{EnLCvsa}.

The behavior of the curves in Fig. \ref{FavPlus} can be explained simply by the go-away setup in the Alice-Rob problem and the Doppler shift. For $F_{av}^{(AB)+}$ and $E_{\cal N}^{(AB)}$ from Alice to Rob with $t_1$ and $\bar{\tau}^{}_2$ fixed, the proper time $\tau_1^{adv}$ in Eq. (\ref{tau1adv}) when Rob receives Alice's signal increases rapidly as the value of $a$ increases, which allows for a much longer duration of decoherence for detector $B$ before Rob's operation. This yields a higher degradation rate in $t_1$ (Alice's clock) for larger $a$ in the evolution of the best averaged FiQT from Alice to Bob. On the other hand, Alice's signal is more redshifted and so Alice's clock looks slower for a larger $a$ in Rob's point of view. When $a$ is not too large, the apparent slowdown of decoherence for detector $A$ can beat the increasing rate of decoherence time for detector $B$ such that the larger $a$ is, the slower is the degradation in $\tau_1^{adv}$ [see the black and gray curves in Fig. \ref{EnLCvsa} (upper-left)]. Similarly, for a fixed value of $a$, Eq. (\ref{tau1adv}) implies that $\tau_1^{adv}$ for Rob grows rapidly as the duration of Rob's acceleration phase $\bar{\tau}^{}_2$ increases, which causes a much faster degradation of $F_{av}^{(AB)+}$ and $E_{\cal N}^{(AB)}$ in $t_1$ also. Indeed, the curves in the upper-right plot ($\bar{\tau}^{}_2=10$) of Fig. \ref{FavPlus} drop faster than those in the upper-left plot ($\bar{\tau}^{}_2=2$) for each value of $a$.
Let $t_{cl}$ be the moment of $t_1$ when $F_{av}^{(AB)+}$ drops to the value $F_{cl}$ for the classical teleportation.
When $a\bar{\tau}^{}_2$ is sufficiently large, $\tau_1^{adv}$ will be so large that $t_{cl}\approx a^{-1}-d$, which is the moment in Alice's clock when Alice crosses the event horizon for Rob as $\bar{\tau}^{}_2\to\infty$. For $F_{av}^{(BA)+}$ and $E_{\cal N}^{(BA)}$ in the opposite teleporting direction, the situations are similar, even though ostensibly there is no event horizon for Alice.

This is not the whole story, though.
If we increase Rob's proper acceleration $a$ further, while the EnLC from Rob to Alice $E_{\cal N}^{(BA)}$ is always an increasing
function of $a$ [Fig. \ref{EnLCvsa} (lower-right)], such a tendency will be altered when $a > O(\Omega)$ in the EnLC
from Alice to Rob $E_{\cal N}^{(AB)}$, as shown in Fig. \ref{EnLCvsa} (upper-right), mainly by the factor $\coth(\pi\Omega/a)$
in the self-correlators of detector $B$, e.g., $\langle (\delta \hat{Q}_B(\tau))^2\rangle_{\rm v}^{\{a\}} \approx (\hbar/2\Omega)
\coth (\pi\Omega/a) ( 1-e^{-\gamma \tau})$, for Eq. (\ref{QBBwc}) when Bob is accelerated \cite{LCH08}. Only in this regime, the Unruh 
effect is significant and dominates over the apparent slowdown of Alice's clock observed in Rob's acceleration phase,
in the sense that a higher Unruh temperature leads to a higher degradation rate of the best averaged FiQT and the EnLC.
After Rob's acceleration phase is over, however, due to the higher relative speed between Alice and Rob causing a stronger redshift of Alice's clock signal with a larger $a$, the degradation later in Rob's point of view can be slower than those with a smaller $a$.
Indeed, we see that the slopes of the black and gray dotted curves ($a=1/4$) are more negative than the slopes of the green and light-green curves ($a=15$), respectively, for $\tau_1^{adv} > \bar{\tau}^{}_2 =10$ in Fig. \ref{EnLCvsa} (left).

Comparing the upper and lower plots in Fig. \ref{FavPlus}, one sees that 
the moment $t_{cl}$ (or $\tau_{cl}$ defined similarly for Rob) when QT from Alice to Rob
(or from Rob to Alice) loses advantage over ``classical" teleportation is always earlier than
the disentanglement time evaluated around the future light cone of the joint measurement by Alice (or Rob) at $t_1$ (or $\tau_1$).
This confirms that the EnLC of the $AB$ pair is a necessary condition for the best averaged FiQT beating the classical one,
as indicated in Eq. (\ref{MVbound}).

\subsection{Beyond ultraweak coupling limit}

Beyond the ultraweak coupling limit, both the averaged fidelity $F_{av}$ and the logarithmic negativity $E_{\cal N}$ are strongly affected by the environment. In the cases in which mutual influences to the first few orders are small compared with the zeroth-order,
quantum entanglement of detectors $A$ and $B$ disappears quickly due to the strong corrosive effects of the environment. We expect that
the best averaged fidelity $F^{(AB)+}_{av}$ and $F^{(BA)+}_{av}$ would drop below $F_{cl}$ even quicker \cite{LSCH12}. Similar results on entanglement were given earlier in Ref. \cite{LCH08}, though the degrees of entanglement in Ref. \cite{LCH08} are evaluated on the time slices in the Minkowski coordinates or Rindler frames rather than those evaluated around the light cones.

\section{Case 3---Quantum twin problem}
\label{QTwin}

In the above results, we have seen that the relativistic effects entering the description of the dynamics of the detector pair can
dominate over the Unruh effect experienced by the accelerated detector in the degradation of the best averaged FiQT and the EnLC
between the pair. The apparent ``slowdown" in the dynamics of the sender in the viewpoint of the receiver in a QT process can be perceived by the receiver in the redshift of the clock signal from the sender. Nevertheless, in the setup of the Alice-Rob problem, since the retarded distances from Alice to Rob and from Rob to Alice are always increasing in time, only the redshift of the clock signal from the other would be observed, and so both Rob and Alice would conceive that their partner's clocks are always slower than their own. One may wonder what will happen when Alice and Rob (Bob) undergo more general motions.

To get a more comprehensive picture, a simple but helpful extension is to consider a setup similar to the classical twin ``paradox"
\cite{DG01}, in which we would have a consistent description of the asymmetric aging, red- and blueshifts of the clock signals, and the inertial and noninertial motions. Indeed, recall that in special relativity the twin paradox originates from the disparity
between Alice the twin at rest and Bob the traveling twin: Alice seeing Bob going away is the same as Bob seeing Alice going away, so each
one is supposed to observe the other with the same time dilation. Why does Bob become younger but not Alice when they meet again?
The resolution is that, for Bob to return to Alice, he must turn around at some point, thus undergoing some period of acceleration, and
the principles of special relativity do not apply to noninertial frames. When coupled to quantum fields, the Unruh effect experienced by
Bob during the periods of acceleration will come into play. With the theoretical tools developed and knowledge gained in the previous sections, luckily, this quantum twin problem becomes straightforward.

\begin{figure}
\includegraphics[width=2cm]{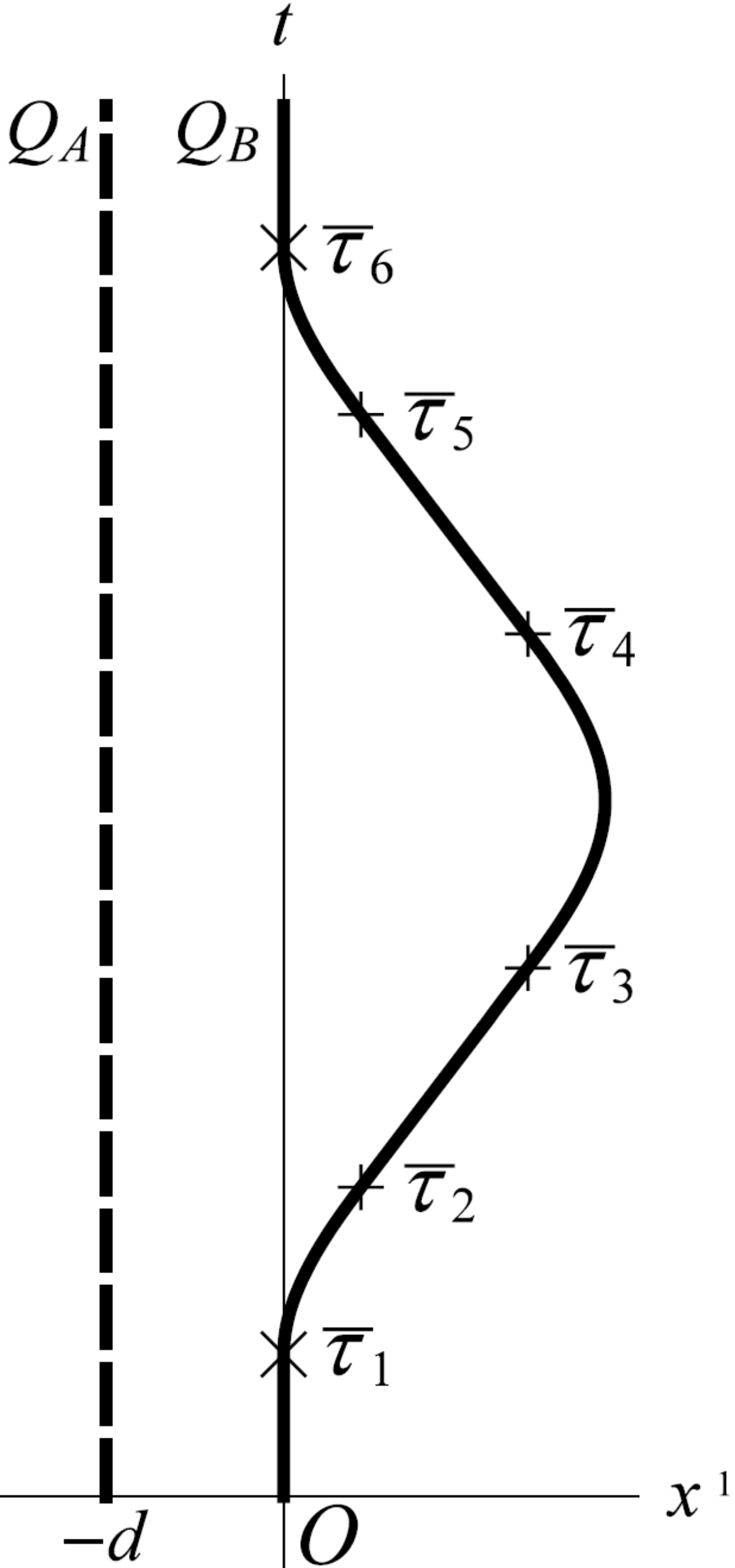}
\caption{QT between Alice (thick dotted worldline) and Rob (thick solid worldline) in a setup of the twin problem, where
the worldline of the traveling twin Bob is given in Eq. (\ref{twinBtraj}).}
\label{twinsetup}
\end{figure}

Suppose Alice is at rest with the worldline $z^\mu_A = (t, -d,0,0)$, $d>0$ and the proper time $\tau_{}^A = t$,
Bob is going along the worldline $z^\mu_B(\tau)$ with $z^2_B=z^3_B=0$ and
\begin{equation}
\left(z^0_B(\tau), z^1_B(\tau) \right) = 
\left\{
  \begin{array}{lll}
    (\tau, \, 0) & & 0 < \tau \le \bar{\tau}^{}_1, \\
    \left({1\over a} \sinh a(\tau-\bar{\tau}^{}_1) + \bar{\tau}^{}_1, \,
		    {1\over a}[\cosh a(\tau-\bar{\tau}^{}_1)-1]\right)  & & \bar{\tau}^{}_1 < \tau \le \bar{\tau}^{}_2, \\
    \left(\gamma_2 (\tau-\bar{\tau}^{}_2) + z^0_B(\bar{\tau}^{}_2), \, 
		     \gamma_2 v_2 (\tau-\bar{\tau}^{}_2)+ z^1_B(\bar{\tau}^{}_2) \right) & & \bar{\tau}^{}_2 < \tau \le \bar{\tau}^{}_3, \\
    \left( {1\over a} [ \sinh a(\tau-\bar{\tau}^{}_{3p} )-\gamma_2 v_2 ] +z^0_B(\bar{\tau}^{}_3),  \,
			{-1\over a}[ \cosh a (\tau-\bar{\tau}^{}_{3p} )- \gamma_2 ]+ z^1_B(\bar{\tau}^{}_3)\right)
			& {\rm for} &\bar{\tau}^{}_3 < \tau \le \bar{\tau}^{}_4, \\
    \left( \gamma_2 (\tau-\bar{\tau}^{}_4)+ z^0_B(\bar{\tau}^{}_4), \,
		    -\gamma_2 v_2 (\tau-\bar{\tau}^{}_4)+ z^1_B(\bar{\tau}^{}_4)\right)
		    & & \bar{\tau}^{}_4 < \tau \le \bar{\tau}^{}_5, \\
    \left({1\over a}[\sinh a(\tau-\bar{\tau}^{}_{5p} )- \gamma_2 v_2 ] + z^0_B(\bar{\tau}^{}_5), \,
		{1\over a} [ \cosh a(\tau- \bar{\tau}^{}_{5p} ) - \gamma_2 ] + z^1_B(\bar{\tau}^{}_5)\right)
		& & \bar{\tau}^{}_5 < \tau \le \bar{\tau}^{}_6, \\
    \left((\tau-\bar{\tau}^{}_6)+ z^0_B(\bar{\tau}^{}_6), \, 0\right)
		    & & \tau>\bar{\tau}^{}_6,
    \end{array}\right.
\label{twinBtraj}
\end{equation}
where $\tau_{}^B = \tau$ is Bob's proper time,
$\bar{\tau}^{}_p\equiv \bar{\tau}^{}_2-\bar{\tau}^{}_1 = (\bar{\tau}^{}_4-\bar{\tau}^{}_3)/2 =\bar{\tau}^{}_6-\bar{\tau}^{}_5$,
$\bar{\tau}^{}_{3p} \equiv \bar{\tau}^{}_3 + \bar{\tau}^{}_p$, $\bar{\tau}^{}_{5p} \equiv\bar{\tau}^{}_5 + \bar{\tau}^{}_p$,
$\bar{\tau}^{}_3-\bar{\tau}^{}_2 = \bar{\tau}^{}_5-\bar{\tau}^{}_4$,
$\gamma_2 = \cosh a \bar{\tau}^{}_p$, and
$\gamma_2 v_2 = \sinh a \bar{\tau}^{}_p$ (see Fig. \ref{twinsetup}).
Here we set the minimal distance between Alice and Bob $d$ to be sufficiently large to avoid the singular behavior of the retarded fields, and thus the mutual influences, when the detectors are too close to each other in the final stage
(for example, see Ref. \cite{LH09}).

\subsection{Evolution of correlators}

Below, we consider a case in the ultraweak coupling limit, with Bob still at his youth $(\gamma\bar{\tau}^{}_6 \ll 1)$ at the moment when he rejoins Alice, who is also in her early age $(\gamma z^0(\bar{\tau}^{}_6) < 1)$ but much advanced in age than Bob at that moment [e.g., $\bar{\tau}^{}_6=16$ for Rob and $z^0(\bar{\tau}^{}_6) =220$ for Alice in Figs. \ref{EntDyn} and \ref{FavTwin}].

As before, suppose the combined system is initially in a product state $\hat{\rho}^{}_{\Phi_{\bf x}}\otimes\hat{\rho}^{}_{AB}
\otimes\hat{\rho}_C^{(\alpha, r_0)}$. On top of the well-studied self-correlators for a detector at rest in Minkowski vacuum \cite{LH07},
the subtracted v-parts of the self-correlators of detector $B$ \cite{OLMH11, DLMH13} in our weak coupling limit, $\delta\langle
{\cal R}^{}_B(\tau), {\cal R}'_B(\tau) \rangle_{\rm v}\equiv \langle {\cal R}^{}_B(\tau),{\cal R}'_B(\tau)\rangle_{\rm v}-
\langle {\cal R}^{}_B(\tau),{\cal R}'_B(\tau)\rangle_{\rm v}|_{a=0}$, ${\cal R}, {\cal R}'= \delta Q, \delta P$ 
have been obtained numerically. We found that $\delta\langle {\cal R},{\cal R}'\rangle_{\rm v}$ starts to oscillate
after the launch of Bob. The oscillations would be amplified whenever the acceleration suddenly changes from one stage to the next
due to the nonadiabatic effect \cite{OLMH11}, while its mean value
grows due to the Unruh effect when detector B is undergoing accelerations and decays during the time intervals in the inertial
motion. Anyway, the amplitude of $\delta\langle {\cal R}^{}_B,{\cal R}'_B\rangle_{\rm v}$ is always as small as $O(\gamma)$ compared
with $\langle  {\cal R}^{}_B,{\cal R}'_B \rangle_{\rm v}$, while $\langle  {\cal R}^{}_B,{\cal R}'_B \rangle_{\rm v}$ is small
compared with  $\langle  {\cal R}^{}_B,{\cal R}'_B \rangle_{\rm a}$ in such an early stage.

We further obtained the numerical results for the cross-correlators between $A$ and $B$, $\langle {\cal R}^{}_A(t),
{\cal R}'_B(\tau_{}^{adv}(t)-\epsilon)\rangle$, and $\langle {\cal R}^{}_A(t_{}^{adv}(\tau)-\epsilon),{\cal R}'_B(\tau)\rangle$,
around the future light cone of Alice and Bob at $\tau_{}^A = t$ and $\tau_{}^B=\tau$, respectively.
We find that they oscillate in time about zero during the whole journey of Bob until he meets Alice again.
The oscillations appear irregular since the motions and the time dilations of the two detectors are asymmetric.
While the amplitudes of the oscillations of the a-parts of the cross-correlators are $O(1)$,
the amplitudes of the v-parts are $O(\gamma)$ and negligible in the weak coupling limit.
After Bob returns and both detectors are at rest, the behavior of the a-parts of the cross-correlators
continues in the same way, but the v-parts of $\langle Q_A, Q_B\rangle$ and $\langle P_A, P_B\rangle$
turn into small oscillations on top of slow growths or decays in time,
similar to those in the cases with two detectors at rest (see Sec. V in Ref. \cite{LH09}).

Corrections from the mutual influences $\langle {\cal R}^{(0)}_i, {\cal R}^{(1)}_j\rangle_{\rm a,v}$, $i,j=A,B$ up to the
first order of $\gamma/d$ have been worked out to check the consistency of our approximation.
There is one correction to each of both the a-part and v-part of the correlators $\langle Q_i^2 \rangle$
and $\langle P_i^2 \rangle$ and two corrections to those for the other correlators. Thus, we have a total of 32 corrections of the first order. We find that, during Bob's journey, the corrections to the v-part and the a-part of each correlator are $O(\gamma)$,
small compared with the zeroth-order results. After Bob returns and stays at rest by Alice, these corrections from the mutual influences
start to grow in magnitude. If the separation $d$ of Bob and Alice is small, these corrections may overtake the zeroth-order results and one has to include higher-order mutual influences \cite{LH09}. Here, we simply terminate our simulation at $\tau_{}^B = \bar{\tau}^{}_f \approx 24$ in Bob's proper time, which is early enough to justify our first-order approximation.

One may worry that the backreaction from detector $B$ to the field during $\tau \in (\bar{\tau}^{}_4, \bar{\tau}^{}_6)$ would form a shock wave and hit detector $A$ in the period when Bob heads back to Earth and decelerates [$t\in (t_{adv}(\bar{\tau}^{}_4), t_{adv}(\bar{\tau}^{}_6))\approx (220.88, 221.43)$ in the left plots of Figs. \ref{EntDyn} and \ref{FavTwin}], analogous to the shock electromagnetic wave along the past horizon of a uniformly accelerated charge in classical electrodynamics \cite{Bo80}. Fortunately, in our results, these mutual influences do not significantly impact on detector $A$ since they are off resonant.

\subsection{Entanglement dynamics}

\begin{figure}
\includegraphics[width=1cm]{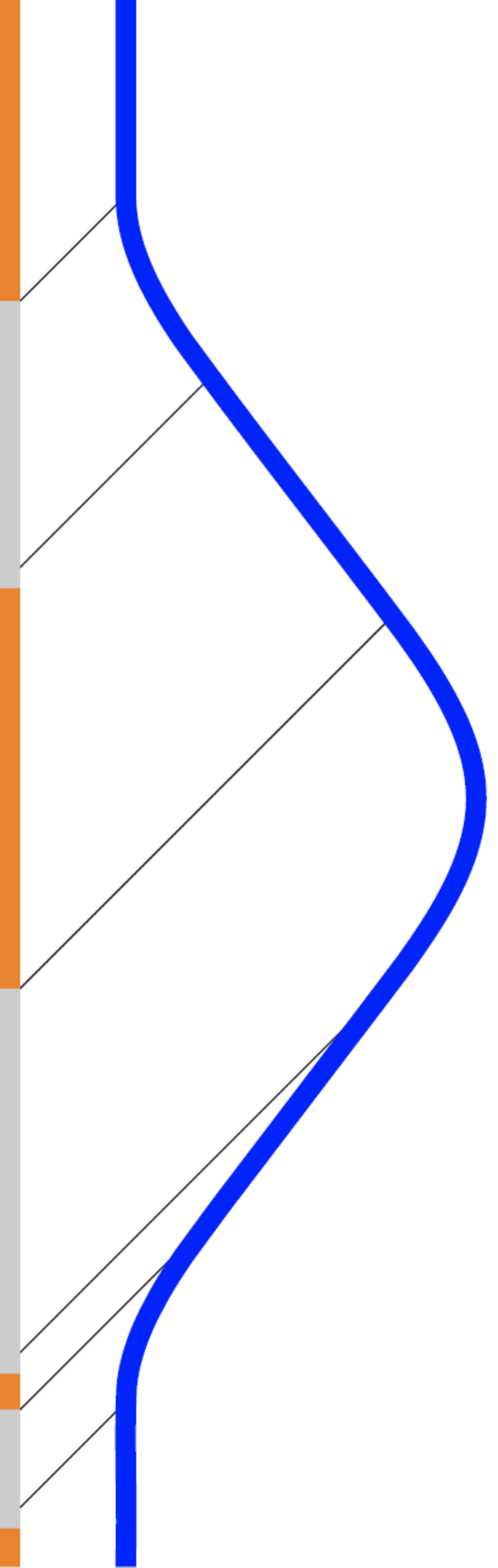} 
\includegraphics[width=5.3cm]{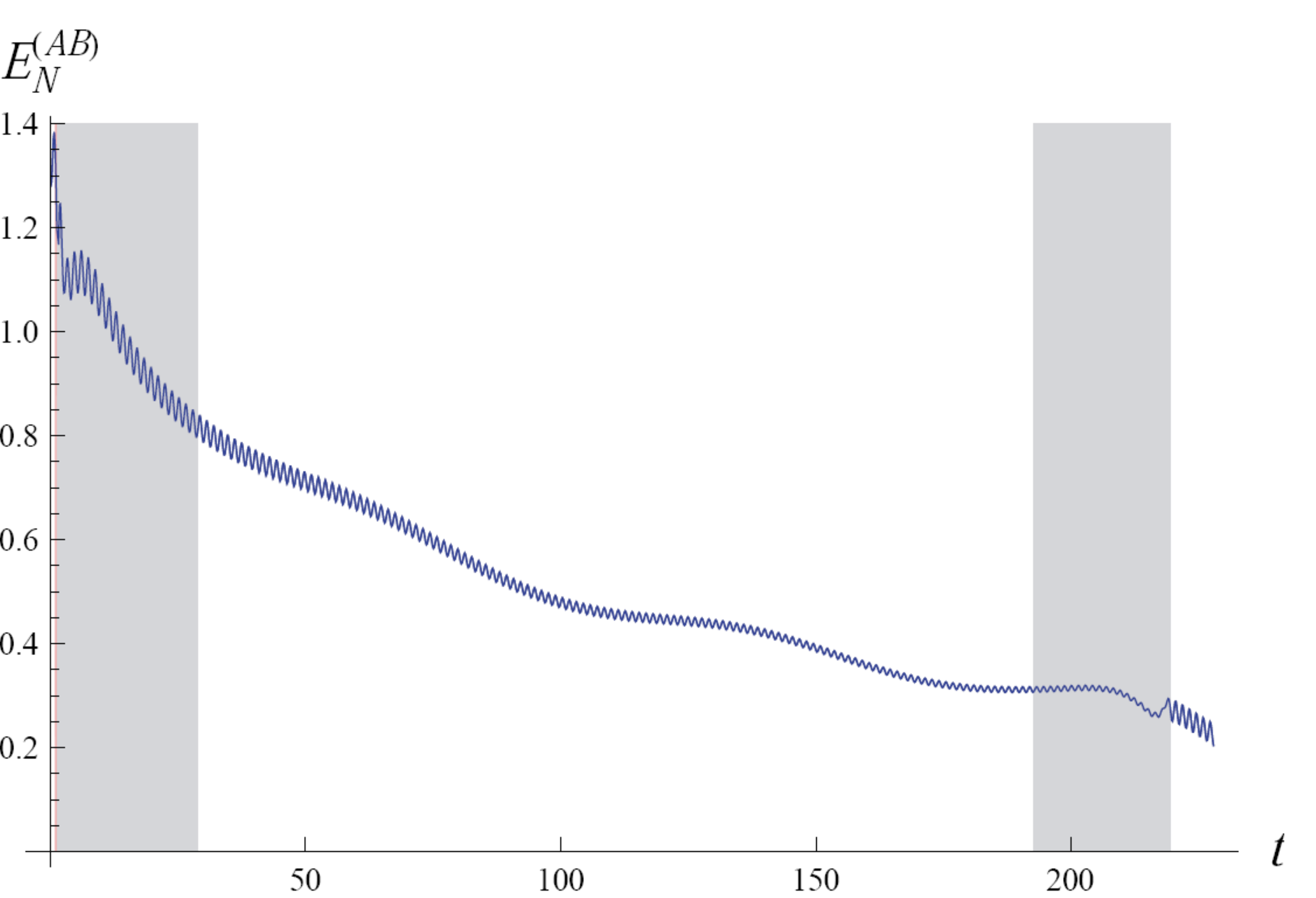} 
\includegraphics[width=5.3cm]{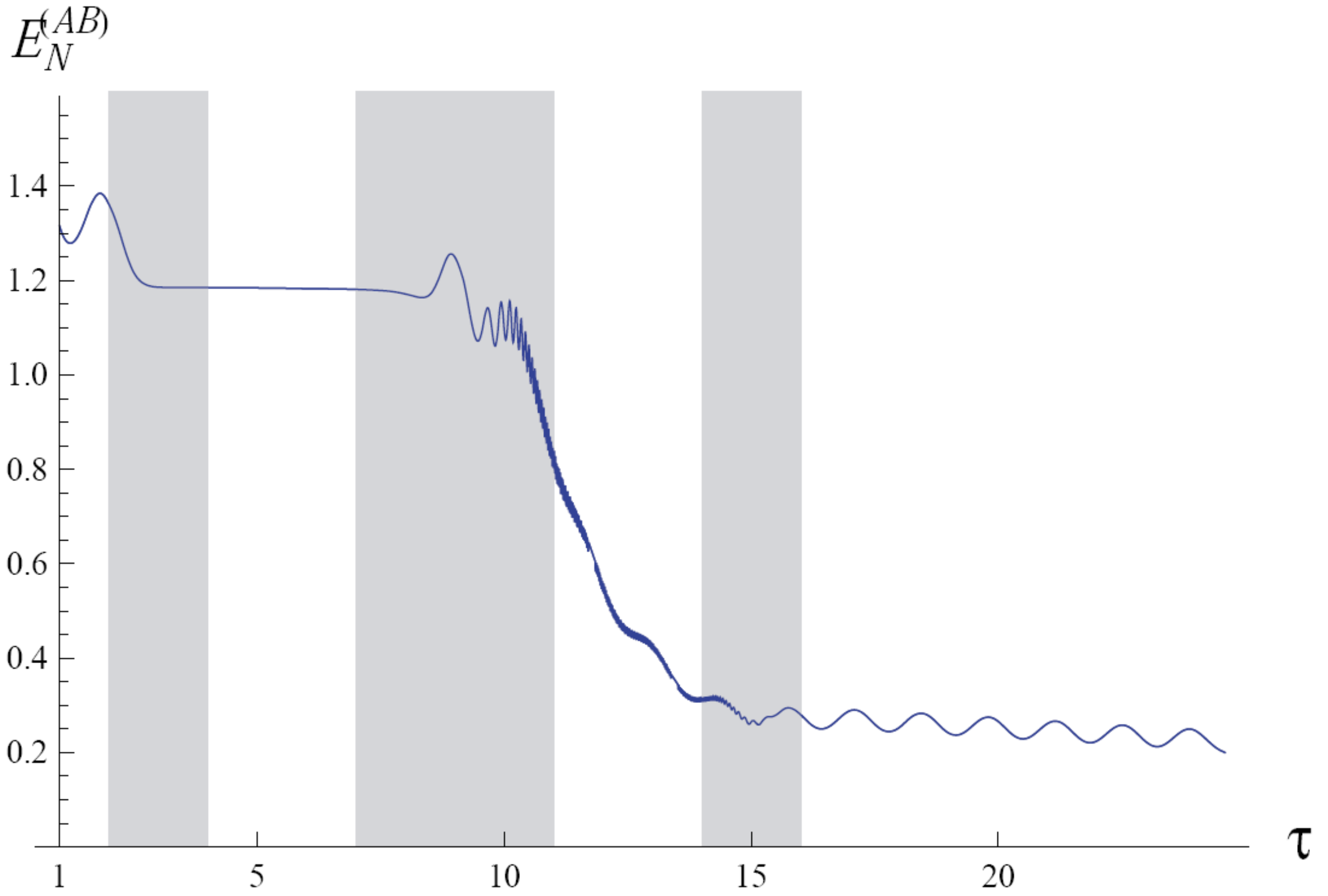} 
\includegraphics[width=5.8cm]{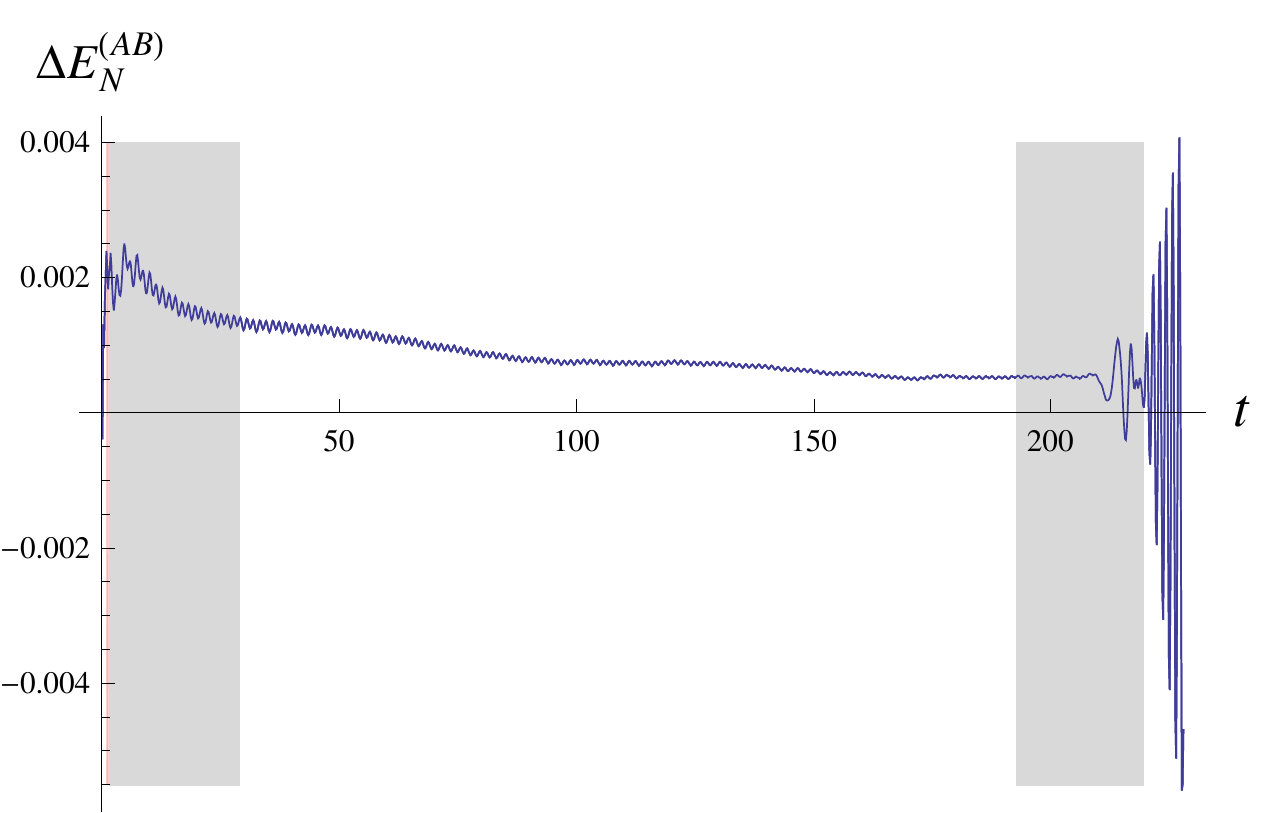} 
\includegraphics[width=1cm]{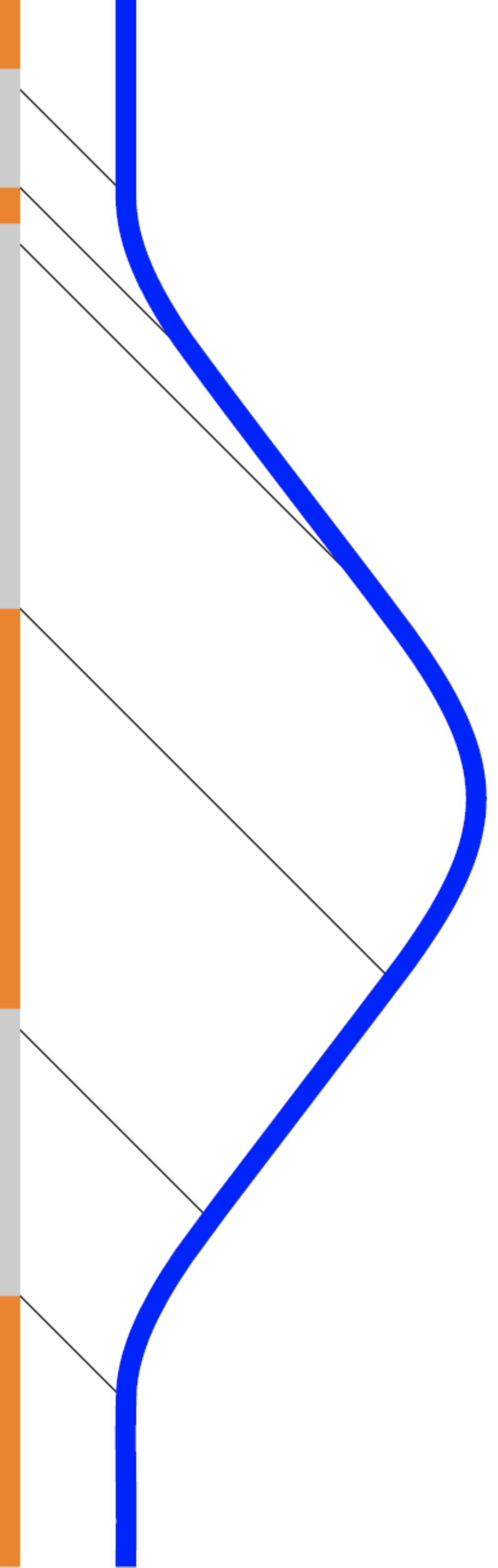} 
\includegraphics[width=5.3cm]{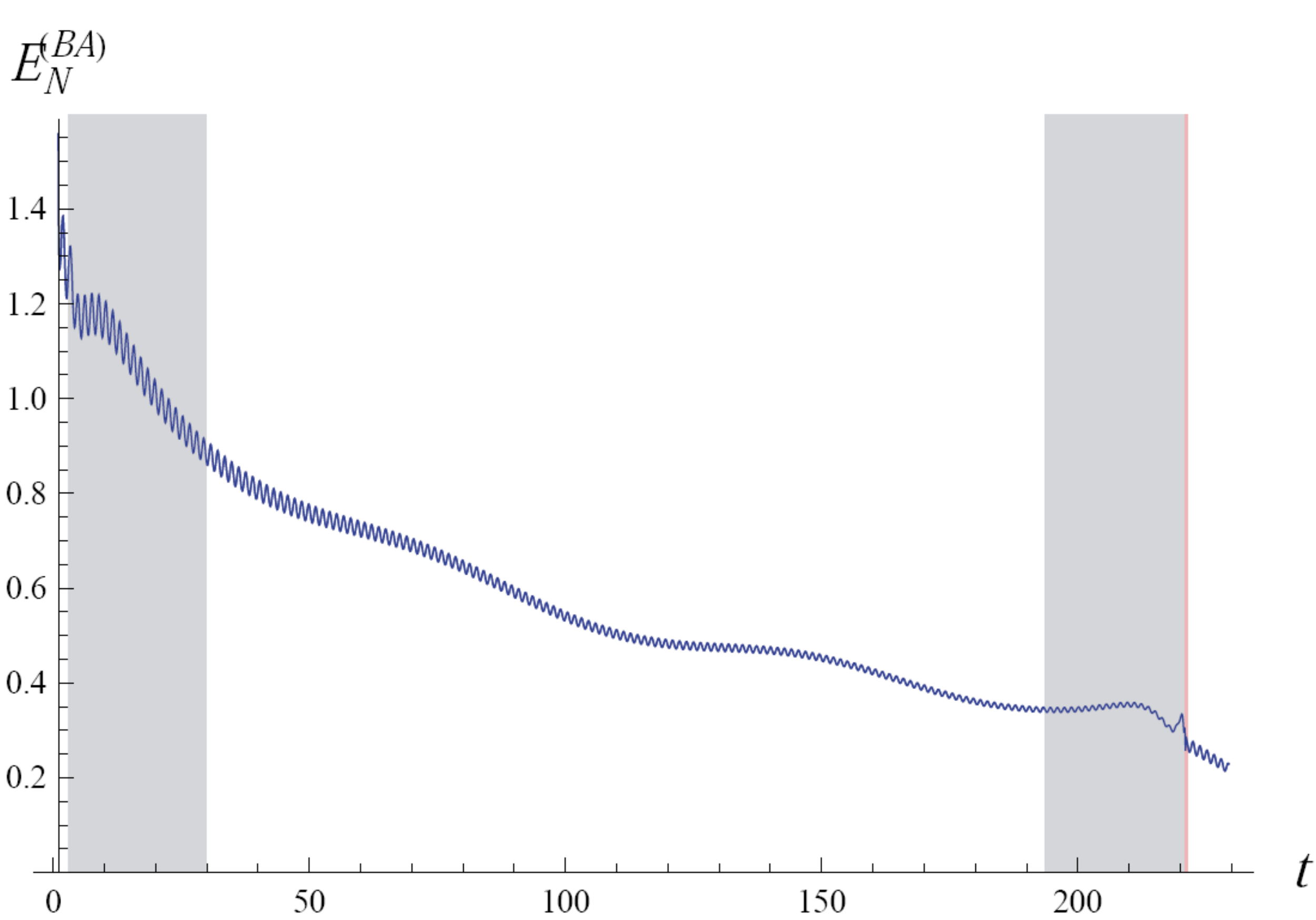} 
\includegraphics[width=5.3cm]{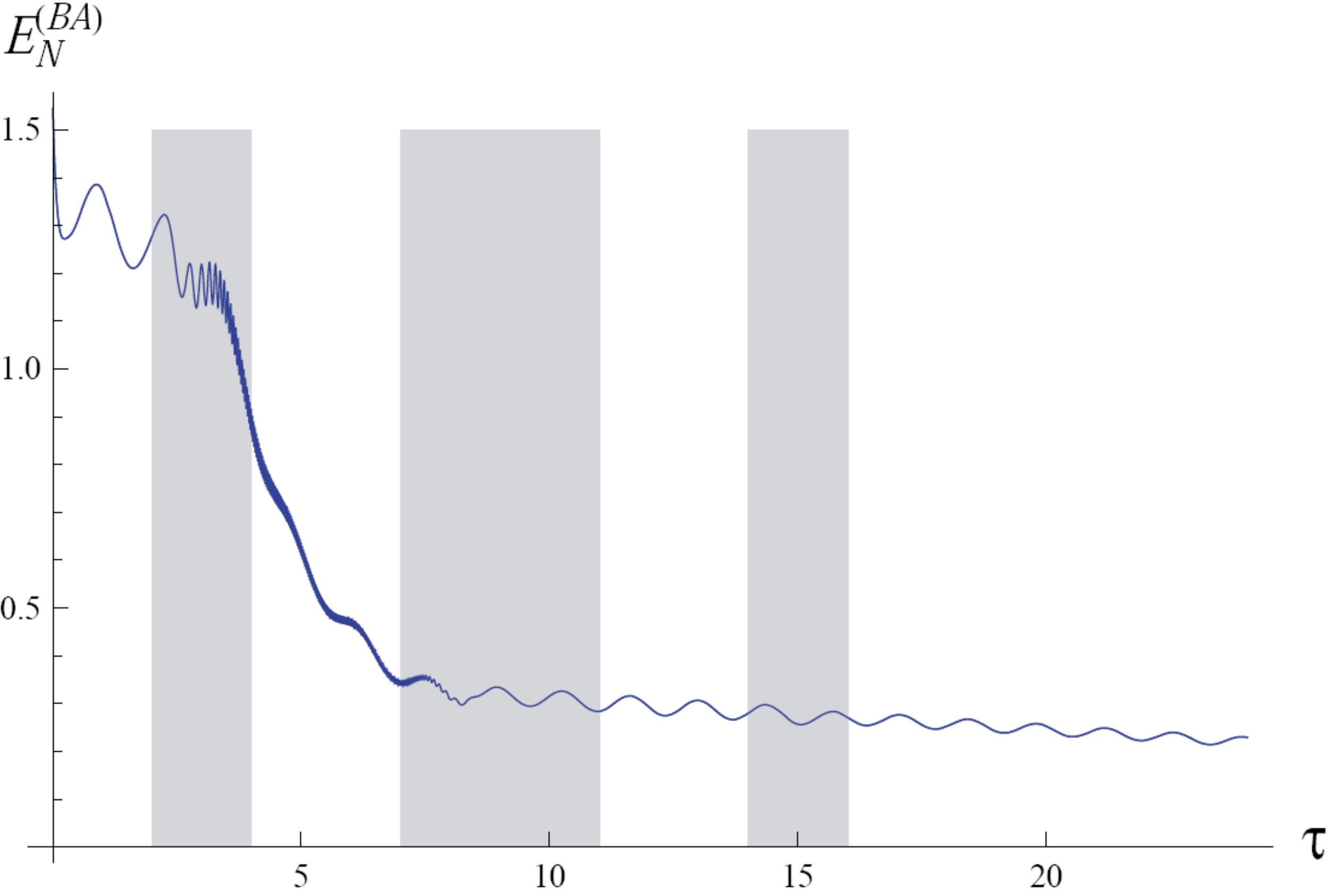} 
\includegraphics[width=5.8cm]{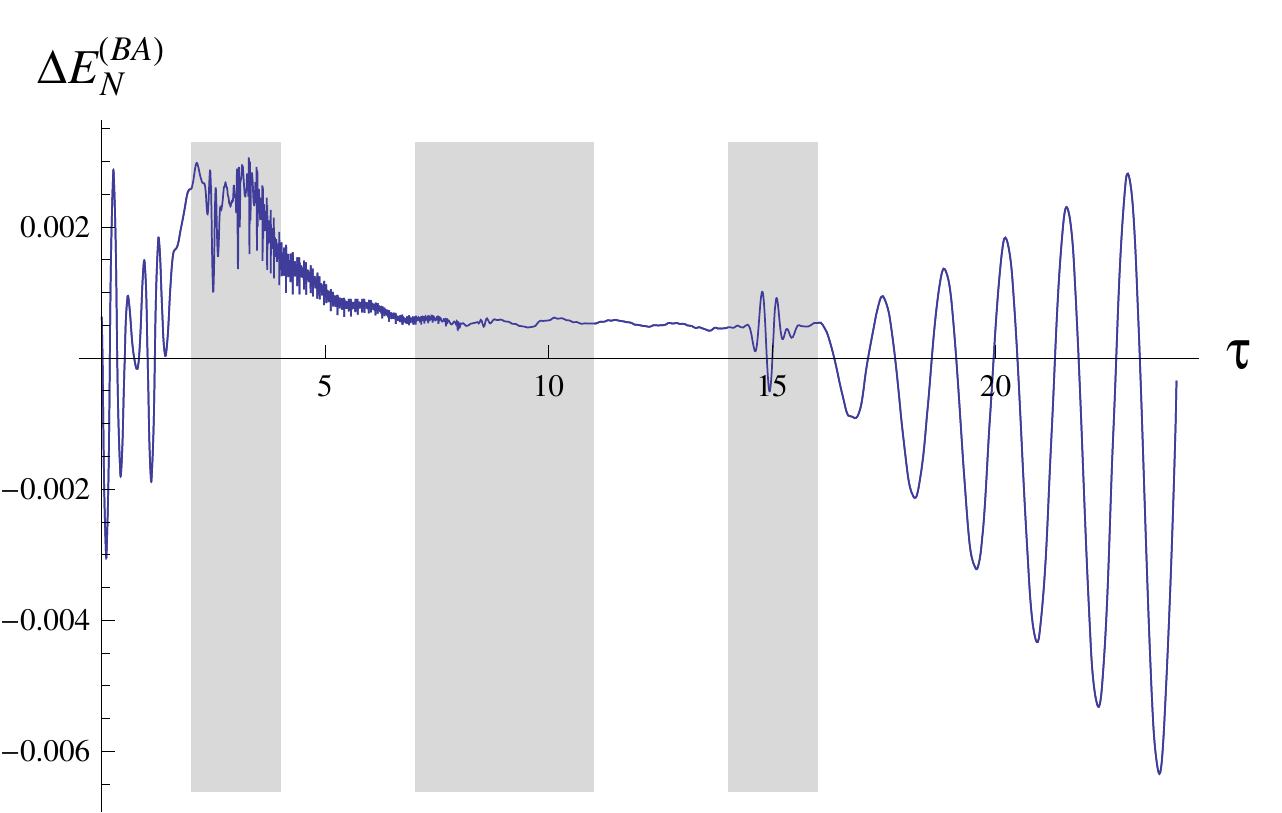} 
\caption{Dynamics of the EnLC in both teleporting directions including first-order correction from the mutual influences, in the clocks and points of view of Alice (upper-left and lower-left plots) and Bob (lower-middle and upper-middle plots).
The gray and pink regions in these plots represent the three time intervals when Alice's signal reaches Bob
or Bob's signal is sent to Alice during Bob's acceleration phase (the leftmost diagrams).
Here, $\gamma=0.001$, $\Omega=2.3$, and $\Lambda_0=\Lambda_1=20$, $a=2$, $(\bar{\alpha}, \bar{\beta})=(1.4, 0.2)$, and the initial or
final spatial separation $d = 1$. For Bob's worldline, we set $(\tau_0, \bar{\tau}^{}_1, \bar{\tau}^{}_2, \bar{\tau}^{}_3, \bar{\tau}^{}_4, \bar{\tau}^{}_5, \bar{\tau}^{}_6) = (0, 2, 4, 7, 11, 14, 16)$. Other parameters have the same values as those in case 2.
(Upper right) $\Delta E_{\cal N}^{(AB)}\equiv E_{\cal N}^{(AB)}-E_{\cal N}^{(AB)(0)}$ is the correction of
entanglement strength to the zeroth-order result $E_{\cal N}^{(AB)(0)}$ from the mutual influences up to the first order.
(Lower right) $\Delta E_{\cal N}^{(BA)}\equiv E_{\cal N}^{(BA)}-E_{\cal N}^{(BA)(0)}$ is similar.}
\label{EntDyn}
\end{figure}

With the results of the correlators we are able to calculate the dynamics of the EnLC in both teleporting directions.
Our first example is shown in Fig. \ref{EntDyn}. In the left plots, one can see similar decays of $E_{\cal N}^{(AB)}$ (corresponding to
the QT from Alice to Bob) in Alice's clock and $E_{\cal N}^{(BA)}$ (from Bob to Alice) in Alice's point of view.
While in the middle plots the two curves in Bob's clock or point of view drop significantly in different periods, the values of
$E_{\cal N}^{(AB)}$ and $E_{\cal N}^{(BA)}$ around the moment when Bob comes back to Alice are quite the same.
Once again, the details of the history depend on the point of view,
but here we further see that different views on the EnLC tend to agree when Bob rejoins Alice. 
The reason is simple. When two detectors are close enough, the amplitudes of the mode functions in
the operators $Q$, $P$ of detectors $A$ and $B$ at $\tau_{}^A=t$ and $\tau_{}^B=\tau_{}^{adv}(t)$, respectively, are relatively close to
the ones at $\tau_{}^A=t_{}^{adv}(\tau)$ and $\tau_{}^B=\tau$ if $d \ll c/\gamma$. So, these operators give similar expectation values of the two-point correlators with respect to the same initial state.
In the case Rob never returns, as in the Alice-Rob problem studied in the previous section, $E_{\cal N}^{(AB)}$ and $E_{\cal N}^{(BA)}$ in different teleporting directions will never be commensurate after the initial moment.

In our example, the mutual influences tend to enhance the entanglement during the space journey of Bob.
Denote the zeroth-order results of the logarithmic negativities for the EnLC as $E_{\cal N}^{(0)}$, and the enhancement by the mutual influences as $\Delta E_{\cal N}^{} \equiv(E_{\cal N}^{} - E_{\cal N}^{(0)})$.  In the right plots of Fig. \ref{EntDyn}, we find both $\Delta E_{\cal N}^{(AB)}$ and $\Delta E_{\cal N}^{(BA)}$ grow from zero to some value when Bob launches ($\tau$,
$\tau_{}^{adv} \approx \bar{\tau}^{}_1$), and then during Bob's journey $\Delta E_{\cal N}^{(AB)}$ and $\Delta E_{\cal N}^{(BA)}$ roughly remain constant between $+0.0014$ to $+0.002$, which is of the same order as $\gamma/d \approx 10^{-3}$. However, when Bob returns to Alice, the corrections to the logarithmic negativity from the mutual influences oscillate between positive and negative values with the amplitudes increasing in time.

Furthermore, in the right plots of Fig. \ref{EntDyn}, one can see that $\Delta E_{\cal N}^{(AB)}$ appears to be slightly ``kicked" at about $t\in (t_{adv}(\bar{\tau}^{}_4), t_{adv}(\bar{\tau}^{}_6))\approx (220.88, 221.43)$ and $\Delta E_{\cal N}^{(BA)}$ at about $\tau \approx 15 \in (\bar{\tau}^{}_5, \bar{\tau}^{}_6)$.
This could be due to the shock waves emitted by detector $B$ during $\tau \in (\bar{\tau}^{}_4, \bar{\tau}^{}_6)$
that all hit detector $A$ at $t \approx 221$. 
In our results the impact of the first-order correction never gets significant compared to the zero-order correlators.

\subsection{Quantum teleportation}

\begin{figure}
\includegraphics[width=7.5cm]{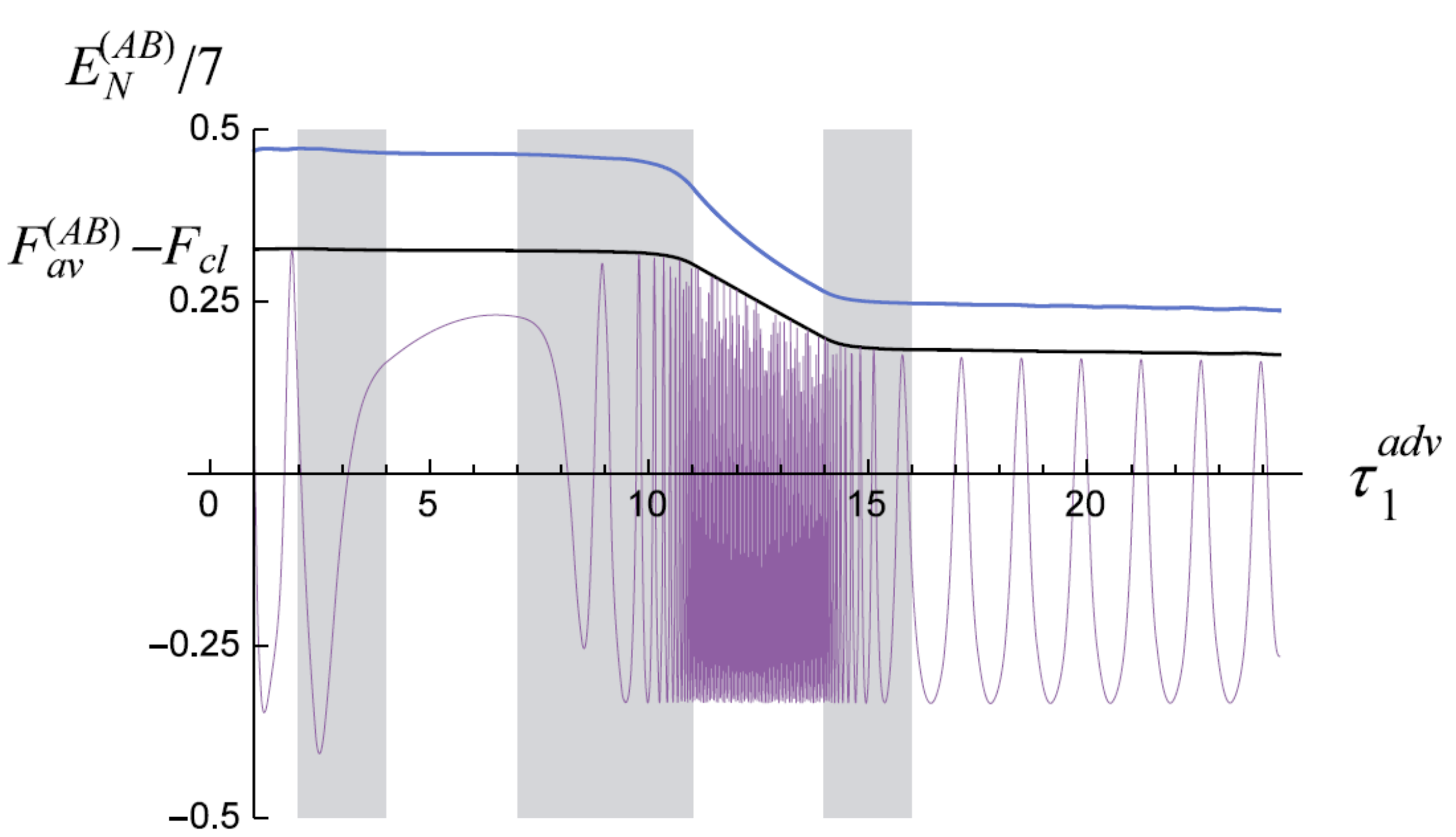}
\includegraphics[width=7.5cm]{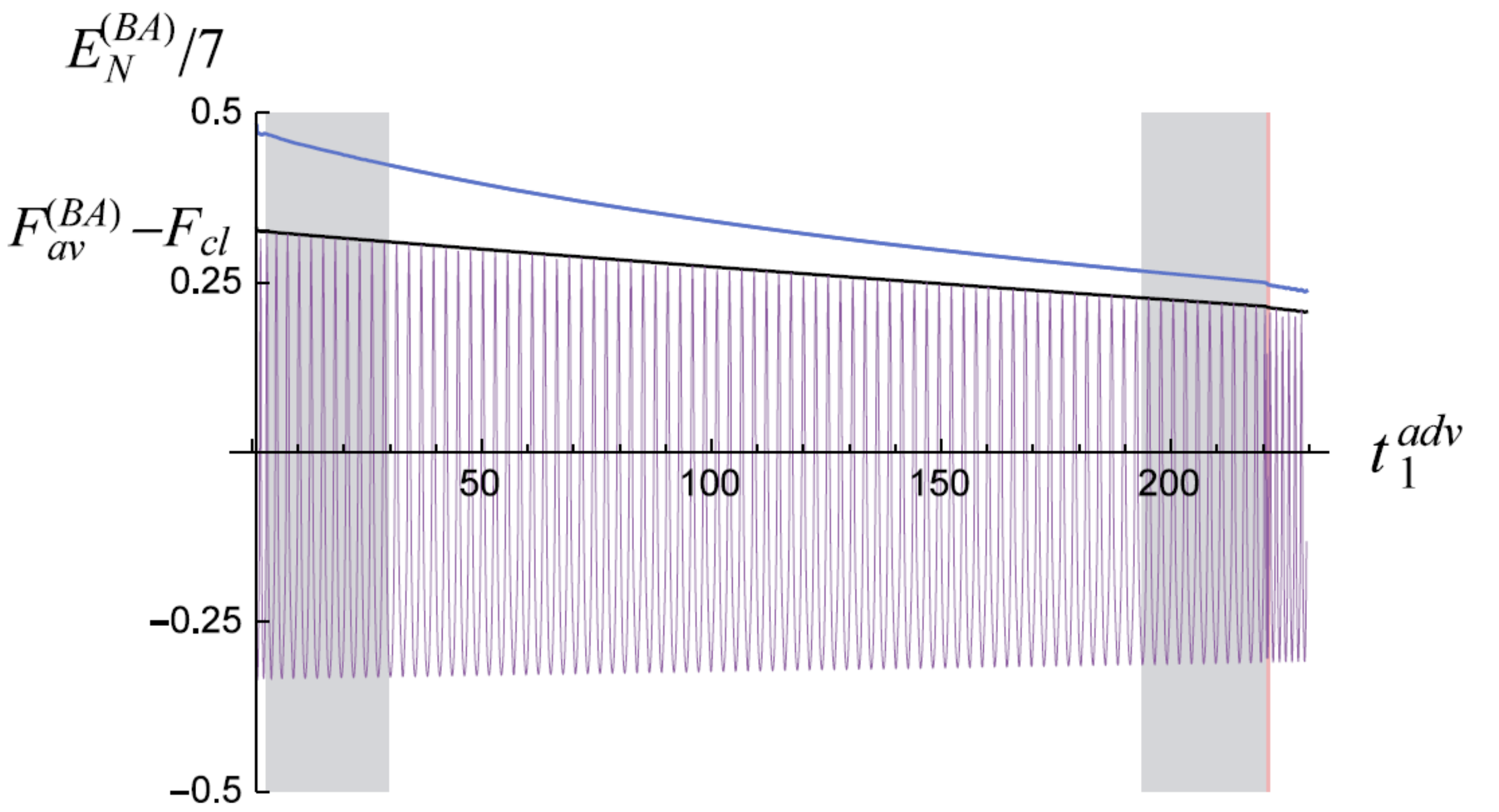}
\caption{The averaged FiQT of a coherent state of detector $C$ from Alice to Bob ($F_{av}^{(AB)}$)
and from Bob to Alice ($F_{av}^{(BA)}$) with (black curves) and without (purple) using the improved protocol
in the viewpoints of Bob (left) and Alice (right), respectively.
Here the entangled pair starts initially with $(\bar{\alpha},\bar{\beta})= (e^{-r_1}/\sqrt{\Omega}, e^{-r_1}\sqrt{\Omega})$, $r_1 = 1.2$,
and we assume the joint measurements of detectors $C$ and $A$ by Alice or $C$ and $B$ by Bob collapse the measured detector pair to
a squeezed state with squeeze paramater $r_2 =1.1$. Other parameter values are the same as in the previous figures.
The scaled logarithmic negativities of EnLC with the same parameters are plotted in blue curves for comparison.
One can see that the evolution of $E_{\cal N}$ in time is similar to $F_{av}^+-F_{cl}$.}
\label{FavTwin}
\end{figure}

Next, to compare the averaged FiQT, we set $(\bar{\alpha},\bar{\beta}) = (e^{-r_1}\sqrt{\hbar/\Omega}, e^{-r_1}\sqrt{\hbar\Omega} )$, $r_1=1.2$ for the initial state of the entangled pair of the detectors as the one in the previous section.
The results are shown in Fig. \ref{FavTwin}. Again one can see that the evolutions of the best averaged FiQT $F_{av}^+$ in either teleporting direction subtracted by $F_{cl}$ is similar to the evolution of the logarithmic negativity
$E_{\cal N}$ of the EnLC of detectors $A$ and $B$.

We keep the curves for the averaged fidelities $F_{av}$ without using the improved protocol in the upper row of
Fig. \ref{FavTwin} to give the readers a flavor how the sender's clock is observed by the receiver [recall Eqs. (\ref{weakFav}) and
(\ref{X2Y2})]. One can see that there is no significant enhancement of decay for $F_{av}^+$ or $E_{\cal N}$ due to
the Unruh effect when Bob is in any acceleration phase (gray or pink regions), since we take the proper acceleration $a=2$ for Bob,
which is not too large there. In contrast, significant drops of $F_{av}^{(AB)+}$ or $E_{\cal N}^{(AB)}$ in Fig. \ref{FavTwin} (left) happen between the second and the third acceleration phases, when Bob sees a strong blueshift in the clock signal emitted by Alice
and so Alice's clock looks much faster than Bob's in his viewpoint during this period
[Alice's signal emitted during $(\bar{t}^{}_4, \bar{t}^{}_5) = (28.836, 192.63)$ reaches Bob during the
period $(\tau^{adv}_{}(\bar{t}^{}_4), \tau^{adv}_{}(\bar{t}^{}_5) )= (\bar{\tau}^{}_4, \bar{\tau}^{}_5) = (11, 14)$].
This implies that quantum coherence of detector $A$ in this period fades much more quickly than any other period in Bob's viewpoint
so that quantum entanglement and the best averaged FiQT are degraded faster in this stage.
The significant drops of the EnLC in the middle plots of Fig. \ref{EntDyn} are due to the same reasons.
For $F_{av}^{(BA)+}$ and $E_{\cal N}^{(BA)}$ in Fig. \ref{FavTwin} (right), the drop is much less significant, though. This is
because the period in which Alice receives similar blueshift clock signal from Bob is much shorter than the time scales of decoherence
($1/\gamma = 1000$) either in Bob's clock ($\bar{\tau}^{}_5-\bar{\tau}^{}_4 =3$) or in Alice's point of view ($t^{adv}(\bar{\tau}^{}_5)-t^{adv}(\bar{\tau}^{}_4) = (\bar{\tau}^{}_5-\bar{\tau}^{}_4)e^{-a \bar{\tau}_p} \approx 0.055$). 

In the above cases we have seen that the relativistic effects play a dominant role in QT. One can ask when the Unruh effect will become more significant in the QT from Alice to Bob. Our results so far show that this happens only in Bob's point of view and only when Bob's proper acceleration $a$ is large enough (see Figure \ref{EnLCvsa} (upper row), for example). In other words, only in a highly accelerated receiver's point of view can this happen.
One can construct setups in which the Unruh effect can be singled out, such as those with both detectors uniformly accelerated or in alternating uniform acceleration [Fig. \ref{AUAtwin} (middle and right)], 
but then the receiver is also accelerated in these setups after all.
Is it possible for a receiver in QT remaining at rest to see the domination of the Unruh effect? With this aim, we construct below a setup with Alice at rest, while the relativitic effects of time dilation and varying retarded distance are suppressed and the Unruh-like effect are significant in QT in both directions.

\section{Case 4---Traveling twin in alternating uniform acceleration}

\begin{figure}
\includegraphics[width=2cm]{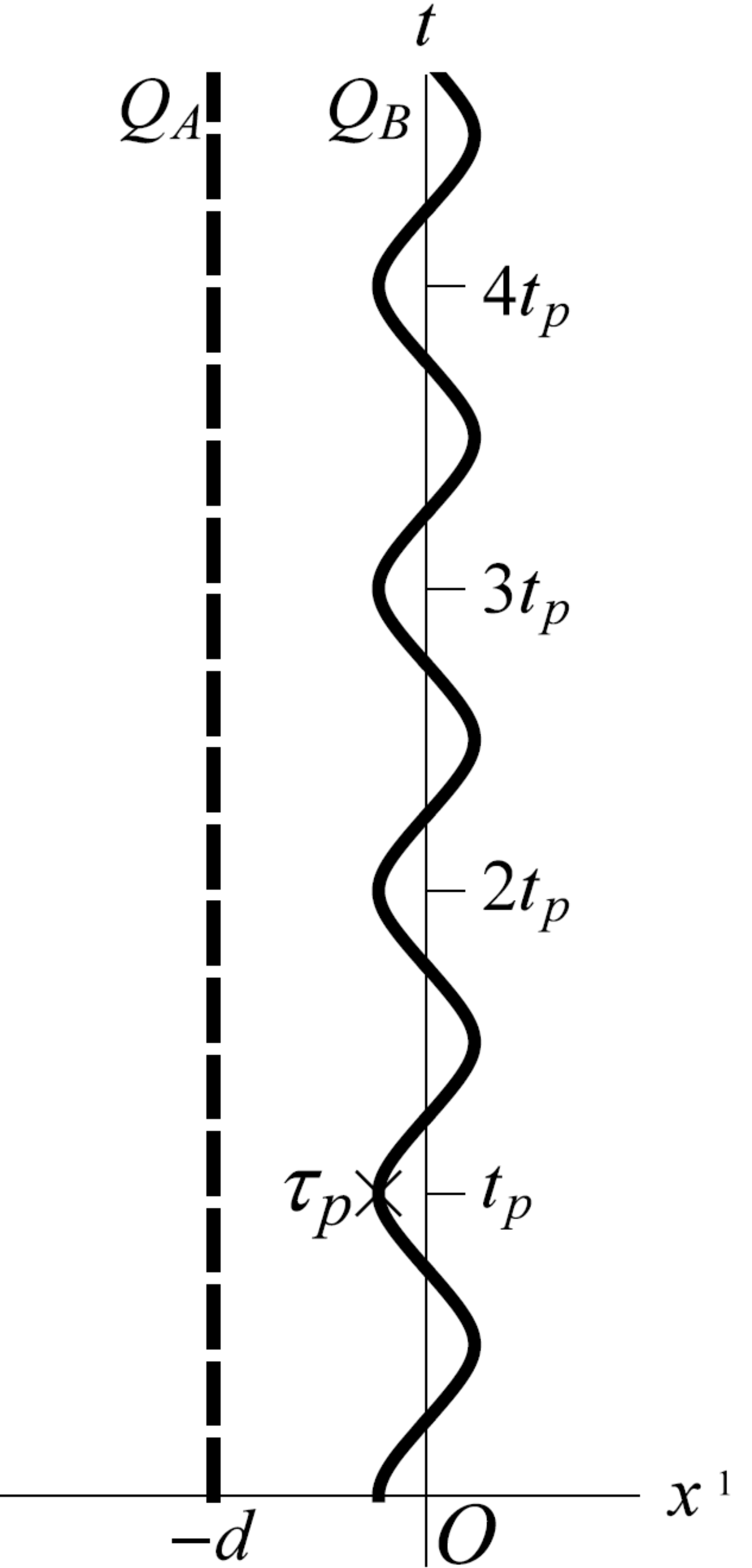}\hspace{1cm}
\includegraphics[width=3.3cm]{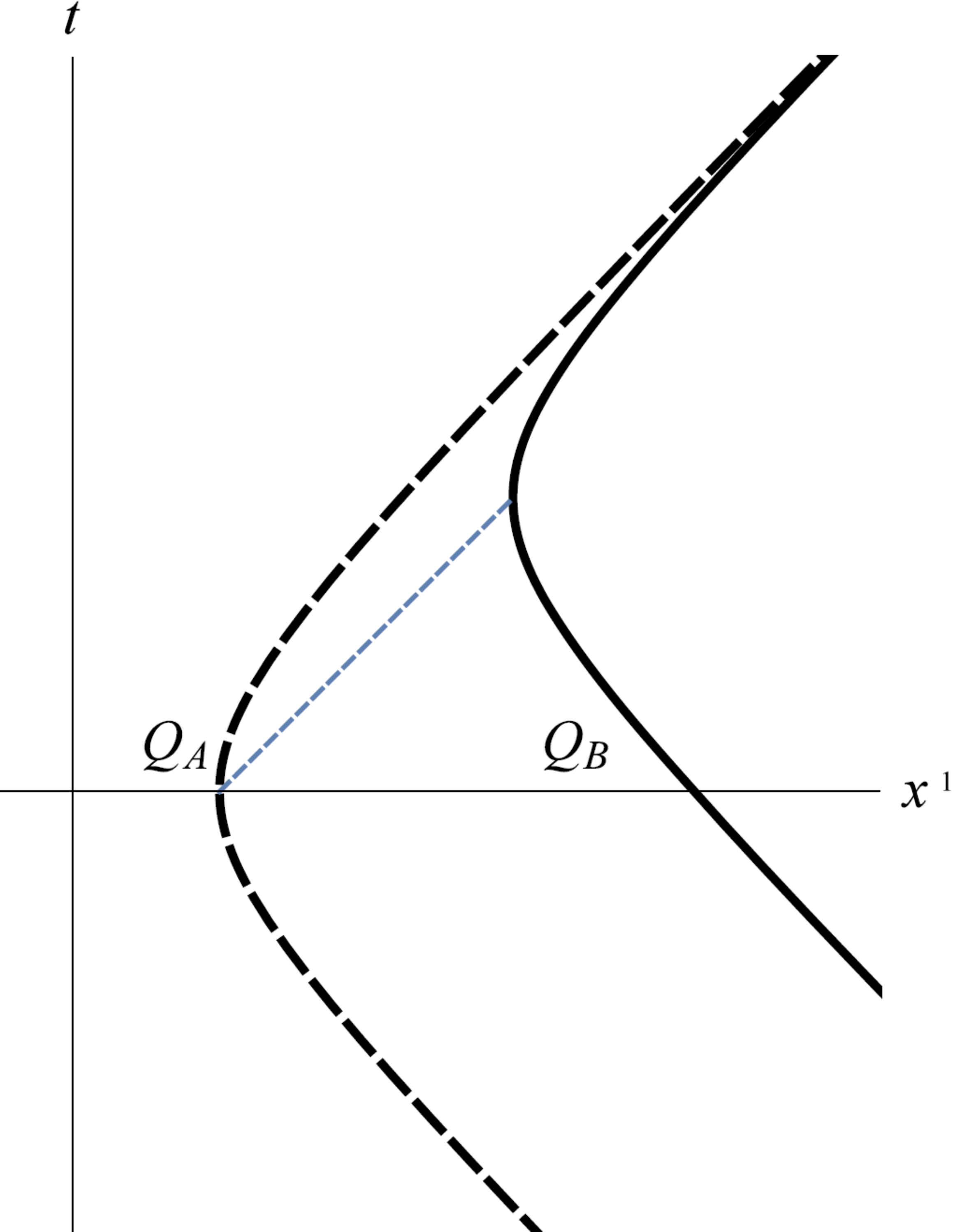}\hspace{1cm}
\includegraphics[width=2.4cm]{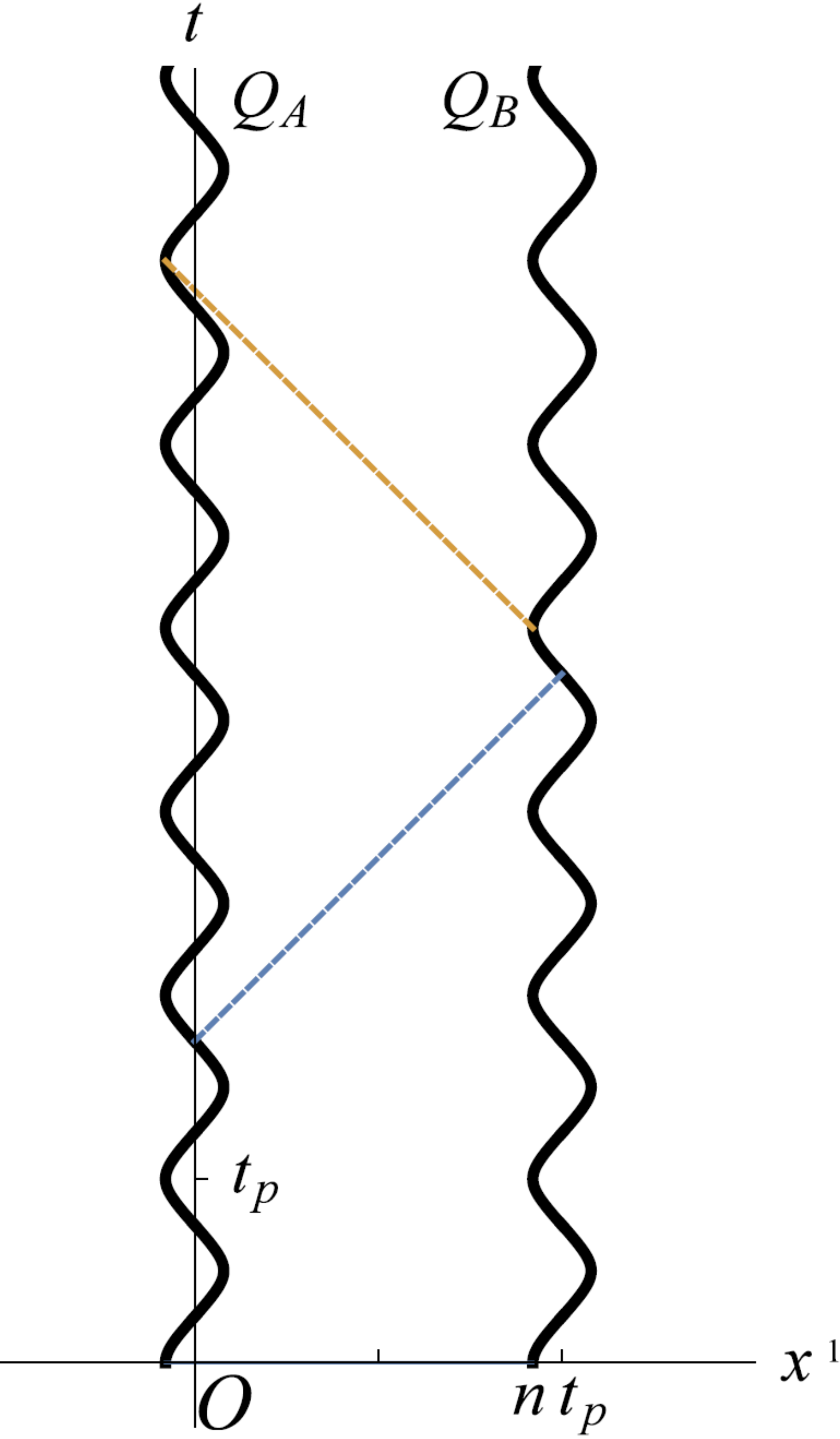}
\caption{(Left) QT from Alice (thick dotted) to Rob (thick solid) in a variation of the twin problem, where
the traveling twin Bob is in alternating uniform acceleration with the worldline (\ref{AUAtraj}).
One can conjure up settings that single out the Unruh effect, such as letting both Alice and Bob be uniformly accelerating (middle) or both in alternating uniform acceleration (right), where $n$ is an integer.
Note that in the middle plot the relativistic effects in affecting the description of the dynamics are totally suppressed only in the one-way QT from Alice to Bob, but not from Bob to Alice.}
\label{AUAtwin}
\end{figure}

To highlight the regimes in which the Unruh effect stands out in comparison with other relativistic effects, we design a case in which Bob the traveling twin undergoes an alternating uniform acceleration (AUA) considered in Ref. \cite{DLMH13} with the period of motion so short that the maximum speed of Bob is low enough and the retarded distance between Alice and Bob does not vary too much, while the proper acceleration can still be very high. Consider the case with Alice at rest along the worldline $(t, -d, 0, 0)$ and Bob going along the worldline
\begin{equation}
    z_B^\mu(\tau)=\left({1\over a}\left[\sinh a\left(\tau- n{\bar{\tau}^{}_p\over 2}\right) +2n\sinh a{\bar{\tau}^{}_p\over 4} \right],
		 {(-1)^n\over a}\left[\cosh a\left(\tau -n{\bar{\tau}^{}_p\over 2}\right) -\cosh a {\bar{\tau}^{}_p\over 4} \right],0,0 \right)
\label{AUAtraj}
\end{equation}
with $n(\tau) \equiv {\rm Floor}\{ (2\tau/\bar{\tau}^{}_p) + (1/2) \}$, linearly oscillating in the $x^1$ axis about the spatial origin
[see Fig. \ref{AUAtwin} (Left)], where $\bar{\tau}^{}_p$ is the period of Bob's oscillatory motion
in his proper time. In this case the classical light signal emitted by Alice at $t$ will reach Bob at
\begin{equation}
  \tau^{adv}(t) = \tilde{n} {\bar{\tau}^{}_p\over 2} - {(-1)^{\tilde{n}}\over a} \log \left\{\cosh a{\bar{\tau}^{}_p\over 4}+
	    (-1)^{\tilde{n}}\left[ 2 \tilde{n} \sinh a {\bar{\tau}^{}_p\over 4}- a(t + d) \right]  \right\},
\end{equation}
where $\tilde{n}(t) \equiv {\rm Floor}\{ (2t/\bar{t}^{}_p) + (1/2) \}$ with $\bar{t}^{}_p \equiv 4a^{-1}\sinh(a \bar{\tau}^{}_p/4)$,
while the classical light signal emitted by Bob at $\tau$ will reach Alice at
$t^{adv}(\tau) = d + z_B^0(\tau) + z_B^1(\tau)$.
To compare with cases 2 and 3 in which the mutual influences are small, the retarded distance between Alice and Bob
is set to be large enough. Also when the period of motion is much less than the natural period of the detector
($\bar{\tau}_p \ll T \equiv 2\pi/\Omega$), the time-averaged subtracted Wightman function will be a good approximation
in calculating the self-correlators of detector $B$ (see Sec. 5.1 in Ref. \cite{DLMH13}).
These assumptions simplify the calculation very much in the weak coupling limit.

\begin{figure}
\includegraphics[width=6.7cm]{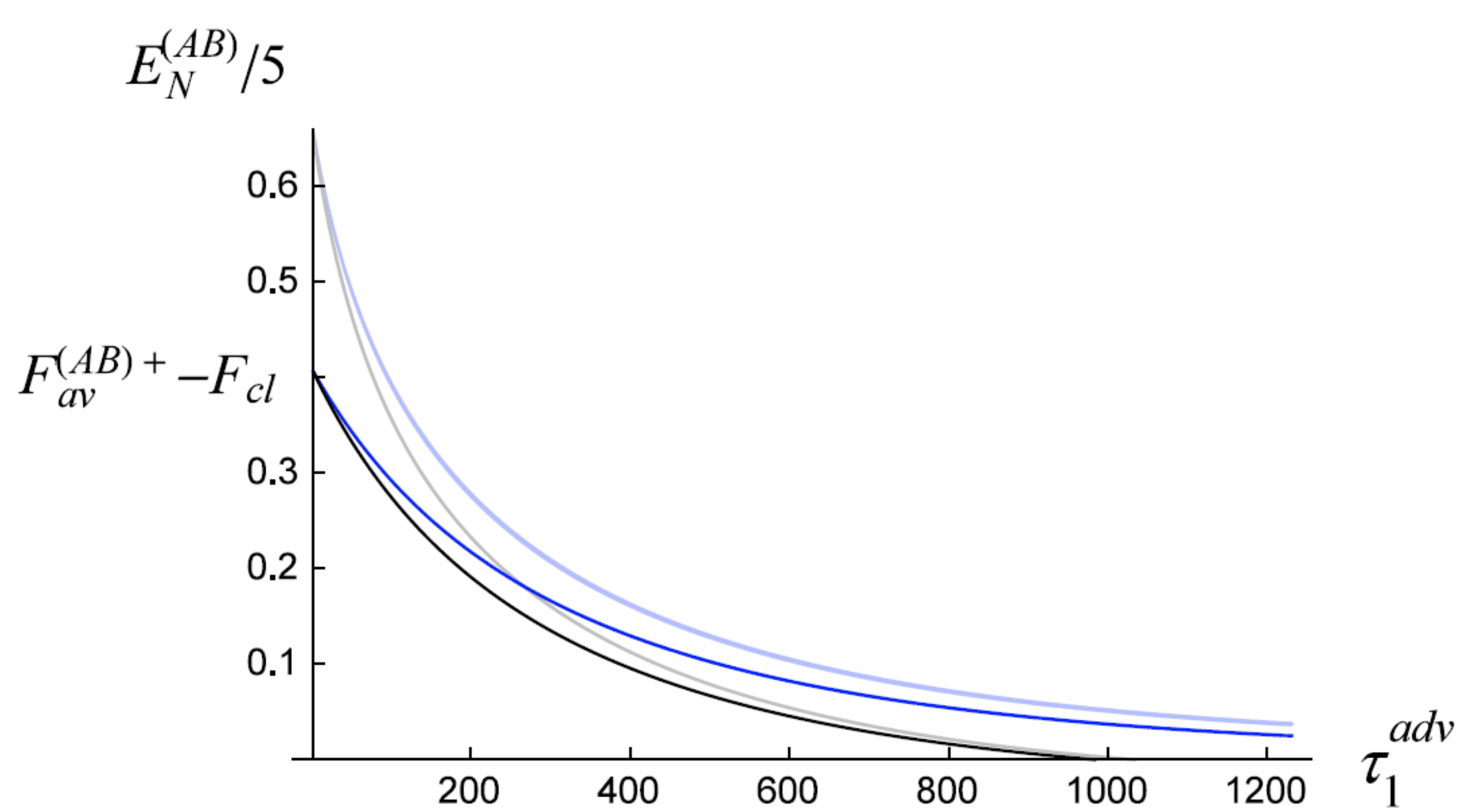}
\includegraphics[width=5.5cm]{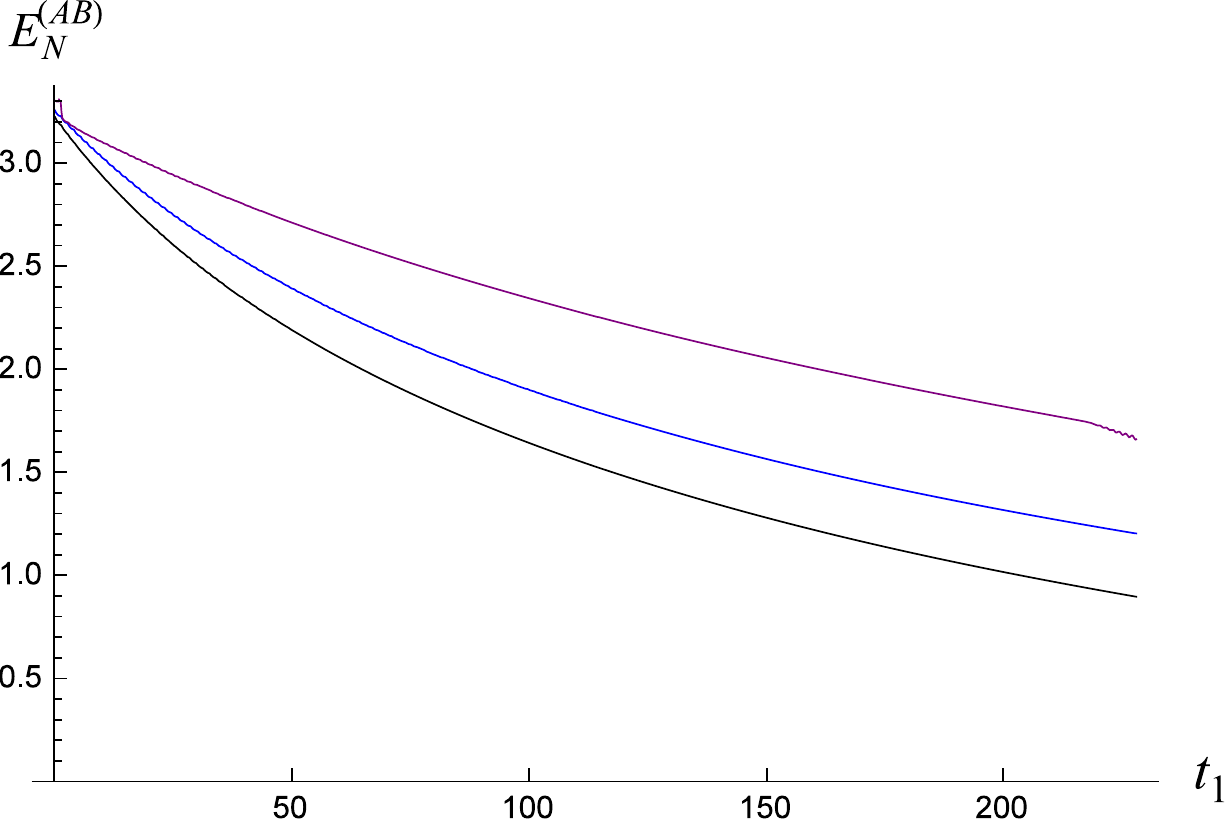}
\includegraphics[width=5.5cm]{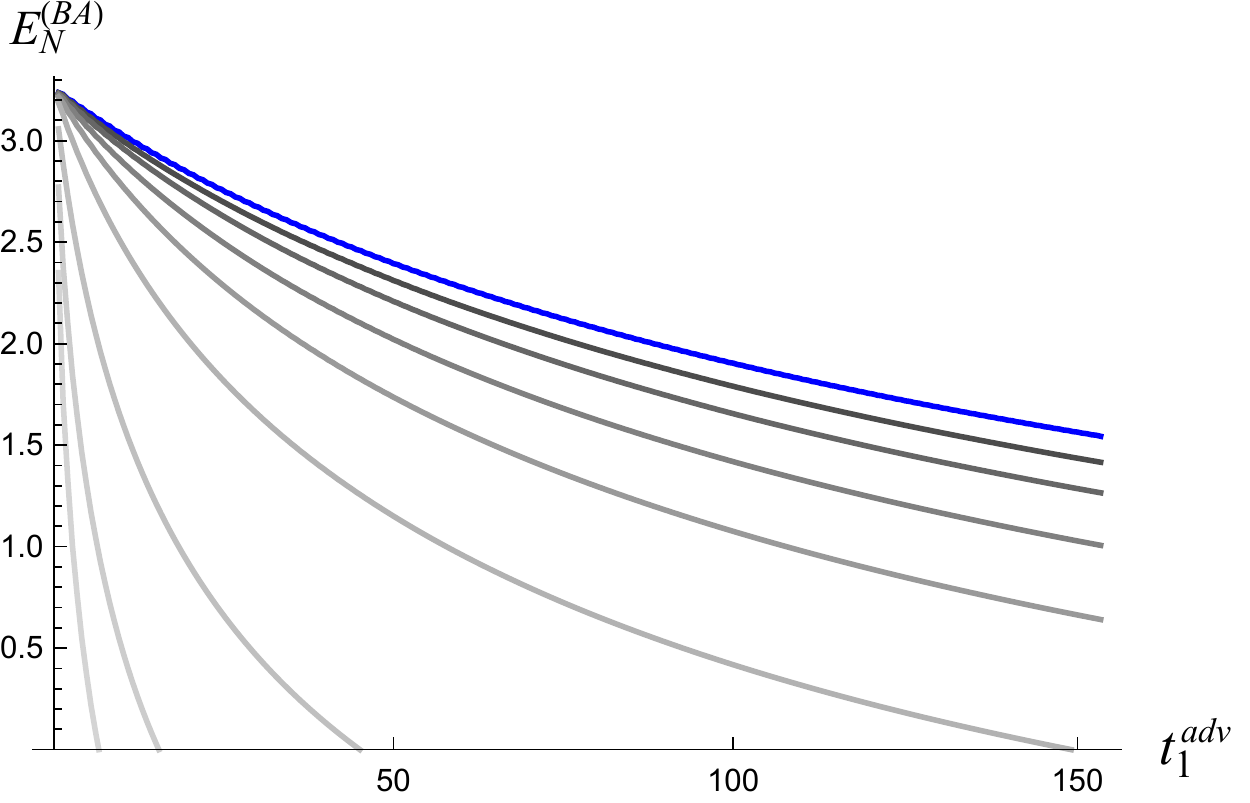}\\
\includegraphics[width=6.7cm]{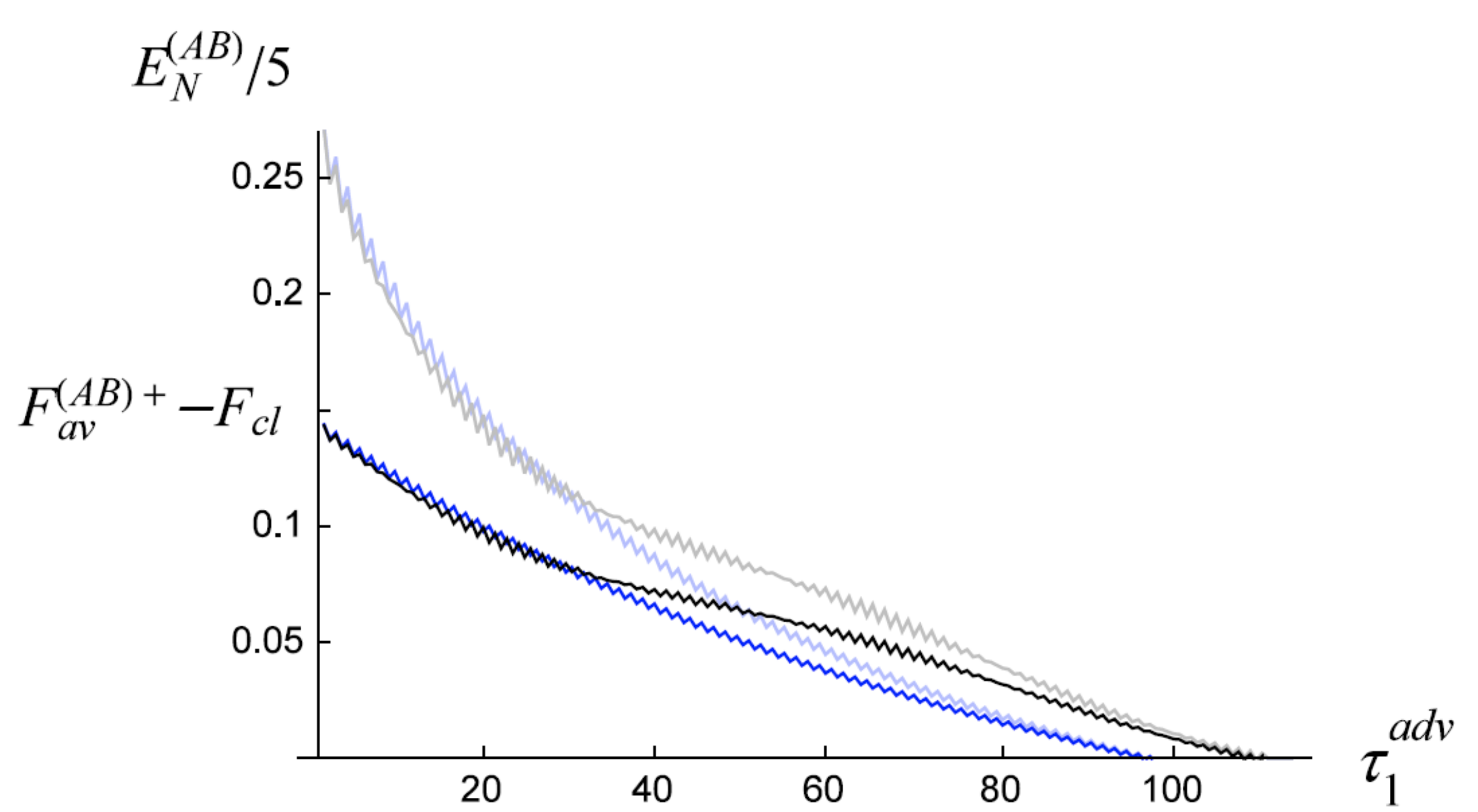}
\includegraphics[width=5.5cm]{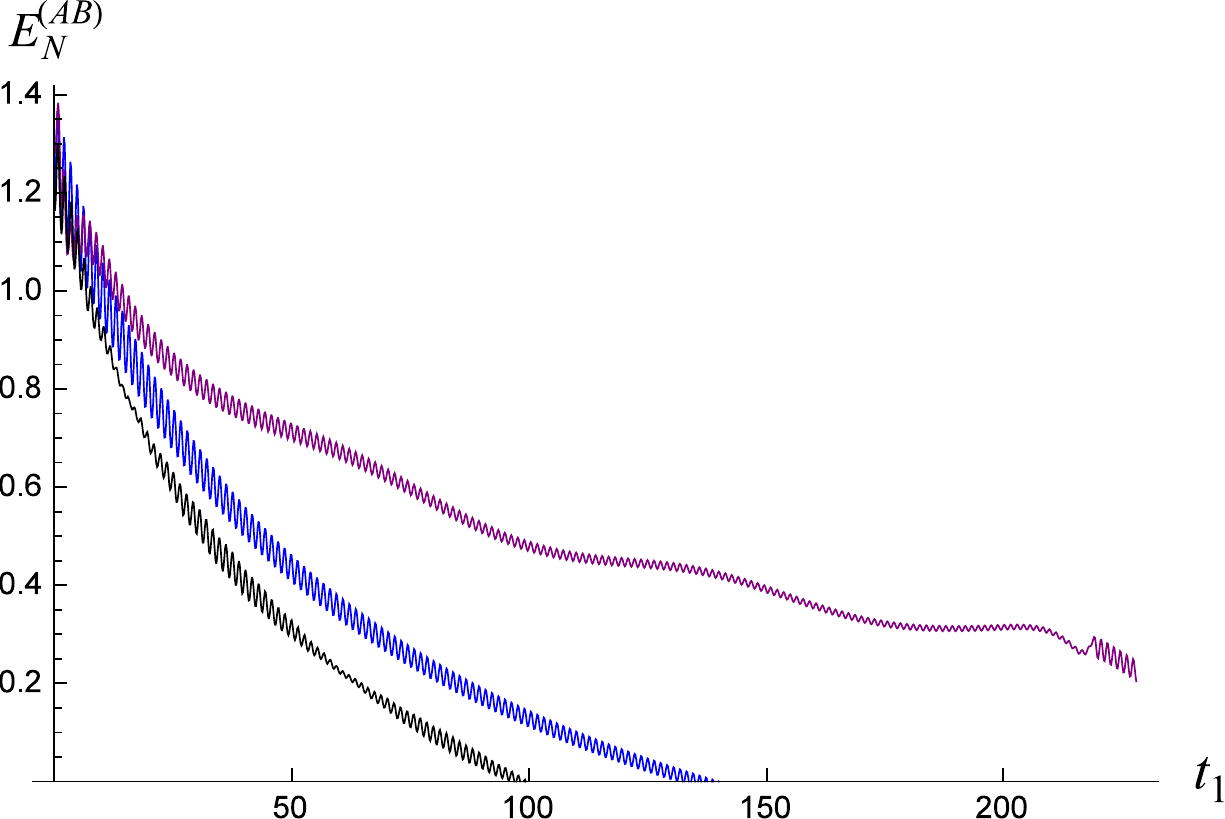}
\includegraphics[width=5.5cm]{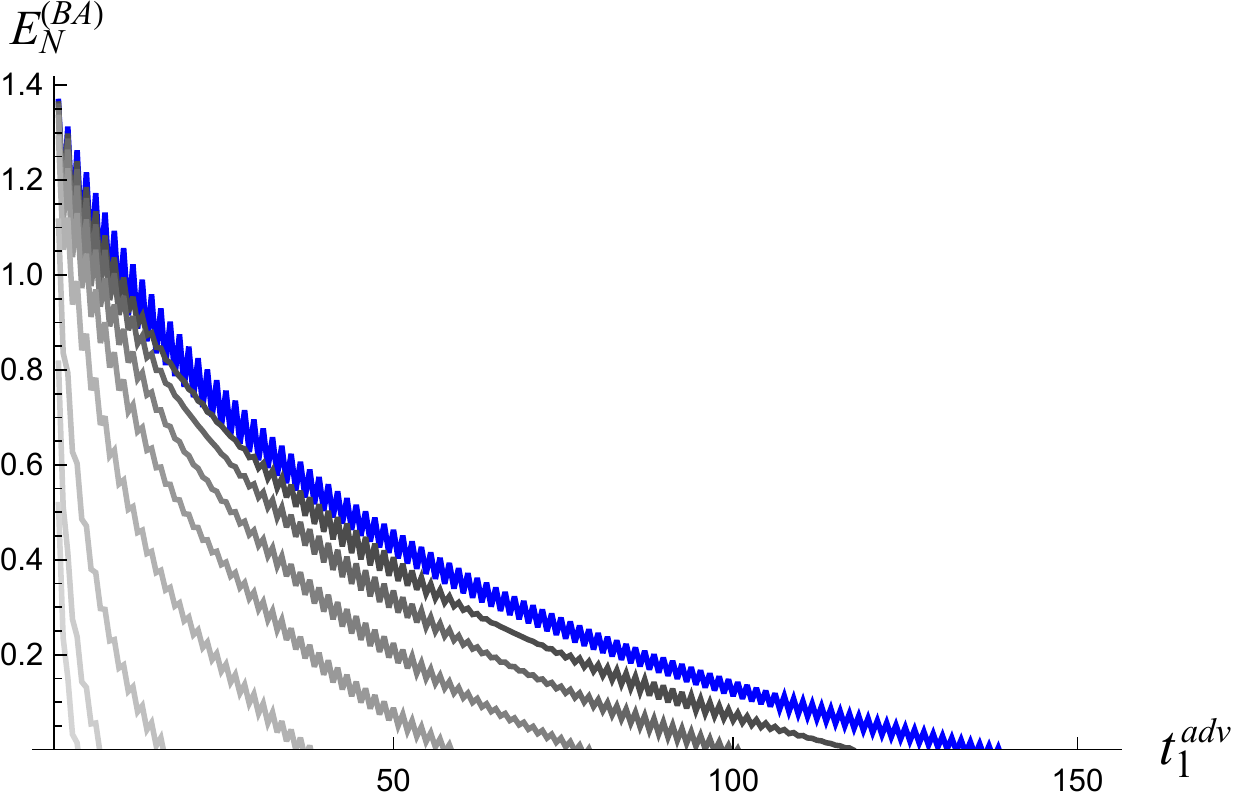}
\caption{Dynamics of the EnLC and the FiQT between Alice and Bob with Bob at rest (blue curves), in AUA (gray and black), and as in the twin problem (purple). The mutual influences are ignored, and the initial state of the $AB$ pair has $(\bar{\alpha},\bar{\beta})= (e^{-r_1}/\sqrt{\Omega}, e^{-r_1}\sqrt{\Omega})$ with $r_1=1.2$ (upper row), or $(1.4, 0.2)$ (lower).
(Left) The scaled $E_{\cal N}^{(AB)}$ of the EnLC (lighter) and $F_{av}^{(AB)+}$ (darker) from Alice to Bob subtracted by the classical fidelity $F_{cl}=1/2$, with $d=1$ both for Bob in AUA and at rest, in Bob's point of view.
Bob in AUA has $a=10$ and the period of his oscillatory motion $\bar{\tau}_p = T/16$, $T\equiv 2\pi/\Omega$.
The squeezed parameter in $\rho_{AC}^{(\beta)}$ is $r_2 =5.1$, and other parameters are the same as before. 
In the lower-left plot the teleported state has $r_0=\log 2$.
(Middle) Comparison of the EnLC between Alice at $t_1$ and Bob at $\tau^{adv}(t_1)$ in different motions in Alice's clock.
Here, $d=4$,
$a=2$ in the twin problem and $a=20$ in the AUA case in which $\bar{\tau}_p = T/32$ for Bob.
(Right) Dynamics of the EnLC between Bob at $\tau_1$ and Alice at $t^{adv}_1\equiv t^{adv}(\tau_1)$ in Alice's point of view,
where Bob is at rest (blue, $a=0$) or undergoes AUA (gray, from dark to light $a=2^n\cdot 10$, $n=0$ to $7$
with $a \bar{\tau}_p = 10T/16$ fixed). Again $d=4$ with other parameters unchanged.}
\label{ENFav134}
\end{figure}

We show some selected results in Fig. \ref{ENFav134}. For the logarithmic negativity $E_{\cal N}^{(AB)}$ of the EnLC and the best averaged FiQT $F_{av}^{(AB)+}$ from Alice to Bob in Alice's clock or in Bob's point of view, when $a$ is small and $\bar{\tau}^{}_p$ is large, the disentanglement time for the EnLC of the joint measurement by Alice is still longer than the one in case 1 with the same parameters except $a=0$. Here, time dilation of detector $B$ dominates.
As $a$ gets larger, with the maximum speed fixed ($a \bar{\tau}^{}_p =$constant), one starts to see the evolution curves for $E_{\cal N}^{(AB)}$ and $F_{av}^{(AB)+}$ drop faster than the ones with $a=0$ in some parameter range of $(\bar{\alpha},\bar{\beta})$ for the initial state (\ref{rhoABI}) [Fig. \ref{ENFav134} (Left) in Bob's point of view; the plots in Alice's clock look similar].
When $a$ is large enough, the initial states with all values of $(\bar{\alpha}, \bar{\beta})$ will see faster degradations of the EnLC and the best averaged FiQT,  both in Alice's clock or in Bob's point of view than those in the $a=0$ case [Fig. \ref{ENFav134} (Middle)].
Now, we can say that the Unruh effect dominates, though the effective temperature experienced by detector $B$ is lower than the Unruh temperature with the averaged proper acceleration $a$ \cite{DLMH13}.
In the reverse teleporting direction, for the logarithmic negativity $E_{\cal N}^{(BA)}$ of the EnLC in Alice's point of view, we see clearly that the larger $a$ is, the shorter the disentanglement time in Fig. \ref{ENFav134} (right), where
the Unruh effect has been dominating the degradation of $E_{\cal N}^{(BA)}$ from $a=10$ for all values of
$(\bar{\alpha}, \bar{\beta})$, while $E_{\cal N}^{(AB)}$ with $a=10$ still has a longer disentanglement time than the one with $a=0$
in a corner of the parameter space around $(\bar{\alpha}, \bar{\beta})\approx (1.4,0.2)$,
as shown in the lower-left plot of Fig. \ref{ENFav134} .

One interesting observation in calculating Fig. \ref{ENFav134} (lower left) is that when $a$ is large enough 
the averaged FiQT of a coherent state using the entangled $AB$ pair initially with $(\bar{\alpha},\bar{\beta})$ in some finite parameter
range will never achieve $F_{av}^{(AB)}$ or $F_{av}^{(BA)} \ge F_{cl} = 1/2$. One has to modify the quantum state to be teleported from a coherent state to a squeezed coherent state with the squeezed parameter $r_0 > 0$ in Eq. (\ref{rhoCI}) and tune the value of $r_0$
to push the averaged FiQT above $F_{cl}$ toward the optimal fidelity $F_{opt}$ in (\ref{MVbound}), so that
the time $t_{cl}$ when $F_{av}-F_{cl}$ touches zero is closer to the disentanglement time $t_{dE}$ of the EnLC.
Note that $r_0$ itself is a-part of the protocol and not among the quantum information to be teleported.

\section{Summary and Discussion}

We have considered the quantum teleportation of continuous variables applied to three Unruh--DeWitt detectors with internal
harmonic oscillators coupled to a common quantum field. The basic properties of relativistic effects in dynamical open quantum systems such as the frame dependence of quantum entanglement, wave functional collapse, Doppler shift, quantum decoherence, and the Unruh effect have all been considered consistently and their linkage manifestly displayed. Below is a summary of what we have learned from these studies.

\subsection{Entanglement around the light cone}

Quantum entanglement of two localized objects at different positions requires the knowledge of spacelike correlations,
while the averaged FiQT involves timelike correlations between two causally connected events. In general these two quantities are incommensurate.
To compare them, in Sec. \ref{FiQTEnLC} we introduced the projection of the wave functional {\it around the future light cone} of the joint-measurement event by the sender, so that right after the wave functional collapse the sender's classical signal of the outcome reaches the receiver, according to which the receiver performs the local operation immediately. The averaged FiQT obtained in this way
can be directly compared with the degree of quantum entanglement in the entangled detector pair evaluated right before the wave functional collapse, namely, the EnLC, which can be easily calculated in the Heisenberg picture.

We have observed that the best averaged FiQT always drops below the fidelity of classical teleportation earlier than the disentanglement
time for the EnLC in each of our numerical results. This confirms the inequality (\ref{MVbound}), which implies that entanglement of
the detector pair is a necessary condition for the averaged FiQT beating the classical fidelity. In Sec. \ref{LT2IHO}, we further
showed that the inequality (\ref{MVbound}) may appear to be violated by the degrees of quantum entanglement
evaluated on a time slice in conventional coordinate systems. This proves that the EnLC, rather than the conventional ones,
is essential in QT in a relativistic open quantum system.

\subsection{Multiple clocks and points of view}

For a relativistic system including both the local and nonlocal objects such as a detector-field interacting system, the Hamiltonian,
quantum states, and quantum entanglement extracted from the states all depend on the choice of the reference frame \cite{LCH08}.
Part of the coordinate dependence can be suppressed by evaluating the physical quantities
around the future or past light cones of a local observer. However, this does not give a unique description on a physical process,
since each local object has a clock reading its own proper time, which is invariant under coordinate transformations.
In particular, a QT process involves two different physical clocks for the sender and the receiver localized in space,
and the degradation of the EnLC and the averaged FiQT in the same process can appear very differently in the sender's clock and in the
receiver's point of view along his/her past light cone. When describing nonlocal physical processes with local objects
in a relativistic open quantum system, one has to first specify which clock or which point of view being used;
otherwise, there will be ambiguity in the statements.

\subsection{time dilation, Doppler shift, and acceleration}

It is easy to understand that the FiQT between localized quantum objects in a field vacuum with one party or both accelerated would be degraded by the Unruh effect because of the thermality appearing in these accelerated objects \cite{AM03}. However the more ubiquitous relativistic effects in inertial frames such as time dilation and Doppler shift (related to the relative speed) that are mixed in with effects due to acceleration have not been understood fully in the context of QT. These effects and their interplay are the focus of this study. What we found that may be surprising is that the relativistic effects in affecting the description of the dynamics can overwhelm 
the Unruh effect. For example, there is degradation of fidelity when both parties are inertial, as shown in our case 1, and a larger acceleration does not always lead to a faster degradation, as shown in our cases 2 and 3.

The averaged FiQT in cases 2, 3, and 4 do depend on the proper acceleration $a$ in Bob's acceleration phase significantly.
In case 2, we find that the larger $a$ is, the higher the degradation rate will be in the sender's clock for
the best averaged fidelities $F_{av}^+$ of QT both from Alice to Rob and from Rob to Alice.
Nevertheless, the increasing redshift as the retarded distance between Alice and Rob increasing indefinitely in time is the key factor
for the $a$ dependence here. In the receiver's point of view, that the degradation rate increases as $a$ increases,  is true only for a
receiver accelerated with proper acceleration large enough, when the Unruh effect fully dominates.
In case 3 a larger $a$ turns out to give a longer disentanglement time of the EnLC in the clock of the sender Alice at rest.
The key factor there is that detector $B$ with the traveling twin Bob ages much slower than detector $A$ with Alice at rest when they
compare their clocks at the same place after Bob rejoins Alice. The acceleration of Bob leads to this asymmetry
of time flows as in the well-known twin paradox and Bob's slower clock helps to keep the freshness of quantum coherence in the
$AB$ pair longer from the view of Alice's clock, while the retarded distance between Alice and Bob is bounded from above.

To suppress the relativistic effects in what is observed by Alice, who is always at rest, we considered case 4 in which Bob is undergoing an alternating uniform acceleration with a small speed and a constant averaged retarded distance. The results indeed show that the larger the $a$,  the shorter the disentanglement time for EnLC,  even in Alice's point of view when $a$ is large enough, although
the Unruh temperature is not well defined in this setup for the lack of a sufficiently long duration of uniform acceleration.
\\

\begin{acknowledgments}
We thank Kazutomu Shiokawa for very helpful input in the early stage and for his collaboration in the earlier version of this work
\cite{LSCH12}. S. Y. L. thanks Tim Ralph for useful discussions.
Part of this work was done while B. L. H. visited the National Center for Theoretical Sciences (South) and the Department of Physics of
National Cheng Kung University, Taiwan, the Center for Quantum Information and Security at Macquarie University,
the Center for Quantum Information and Technology at the University of Queensland, Australia, in January--March, 2011
and the National Changhua University of Education, Taiwan, in January 2012. He wishes to thank the hosts of these institutions
for their warm hospitality. This work is supported by the Ministry of Science and Technology of Taiwan under Grants
No. MOST 102-2112-M-018-005-MY3 and No. MOST 103-2918-I-018-004 and in part by the National Center for Theoretical Sciences, Taiwan,
and by USA NSF PHY-0801368 to the University of Maryland.
\end{acknowledgments}

\begin{appendix}

\section{Reduced state of a detector with its entangled partner being measured}
\label{MeasRedSt}

In our linear system the operators of the dynamical variables at some coordinate time $x^0=T$ of an observer's frame after
the initial moment $T_0$ are linear combinations of the operators defined at the initial moment \cite{LH06}:
\begin{eqnarray}
  & &\hat{Q}^{}_{\bf d}(\tau_{}^{\bf d}(T)) = 
    \sum_{{\bf d}'}\left[\phi^{{\bf d}'}_{\bf d}(\tau_{}^{\bf d})\hat{Q}^{[0]}_{{\bf d}'} +
    f^{{\bf d}'}_{\bf d}(\tau_{}^{\bf d})\hat{P}^{[0]}_{{\bf d}'} \right] +
    \int d^3y \left[ \phi^{\bf y}_{\bf d}(\tau_{}^{\bf d})\hat{\Phi}^{[0]}_{\bf y} +
    f^{\bf y}_{\bf d}(\tau_{}^{\bf d})\hat{\Pi}^{[0]}_{\bf y} \right], \label{Qexp}\\
  & &\hat{\Phi}^{}_{\bf x}(T) = 
    \sum_{{\bf d}'} \left[\phi^{{\bf d}'}_{\bf x}(T)\hat{Q}^{[0]}_{{\bf d}'} +
    f^{{\bf d}'}_{\bf x}(T)\hat{P}^{[0]}_{{\bf d}'}\right] + \int d^3y \left[
    \phi^{\bf y}_{\bf x}(T)\hat{\Phi}^{[0]}_{\bf y} + f^{\bf y}_{\bf x}(T)\hat{\Pi}^{[0]}_{\bf y} \right], \label{Phiexp}
\end{eqnarray}
from which the conjugate momenta $\hat{P}^{}_{\bf d}(T)$ and $\hat{\Pi}^{}_{\bf x}(T)$ to $\hat{Q}^{}_{\bf d}(T)$ and
$\hat{\Phi}^{}_{\bf x}(T)$, respectively, can be derived according to the action (\ref{Stot1}).
Here we denote $\hat{\cal O}_{\zeta}^{[n]} \equiv \hat{\cal O}_{\zeta}(T_n)$ (e.g., $\hat{\Phi}_{\bf y}^{[n]}\equiv \hat{\Phi}
(T_n, {\bf y})$ and $\hat{\Pi}_{\bf y}^{[n]}\equiv \hat{\Pi}(T_n, {\bf y})$), and
all the ``mode functions" $\phi^{\zeta}_\xi(T)$ and $f^{\zeta}_\xi(T)$ are real functions of time
($\zeta, \xi, \nu \in \{A, B, C\}\cup \{{\bf x}\}$, ${\bf x}\in {\bf R}^3$ in (3+1)-dimensional Minkowski space),
which can be related to those in $k$ space in Ref. \cite{LH06}.
Then from Eqs. $(\ref{Qexp})$ and $(\ref{Phiexp})$, those correlators in Eqs. (\ref{rhoABC})--(\ref{Rdd1T})
can be expressed as combinations of
the mode functions and the initial data, e.g.,
\begin{eqnarray}
  \langle \hat{Q}_A^2(\tau^{}_A)\rangle &=&
  \phi_A^A(\tau^{}_A)\phi_A^A(\tau^{}_A)\langle (\hat{Q}_A^{[0]})^2\rangle^{}_0 +\nonumber\\ & &
   \int d^3x d^3y\, \phi_A^{\bf x}(\tau^{}_A)\phi_A^{\bf y}(\tau^{}_A)\langle \hat{\Phi}_{\bf x}^{[0]},\hat{\Phi}_{\bf y}^{[0]}
   \rangle^{}_0 + \ldots,
\label{QA2examp}
\end{eqnarray}
where $\langle \cdots \rangle^{}_n$ denotes that the expectation values are taken from the quantum state right after $x^0=T_n$.

Comparing the expansions $(\ref{Qexp})$ and $(\ref{Phiexp})$ of two equivalent continuous evolutions, one from
$x^0= T_0$ to $x^0=T_1$ then from $x^0=T_1$ to $x^0=T_2$ and the other from $x^0=T_0$ all the way to
$x^0=T_2$, one can see that the mode functions have the identities,
\begin{eqnarray}
  \phi^{\zeta[20]}_\xi  &=& \sum_{{\bf d}'}\left[\phi_\xi^{{\bf d}'[21]}\phi_{{\bf d}'}^{\zeta[10]} +
    f_\xi^{{\bf d}'[21]} \pi^{\zeta[10]}_{{\bf d}'}\right] +
    \int d^3x' \left[ \phi^{{\bf x'}[21]}_\xi \phi^{\zeta[10]}_{\bf x'} + f^{{\bf x'}[21]}_\xi \pi^{\zeta[10]}_{\bf x'}\right]
  \nonumber\\
  &\equiv& \phi_\xi^{\nu[21]}\phi_{\nu}^{\zeta[10]} + f_\xi^{\nu[21]} \pi^{\zeta[10]}_{\nu}, \label{id1}\\
  f^{\zeta[20]}_\xi  &=& \phi_\xi^{\nu[21]}f_{\nu}^{\zeta[10]} + f_\xi^{\nu[21]} p^{\zeta[10]}_{\nu},
  \label{id2}
\end{eqnarray}
where the DeWitt--Einstein notation with $\nu \in \{A, B, C\}\cup \{{\bf x}\}$ is understood, $F^{[mn]} \equiv F(T_m-T_n)$, and
$\pi^{\zeta}_{\bf d}(\tau_{}^{\bf d}(T)) \equiv \partial^{}_{\bf d}\phi^{\zeta}_{{\bf d}}(\tau_{}^{\bf d}(T))$,
$\pi^{\zeta}_{\bf x}(T) \equiv \partial^{}_0 \phi^{\zeta}_{\bf x}(T)$,
$p^{\zeta}_{\bf d}(\tau_{}^{\bf d}(T)) \equiv \partial^{}_{\bf d}f^{\zeta}_{{\bf d}}(\tau_{}^{\bf d}(T))$, and
$p^{\zeta}_{\bf x}(T) \equiv \partial^{}_0 f^{\zeta}_{\bf x}(T)$ in the momentum operators.
Similar identities for $\pi^\zeta_\xi$ and $p^\zeta_\xi$ can be derived straightforwardly from Eqs. $(\ref{id1})$ and $(\ref{id2})$.
Such identities can be interpreted as embodying the Huygens principle of the mode functions
and can be verified by inserting particular solutions of the mode functions into the identities.

In Ref. \cite{Lin11a} one of us has explicitly shown that in a Raine--Sciama--Grove detector--field system in (1+1)-dimensional Minkowski space, quantum states in different frames, starting with the same initial state defined on the same fiducial time slice and then collapsed by the same {\it spatially local} measurement on the detector at some moment, evolve to the same quantum state on the same final time slice (up to a coordinate transformation), no matter which frame is used by the observer or which time slice is the wave functional collapsed on between the initial and the final time slices. This implies that the reduced state of detector $B$ at the final time is coordinate independent even in the presence of spatially local projective measurements.
For the Unruh--DeWitt detector theory in (3+1)-dimensional Minkowski space considered here, the argument is similar, as follows.

Right after the local measurement on detectors $A$ and $C$ at $T^{}_1$ (for a simpler case with the local measurement
only on detector $A$, see Ref. \cite{Lin11b}), the quantum state at $T^{}_1$ 
collapses to $\tilde{\rho}^{}_{AC} \otimes \tilde{\rho}^{}_{B\Phi_{\bf x}}$ on the $T_1$ slice 
of the observer's frame. Similar to Eq. $(\ref{rhoB})$, here $\tilde{\rho}^{}_{B\Phi_{\bf x}}$ for detector $B$ and the field 
$\Phi_{\bf x}$ in the postmeasurement state is obtained by
\begin{equation}
  \tilde{\rho}^{}_{B\Phi_{\bf x}}(K^{\bar{\sigma}},\Delta^{\bar{\sigma}}) = N \int
	 {d^2 {\cal K}^A \over 2\pi\hbar} {d^2 {\cal K}^C \over 2\pi\hbar}
  \tilde{\rho}_{AC}^* ({\cal K}^A, {\cal K}^C) \rho({\cal K}^{\bf d}, {\cal K}^{\bf x}; T_1)
\label{rhoBPhi0}
\end{equation}
where $\rho$ is the quantum state of the combined system evolved from $T_0$ 
to $T_1$ and ${\bar{\sigma}}\in \{ B\}\cup \{ {\bf x} \}$. Since $\tilde{\rho}^{}_{AC}$ is Gaussian, a straightforward
calculation shows that $\tilde{\rho}^{}_{B\Phi_{\bf x}}$ 
has the form
\begin{eqnarray}
  & &\tilde{\rho}^{}_{B\Phi_{\bf x}}(K^{\bar\sigma},\Delta^{\bar{\sigma}}) =\nonumber\\ & &
  \exp \left[ {i\over \hbar} \left( {\cal J}^{(0)}_{\bar{\zeta}} K^{\bar{\zeta}} -{\cal M}^{(0)}_{\bar{\zeta}}\Delta^{\bar{\zeta}}\right)
  -{1\over 2\hbar^2} \left( K^{\bar{\zeta}} {\cal Q}_{{\bar{\zeta}}{\bar{\xi}}} K^{\bar{\xi}} +
  \Delta^{\bar{\zeta}} {\cal P}_{{\bar{\zeta}}{\bar{\xi}}} \Delta^{\bar{\xi}}-
  2 K^{\bar{\zeta}} {\cal R}_{{\bar{\zeta}}{\bar{\xi}}} \Delta^{\bar{\xi}} \right) \right.\nonumber\\
  & & \hspace{.5cm} \left.  +{1\over 2\hbar^2}\sum_{n=1}^4 {1\over {\cal W}^{(n)}}
  \left( K^{\bar{\zeta}}{\cal J}^{(n)}_{\bar{\zeta}}- \Delta^{\bar{\zeta}}{\cal M}^{(n)}_{\bar{\zeta}} \right)
  \left( {\cal J}^{(n)}_{\bar{\xi}} K^{\bar{\xi}}- {\cal M}^{(n)}_{\bar{\xi}}\Delta^{\bar{\xi}}\right) \right].
\label{rhoBPhi}
\end{eqnarray}
Again we use the DeWitt--Einstein notation for $\bar{\zeta}, \bar{\xi} \in\{ B\}\cup \{ {\bf x} \} $, which run
over the degrees of freedom of detector $B$ and the field defined at ${\bf x}$ on the whole time slice.
$n$ running from $1$ to $4$ corresponds to the four-dimensional Gaussian integrals in Eq. $(\ref{rhoBPhi0})$.
${\cal W}^{(n)}$ depends only on the two-point correlators of detectors $A$ and $C$ at the moment of measurement,
while ${\cal J}^{(n)}_{\bar{\zeta}}(\hat{\Phi}_{\bar{\zeta}})$ and ${\cal M}^{(n)}_{\bar{\zeta}}(\hat{\Pi}_{\bar{\zeta}})$ are linear
combinations of the terms with a cross-correlator between detector $A$ or $C$ and the operators
$\hat{\Phi}_{\bar{\zeta}}$ or $\hat{\Pi}_{\bar{\zeta}}$ ($\hat{\Phi}_{B}\equiv \hat{Q}_B$ and $\hat{\Pi}_{B}\equiv \hat{P}_B$),
respectively, multiplied by a few correlators of $A$ and/or $C$,
all of which are the correlators of the operators evolved from $T_0$ to $T_1$ with respect to the initial state given at $T_0$.
This implies that the two-point correlators right after the wave functional collapse become
\begin{eqnarray}
  \langle \delta\hat{\Phi}^{[1]}_{\bar{\zeta}},\delta\hat{\Phi}^{[1]}_{\bar{\xi}}\rangle^{}_1
    &=& \langle\delta\hat{\Phi}^{[10]}_{\bar{\zeta}}, \delta\hat{\Phi}^{[10]}_{\bar{\xi}}\rangle^{}_0 -
    \sum_{n=1}^4  {{\cal J}^{(n)}_{\bar{\zeta}}(\hat{\Phi}^{[10]}_{\bar{\zeta}})
    {\cal J}^{(n)}_{\bar{\xi}}(\hat{\Phi}^{[10]}_{\bar{\xi}}) \over {\cal W}^{(n)}}, \label{Q2PM} \\
  \langle\delta \hat{\Pi}^{[1]}_{\bar{\zeta}},\delta\hat{\Pi}^{[1]}_{\bar{\xi}}\rangle^{}_1
    &=& \langle\delta\hat{\Pi}^{[10]}_{\bar{\zeta}}, \delta\hat{\Pi}^{[10]}_{\bar{\xi}}\rangle^{}_0 -
    \sum_{n=1}^4  {{\cal M}^{(n)}_{\bar{\zeta}}(\hat{\Pi}^{[10]}_{\bar{\zeta}})
    {\cal M}^{(n)}_{\bar{\xi}}(\hat{\Pi}^{[10]}_{\bar{\xi}}) \over {\cal W}^{(n)}}, \label{P2PM}\\
  \langle \delta\hat{\Phi}^{[1]}_{\bar{\zeta}},\delta\hat{\Pi}^{[1]}_{\bar{\xi}}\rangle^{}_1
    &=& \langle\delta\hat{\Phi}^{[10]}_{\bar{\zeta}}, \delta\hat{\Pi}^{[10]}_{\bar{\xi}}\rangle^{}_0 -
    \sum_{n=1}^4  {{\cal J}^{(n)}_{\bar{\zeta}}(\hat{\Phi}^{[10]}_{\bar{\zeta}})
    {\cal M}^{(n)}_{\bar{\xi}}(\hat{\Pi}^{[10]}_{\bar{\xi}}) \over {\cal W}^{(n)}}. \label{PQPM}
\end{eqnarray}
For example,
$\langle(\delta\hat{Q}_B^{[1]})^2\rangle^{}_1= {\cal Q}_{BB}(T_1)  -
\sum_{n=1}^4 [{\cal J}^{(n)}_{B}(\hat{Q}_B^{[10]}) {\cal J}^{(n)}_{B}(\hat{Q}_B^{[10]})/{\cal W}^{(n)}]$
where ${\cal Q}_{BB}(T_1) =\langle(\delta\hat{Q}_B^{[10]})^2\rangle^{}_0$.
Here $\hat{\cal O}_B^{[1]}$ refers to the operator $\hat{\cal O}_B$ defined at $T_1$ and
$\hat{\cal O}_B^{[10]}$ refers to the operator $\hat{\cal O}_B(T_1-T_0)$ in the Heisenberg picture.

Suppose the future and past light cones of the measurement event by Alice at $x^0=T_1$ crosses the worldline of Bob at
his proper times $\tau^{adv}_1$ and $\tau^{ret}_1$, respectively. At some moment 
in the coordinate time $x^0 = T^{}_M$ of the observer's frame before detector $B$ enters the future light cone of the measurement
event on detector $A$, namely, when Bob's proper time $\tau_{}^B= 
\tau(T^{}_M) \in (\tau^{ret}_1, \tau^{adv}_1)$, the two-point correlators of detector $B$ are either
in the original, uncollapsed form, e.g., $\langle (\delta\hat{Q}_B)^2(T_M-T_0) \rangle^{}_0$, if the wave functional collapse does not
happen yet in some observers' frames, or in the collapsed form evolved from the postmeasurement state, e.g.,
\begin{eqnarray}
  & & \langle (\delta\hat{Q}_B)^2(T^{}_M) \rangle = -\left(\langle \hat{Q}_B^{[M1]} \rangle_1^{}\right)^2 + \nonumber\\ & &
  \left< \left[ \sum_{{\bf d}} \left(\phi_B^{{\bf d}[M1]}\hat{Q}_{\bf d}^{[1]}+
  f_B^{{\bf d}[M1]}\hat{P}_{\bf d}^{[1]}\right)
  +\int dx \left(\phi_B^{x[M1]}\hat{\Phi}_x^{[1]}+f_B^{x[M1]}\hat{\Pi}_x^{[1]}\right) \right]^2 \right>_1
  \nonumber\\
 & & = \langle (\hat{\Upsilon}_B^{[M0]})^2 \left. \right>_0 -\sum_{n=1}^4
  {{\cal I}^{(n)} [\hat{\Upsilon}_B^{[M0]}, \hat{\Upsilon}_B^{[M0]} ]\over {\cal W}^{(n)}} ,
\label{QB2clpsed}
\end{eqnarray}
in other observers' frames. Here we have used the Huygens principles $(\ref{id1})$ and $(\ref{id2})$ and defined
\begin{eqnarray}
  \hat{\Upsilon}^{[M0]}_B &\equiv& \hat{\Phi}_\zeta^{[0]}\left[\phi_B^{\zeta[M0]}-\phi_B^{A[M1]}\phi_A^{\zeta[10]}-
  f_B^{A[M1]}\pi_A^{\zeta[10]}\right] + \nonumber\\ & & \hat{\Pi}_\zeta^{[0]}\left[f_B^{\zeta[M0]}-\phi_B^{A[M1]}f_A^{\zeta[10]}-
  f_B^{A[M1]}p_A^{\zeta[10]}\right]
\label{Updef}
\end{eqnarray}
with $\hat{\Phi}_{A,C}\equiv \hat{Q}_{A,C}$ and $\hat{\Pi}_{A,C}\equiv \hat{P}_{A,C}$, while ${\cal I}^{(n)}$ is derived from
those ${\cal J}^{(n)}_{\bar{\zeta}}$ and ${\cal J}^{(n)}_{\bar{\xi}}$  pairs in Eqs. $(\ref{Q2PM})$-$(\ref{PQPM})$.
Note that before  detector $B$ enters the light cone, one has $\phi_B^{A[M1]} = f_B^{A[M1]}=0$, such that $\hat{\Upsilon}^{[M0]}_B$
reduces to $\hat{Q}^{[M0]}_B$. So at the moment $T_M$, the correlators of detector $B$ do not depend on the data on the $T_1$ slice
except those right at the local measurement event on detectors $A$ and $C$. This means that, once we discover the reduced state of
detector $B$ has been collapsed, the form of the reduced state of $B$ will be independent of the moment when the collapse occurs
in the history of detector $B$ (e.g., $\tau_{}^B=\tau_1^B$ or ${\tau'}^B_1$ in Fig. \ref{IHO2}), 
namely, the moment at which the worldline of detector $B$ intersects the time slice that the wave functional collapsed on.

No matter in which frame the system is observed, the correlators in the reduced state of detector $B$ must have become
the collapsed ones like Eq. $(\ref{QB2clpsed})$ exactly when detector $B$ was entering the future light cone of
the measurement event by Alice, namely, $\tau_{}^B = \tau^{adv}_1$, 
after which the reduced states of detector $B$ observed in different frames became consistent. Also after this moment,
the retarded mutual influences reached $B$ such that $\phi_B^{A[M1]}$ and $f_B^{A[M1]}$ would become nonzero and
get involved in the correlators of $B$. In fact, some information of measurement had entered the correlators of $B$ via the
correlators of $A$ and $C$ at $t_1$ at the position of Alice in ${\cal J}^{(n)}$, ${\cal M}^{(n)}$ and ${\cal W}^{(n)}$ much earlier.
Nevertheless, just like what we learned in QT, that information was protected by the randomness of measurement
outcome and could not be recognized by Bob before he has causal contact with Alice.

Thus, we are allowed to choose a coordinate system with the $T^{}_M$ in Eq. $(\ref{QB2clpsed})$ giving $\tau_{}^B(T_M)=\tau^{adv}_1-\epsilon$, $\epsilon\to 0+$ and to collapse or project the wave functional right before $T^{}_M$, namely, collapse on a time slice almost overlapping the future light cone of the measurement event by Alice. It is guaranteed that there exists some coordinate system having such a spacelike hypersurface that intersects the worldline of Alice at $\tau_{}^A(T_1)$ and the worldline of Bob at $\tau_{}^B =
\tau^{adv}_1-\epsilon$ in a relativistic system.

If we further assume that the mutual influences are nonsingular and Bob performs the local operation right after the classical
information from Alice is received, namely, at $\tau_{}^B = \tau^{adv}_1+\epsilon$ with $\epsilon\to 0+$, then the continuous
evolution of the reduced state of detector $B$ from $\tau_{}^B(T_M)=\tau^{adv}_1-\epsilon$ to $\tau^{adv}_1+\epsilon$
is negligible. In this case we can calculate the best averaged FiQT using Eq. $(\ref{Favformula})$.

\end{appendix}

\hspace{2cm}

\end{document}